\documentclass[12pt,a4paper]{article}
\usepackage{graphicx}
\usepackage{amssymb}
\usepackage{amsmath}
\usepackage{bm}
\usepackage{color}
\usepackage{theorem}
\usepackage{subfigure}
\usepackage{caption}

\usepackage[sort&compress,numbers, merge]{natbib}

\definecolor{Orange}{cmyk}{0,0.61,0.87,0}
\definecolor{JungleGreen}{cmyk}{0.99,0,0.52,0}
\definecolor{OliveGreen}{cmyk}{0.64,0,0.95,0.40}
\definecolor{Brown}{cmyk}{0,0.81,1,0.60}
\definecolor{RoyalBlue}{cmyk}{0.71,0.53,0,0.12}

\newcommand{\eq}[1]{Eq.~\eqref{#1}}
\newcommand{\DRbar}{\ensuremath{\overline{\rm DR}}}

\allowdisplaybreaks[1]

\usepackage{hyperref}

\begin{document}

\begin{titlepage}
\pagestyle{empty}
{\tt
\rightline{KCL-PH-TH/2019-91, CERN-TH-2019-216}
\rightline{UT-19-30, UMN-TH-3907/19, FTPI-MINN-19/28}
}
\vspace{1cm}
\begin{center}
{\bf {\LARGE
Supersymmetric Proton Decay Revisited} }
\end{center}

\vspace{0.05cm}
\begin{center}
{\bf John~Ellis}~$^{1}$,
{\bf Jason~L.~Evans}~$^2$,
 {\bf Natsumi~Nagata}~$^3$\\
\vskip 0.1in
{\bf Keith~A.~Olive}~$^{4}$ and {\bf Liliana~Velasco-Sevilla}~$^5$
\vskip 0.2in
{\small {\it
$^1${Theoretical Particle Physics and Cosmology Group, Department of
  Physics, King's~College~London, London WC2R 2LS, United Kingdom;\\
Theoretical Physics Department, CERN, CH-1211 Geneva 23,
  Switzerland;\\
  National Institute of Chemical Physics \& Biophysics, R{\" a}vala 10, 10143 Tallinn, Estonia}\\
\vspace{0.2cm}
$^2${School of Physics, KIAS, Seoul 130-722, Korea;\\
T. D. Lee Institute, Shanghai 200240, China}\\
\vspace{0.2cm}
$^3${Department of Physics, University of Tokyo, 
Tokyo 113--0033,
Japan}\\
\vspace{0.2cm}
$^4${William I.\ Fine Theoretical Physics Institute, School of Physics and
 Astronomy, University of Minnesota, Minneapolis, Minnesota 55455, USA}\\
 \vspace{0.2cm}
$^5${\it University of Bergen, Department of Physics and Technology,\\
PO Box 7803, 5020 Bergen, Norway}\\
}}

\vspace{0.5cm}
{\bf Abstract}\\

\end{center}
{Encouraged by the advent of a new generation of underground detectors---JUNO, DUNE and Hyper-Kamiokande---that are projected to improve significantly on the present sensitivities to various baryon decay modes, we revisit baryon decay in the minimal supersymmetric SU(5) GUT.
We discuss the phenomenological uncertainties associated with hadronic matrix elements and the value of the strong coupling $\alpha_s$---which are the most important---the weak mixing angle $\theta_W$, quark masses including one-loop renormalization effects, quark mixing and novel GUT phases that are not visible in electroweak interaction processes. We apply our analysis to a variety of CMSSM, super- and sub-GUT scenarios in which soft supersymmetry-breaking parameters are assumed to be universal at, above and below the GUT scale, respectively. In many cases, we find that the next generation of underground detectors should be able to probe models with sparticle masses that are ${\cal O}(10)$~TeV, beyond the reach of the LHC.}


\vfill
\leftline{December 2019}
\end{titlepage}

\section{Introduction}

The time is ripe to review and re-evaluate the prospects for observing proton decay in supersymmetric models.

On the one hand, we anticipate that a new generation of large underground neutrino detectors will be constructed
in the coming years~\cite{JUNO,DUNECDR, HKTDR}, with heightened sensitivity to a wide range of possible proton
(and neutron) decay modes, including many suggested by minimal supersymmetric grand unified theory (GUT) models. As reviewed in~\cite{Kobayashi} and discussed in more detail in Section~\ref{sec:BDKexp}, the JUNO experiment is expected to have 90\% CL sensitivity to the $p \to K^+ \overline{\nu}$ decay mode at the level of a lifetime $> 2 \times 10^{34}$~yr after operating for 10 years---see Section~10 of~\cite{JUNO},
the DUNE experiment is expected to have 90\% CL exclusion sensitivity to
$p \to K^+ \overline{\nu}$ at the level of $\sim 6.5 \times 10^{34}$~yr after 20 years---see Chapter~4 of Volume~2 of the
DUNE CDR~\cite{DUNECDR}, and the Hyper-Kamiokande (Hyper-K) experiment is expected to have 90\% CL exclusion sensitivity to
$p \to K^+ \overline{\nu}$ at the level of $\sim 5 \times 10^{34}$~yr after 20 years--- see Section~III.2 of the Hyper-Kamiokande TDR~\cite{HKTDR}.
These sensitivities reach around an order of magnitude better than the current experimental limit \cite{Abe:2014mwa, Takhistov:2016eqm}. 

On the other hand,
searches at the LHC have pushed up the experimental lower limits on the masses of many sparticle species \cite{nosusy}. For example, searches for  strongly-interacting sparticles by the ATLAS and CMS Collaborations have excluded
them in mass ranges to $\sim 2$~TeV or more~\cite{nosusy2} when their data are interpreted in simplified supersymmetric models.
In parallel,
direct searches have improved upper limits on dark matter scattering cross sections \cite{LUX,PANDAX,XENON},
at least in models with simplifying assumptions about the input soft supersymmetry-breaking masses. These
developments have two relevant implications: the absence of direct evidence of supersymmetry gives
added motivation to indirect searches, e.g., via proton decay, and proton decay matrix elements are generically
more suppressed if the sparticle spectrum is heavier.

When interpreting these broad changes in the experimental perspectives for detecting supersymmetric proton decay, there are several
phenomenological considerations that need to be taken into account in calculating the proton decay rate in any
specific supersymmetric scenario, as we discus in this paper. Although our analysis is in the context of one specific model, namely the minimal superysmmetric SU(5) GUT, we expect our considerations to have broad applications to other models. First of all, we provide a treatment of the uncertainties in the hadronic matrix elements of the
relevant nucleon decay operators, for which we use updated lattice calculations~\cite{AISS}. Secondly, we update the GUT matching of couplings, combining the current best estimate of $\alpha_s(M_Z)$ and
the latest experimental estimate of $\sin^2 \theta_W$. As we discuss in more detail below, the masses of the
coloured GUT Higgses, $H_C$, are particularly sensitive to the value of $\sin^2 \theta_W$, and the current value leads to a smaller value of $M_{H_C}$ than some previous estimates, accelerating proton decay. We also consider the impacts of uncertainties in quark masses, and stress the importance of
including one-loop mass-renormalization effects, which have an effect that is not negligible. We also review the implications of the present uncertainties in
quark-mixing effects. The conventional Cabibbo-Kobayashi-Maskawa (CKM) matrix elements have experimental 
uncertainties that are small enough to be relatively unimportant for our purposes. However, we recall that in even the minimal
supersymmetric GUT there are two additional CP-violating phases that are unconstrained by electroweak-scale
measurements~\cite{Ellis:1979hy}. These can have significant effects on the magnitudes of the nucleon decay matrix elements, possibly even
altering the expected hierarchy of proton decay modes \cite{eemno}.

The layout of this paper is as follows. After our discussion of the present and prospective experimental sensitivities to various possible
nucleon decay modes in Section~\ref{sec:BDKexp}, we
present a review of the minimal supersymmetric SU(5) model in Section~\ref{sec:modelbasics},
followed by a discussion of the basic elements in our proton decay calculation in Section~\ref{sec:decaybasics}. In Section~\ref{sec:uncertainties} we discuss various
phenomenological aspects of proton decay calculations and the relative uncertainties they introduce into the calculation of the proton lifetime. 

We begin Section~\ref{sec:uncertainties} with discussions of the dominant uncertainties originating in the hadronic matrix elements 
that feature in different GUT models for nucleon decay~\cite{AISS},
and in the experimental input value of $\alpha_s$, in Sections~\ref{sec:ME} and \ref{sec:alphas} respectively.
The latter strongly affects the mass of the colour-triplet Higgs, whose strong sensitivity to $\sin^2 \theta_W$ through the GUT matching conditions is discussed in
Section~\ref{sec:sin2thetaW}.
The sensitivities to the second-generation quark
masses, $m_s$ and $m_c$, are considered in Section~\ref{sec:mq}, and
we discuss the effects of their one-loop mass renormalizations in Section~\ref{sec:charm}.
Then, in Section~\ref{sec:mixing} we review the status
of quark mixing, and emphasize in Section~\ref{sec:GUTphi} the important uncertainties associated with the additional CP-violating phases that appear
even in the minimal SU(5) GUT. Possible implications of 
the non-unification of Yukawa couplings in this model are addressed in Section~\ref{sec:yukawauni}.

We apply these considerations to various GUT models in Section~\ref{sec:results}, including the constrained minimal supersymmetric Standard Model (CMSSM)
in which the soft supersymm- etry-breaking sfermion and gaugino masses are assumed to be universal at the GUT scale in
Section~\ref{sec:CMSSM} \cite{cmssm, ELOS, eelnos,eemno,eeloz, ehow++}, super-GUT models in which universality is imposed at a scale higher than the GUT scale \cite{super-GUT,emo,eemno} in
Section~\ref{sec:super-GUT}, and sub-GUT models in which universality is imposed a lower scale
\cite{sub-GUT,ELOS,eelnos,eeloz,mcsub-GUT} in Section~\ref{sec:sub-GUT}.

Finally, Section~\ref{sec:sowhat} summarizes our analysis and discusses our conclusions. The main part of our analysis concentrates on proton decay mediated by
dimension-five operators, but we add for completeness an Appendix concerning the dimension-six operators that mediate $p \to \pi^0 e^+$.

\section{Baryon Decay: Experimental Status and Prospects}
\label{sec:BDKexp}

The current lower limits on the lifetimes for many possible proton and neutron decay modes provided by the Super-Kamiokande experiment are summarized in Refs.~\cite{HKTDR, Takhistov:2016eqm, Babu:2013jba, PDG}. As we review below, proton decay in the minimal supersymmetric SU(5) GUT is predominantly induced by the dimension-five effective operators \cite{Sakai:1981pk} generated by color-triplet Higgs exchange. In this case, the most important decay mode is generally $p \to K^+ \overline{\nu}$, for which the current lower decay lifetime limit is $6.6 \times 10^{33}$~yr \cite{Abe:2014mwa, Takhistov:2016eqm}.~\footnote{A preliminary update to the limit on this decay channel is reported in \cite{Tanaka}: $\tau (p \to K^+ \overline{\nu}) > 8.2 \times 10^{33}$~yr.} In the models we study, the decay $n \to K^0 \overline{\nu}$ is expected to occur at a similar rate, as these two decay channels are related through an SU(2) isospin rotation. However, the current lifetime limit on this decay mode is much weaker, $\tau(n \to K^0 \overline{\nu}) > 2.6 \times 10^{32}$~yr \cite{PDG, Kobayashi:2005pe}, and we expect the future sensitivity also to be much less than that for $p \to K^+ \overline{\nu}$. As we discuss later, there are (small) regions of parameter space where the
decay $p \to \pi^+ \overline{\nu}$ may occur at a rate comparable to $p \to K^+ \overline{\nu}$, but the current sensitivity to this decay mode, $\tau (p \to \pi^+ \overline{\nu}) > 3.9 \times 10^{32}$~yr \cite{Abe:2013lua}, is also considerably less than that for $p \to K^+ \overline{\nu}$. Although there have been no dedicated studies of the sensitivities of future detectors to this decay mode, we expect them also to be significantly weaker than those to $p \to K^+ \overline{\nu}$, and hence uncompetitive.

There are other decay channels that are suppressed in our analysis but can be sizeable in other situations. Among them, $p \to e^+ \pi^0$ is of great importance as this decay mode is generally present in GUTs, particularly non-supersymmetric ones (see, for instance, Ref.~\cite{Mambrini:2015vna}). This decay mode can dominate over $p \to K^+ \overline{\nu}$ even in supersymmetric GUT models if the supersymmetric particles are very heavy (see, e.g., Refs.~\cite{Ellis:2018khn, Evans:2019oyw}) or if the dimension-five proton decay operators are suppressed by some symmetry, as is the case in flipped ${\rm SU}(5) \times {\rm U}(1)$ GUT models \cite{Barr,DKN,flipped2,AEHN, Ellis:2017jcp,shafi1}.~\footnote{There is a short discussion of $p \to e^+ \pi^0$ in the Appendix.} In addition, the rate of this decay channel is enhanced if there exist vector-like SU(5) multiplets below the GUT scale \cite{Hisano:2012wq}. The current limit on this decay mode is 
$\tau (p \to e^+ \pi^0) > 1.6 \times 10^{34}$~yr \cite{Miura:2016krn}.~\footnote{A recent update of this limit is reported in Ref.~\cite{Tanaka}: $\tau (p \to e^+ \pi^0) > 2.0 \times 10^{34}$~yr.}
Other decay modes including charged leptons, such as $p \to \mu^+ \pi^0$ and $n \to e^+/\mu^+ \pi^-$, are usually subdominant, but can be important if there is flavour violation in sfermion masses \cite{Nagata:2013sba} or if the GUT group is different from SU(5) \cite{Ellis:1988tx, Lucas:1996bc,Babu:1998ep,Maekawa:2016tqi, Buchmuller:2019ipg,shafi2}. The present limits on these decay modes are $\tau(p \to \mu^+ \pi^0) > 7.7 \times 10^{33}$~yr \cite{Miura:2016krn},~\footnote{A preliminary update, $\tau (p \to \mu^+ \pi^0) > 1.2 \times 10^{34}$~yr, is given in Ref.~\cite{Tanaka}.}
$\tau(n \to e^+ \pi^-) > 5.3 \times 10^{33}$~yr \cite{TheSuper-Kamiokande:2017tit}, and $\tau(n \to \mu^+ \pi^-) > 3.5 \times 10^{33}$~yr \cite{TheSuper-Kamiokande:2017tit}.  

In the second column in Table~\ref{tab:baryondecaylimits} we summarize the current 90\% CL limits on nucleon decay modes in units of $10^{33}$~yr. For other decay modes, see Refs.~\cite{HKTDR, Takhistov:2016eqm, Babu:2013jba, PDG}.

\begin{table}[!ht]
\centering
\captionsetup{justification=centering}
\caption{\it{Current baryon decay 90\% CL limits and prospective future 90\% CL limits and 3-$\sigma$ discovery sensitivities in units of $10^{33}$~yrs.  
For future prospects, we assume detector operations for 10 (20) years. 
}
}
\vspace{-7mm}
\label{tab:baryondecaylimits}
\begin{tabular}{|l| l| l|l|}
\multicolumn{4}{c}{}\\
\hline \hline
Decay Mode & Current (90\% CL) & Future (Discovery) & Future (90\% CL) \\
\hline
$p \to K^+ \bar{\nu}$ & 6.6 \cite{Takhistov:2016eqm}& JUNO: 12 (20)  \cite{HKTDR} & JUNO: 19 (40)  \cite{JUNO}   \\
& & DUNE: 30 (50) \cite{HKTDR} & DUNE: 33 (65) \cite{DUNECDR} \\
 &  & Hyper-K: 20 (30) \cite{HKTDR} & Hyper-K: 32 (50) \cite{HKTDR} \\
 \hline
  $p \to \pi^+ \bar{\nu}$ & 0.39 \cite{Abe:2013lua}&  & \\
 \hline
 $p \to e^+ \pi^0$ & 16 \cite{Miura:2016krn} & 
 DUNE: 15 (25) \cite{HKTDR}& DUNE: 20 (40)  \cite{HKTDR}\\
 & & Hyper-K: 63 (100) \cite{HKTDR} & Hyper-K: 78 (130) \cite{HKTDR}\\
 \hline
  $p \to \mu^+ \pi^0$ & 7.7 \cite{Miura:2016krn} & Hyper-K: 69 \cite{HKTDR} & Hyper-K: 77 \cite{HKTDR}\\
 \hline
 \hline
 $n \to K^0_S \bar{\nu}$ & 0.26 \cite{PDG}& & \\
 \hline 
 $n \to \pi^0 \bar{\nu}$ & 1.1 \cite{Abe:2013lua}& & \\
 \hline
 $n \to e^+ \pi^-$ & 5.3 \cite{TheSuper-Kamiokande:2017tit} &Hyper-K: 13 \cite{HKTDR} &Hyper-K: 20 \cite{HKTDR}\\
  \hline
 $n \to \mu^+ \pi^-$ & 3.5 \cite{TheSuper-Kamiokande:2017tit} &Hyper-K: 11 \cite{HKTDR} &Hyper-K: 18 \cite{HKTDR}\\
 \hline
\hline
\end{tabular}
\end{table}

JUNO is a liquid scintillator (linear alkylbenzene doped with 3~g/L 2,5-diphenyloxazole and 15~mg/L p-bis-(o-methylstyryl)-benzene) detector of 20 kton fiducial mass, located in Guangdong province, China \cite{JUNO}. This experiment was approved in 2013, construction started in 2015 \cite{Giaz:2018gdd}, and operation is expected to start in 2021 \cite{Miao}. The prospective 90\% CL limit on the $p \to K^+ \bar{\nu}$ channel in 10 years is estimated to be $\tau (p \to K^+ \bar{\nu}) > 1.9 \times 10^{34}$~yrs \cite{JUNO}, while the 3-$\sigma$ discovery reach is estimated in Ref.~\cite{HKTDR} to be $\sim 1.2 \times 10^{34}$~yrs in 10 years. If JUNO starts in 2021, its sensitivity is expected to exceed the Super-Kamiokande bound by $\sim 2026$ \cite{Miao}. 

There are several approved experiments with interesting prospects for detecting nucleon decay. Among them, the DUNE experiment \cite{DUNECDR} is expected to offer the best sensitivity to $p \to K^+ \bar{\nu}$. The DUNE far detector is a liquid-argon time-projection chamber (LArTPC) with a 40 kt fiducial mass, located $\sim 1.5$~km underground at the Sanford Underground Research Facility in South Dakota, USA. This type of detector is advantageous for identifying $K^+$, compared to water Cherenkov detectors such as Super-Kamiokande, since a slow $K^+$ produced in the decay of a proton has a high ionization density. This information, as well as the reconstruction of the decay products associated with the $K^+$, enables a LArTPC to identify the $K^+$ track with high efficiency and thus to have a good sensitivity to nucleon decay channels with a $K^+$ in the final state. The expected 90\% CL limit on the $p \to K^+ \bar{\nu}$ channel in 10 (20) years is $\sim 3.3 ~(6.5)\times 10^{34}$~yrs \cite{DUNECDR}, where the initial run with 10kt and adding another 10 kt each year for four years is assumed. The 3-$\sigma$ discovery reach of DUNE is estimated in Ref.~\cite{HKTDR} based on the expected signal efficiency and background rates given in Ref.~\cite{DUNECDR}: $\tau (p \to K^+ \bar{\nu}) \sim 3 ~(5) \times 10^{34}$~yrs in 10 (20) years.

Hyper-Kamiokande (Hyper-K) is a water Cherenkov detector with a fiducial mass of 187~kt \cite{HKTDR}. The candidate site of the Hyper-Kamiokande detector is in the Tochibora mine in Gifu prefecture, Japan, which is located 8~km south of Super-Kamiokande. It is challenging to detect the $K^+$ from nucleon decays in a water Cherenkov detector since its momentum is below the Cherenkov light threshold, and thus only the decay products of $K^+$ can be used for the identification of $K^+$. The expected 90\% CL limit on the $p \to K^+ \bar{\nu}$ channel in 10 (20) years is $\sim 3.2 ~(5)\times 10^{34}$~yrs \cite{HKTDR}; if a second tank is installed after 6 years, the reach becomes $\sim 4 ~(7)\times 10^{34}$~yrs in 10 (20) years. The 3-$\sigma$ discovery reach is $\sim 2 ~(3) \times 10^{34}$~yrs in 10 (20) years. We note that Hyper-K has not yet reported prospective sensitivities for $p,n\to \pi^{+,0} \bar \nu$.

In the third and fourth columns in Table~\ref{tab:baryondecaylimits}, we summarize the future 3-$\sigma$ discovery and 90\% CL limit sensitivities, respectively, for 10 (and, where available, 20) year operations of the future experiments. As can be seen, considerable improvements in sensitivities are expected for many decay channels, including $p \to K^+ \bar{\nu}$ in particular.

\section{Proton Decay Basics}
\label{sec:basics}

\subsection{Minimal Supersymmetric SU(5) GUT Model}
\label{sec:modelbasics}

Minimal supersymmetric SU(5)~\cite{Dimopoulos:1981zb} is the
simplest supersymmetric extension of the original SU(5) GUT model \cite{Georgi:1974sy}. 
Matter superfields are embedded into
three sets of $\bar{\bf 5}\oplus {\bf 10}$ representations, $\Phi_i$ and $\Psi_i$, of the SU(5) gauge group, one per generation. 
The left-handed
down-type antiquark and the left-handed lepton chiral superfields, $\overline{D}_i$
and ${L}_i$, respectively, reside in ${\Phi}_i$, while the left-handed quark doublet, left-handed
up-type antiquark, and left-handed charged-lepton chiral superfields,
${Q}_i$, $\overline{U}_i$, and $\overline{E}_i$, respectively, are in the
${\Psi}_i$. Two chiral Higgs superfields
$H_u$ and $H_d$ belong to ${\bf 5}$ and $\overline{\bf 5}$
representations, $H$ and $\overline{H}$, respectively, which contain ${\bf 3}$ and $\overline{\bf 3}$ coloured Higgs superfields $H_C$ and $\overline{H}_C$, respectively.
Their vacuum expectation values (vevs) break the electroweak SU(2)$\times$U(1) gauge group down spontaneously to U(1)$_{\rm EM}$.
The SU(5) GUT gauge group is spontaneously broken by the vev of a ${\bf 24}$ chiral superfield,
$\Sigma \equiv \sqrt{2}\Sigma^A T^A$, 
to the Standard Model (SM) gauge group, where $T^A$ ($A=1, \dots, 24$) are the generators of SU(5) with ${\rm Tr}(T^A T^B) =
\delta_{AB}/2$. The full renormalizable superpotential of the minimal supersymmetric SU(5) GUT (assuming $R$-parity conservation) is:
\begin{align}
 W_5 &=  \mu_\Sigma {\rm Tr}\Sigma^2 + \frac{1}{6} \lambda^\prime {\rm
 Tr} \Sigma^3 + \mu_H \overline{H} H + \lambda \overline{H} \Sigma H
\nonumber \\
&+ \left(h_{\bf 10}\right)_{ij} \epsilon_{\alpha\beta\gamma\delta\zeta}
 \Psi_i^{\alpha\beta} \Psi^{\gamma\delta}_j H^\zeta +
 \left(h_{\overline{\bf 5}}\right)_{ij} \Psi_i^{\alpha\beta} \Phi_{j \alpha}
 \overline{H}_\beta ~,
\label{W5}
\end{align}
where Greek sub- and superscripts denote SU(5) indices, and $\epsilon_{\alpha\beta\gamma\delta\zeta}$ is the totally
antisymmetric tensor with $\epsilon_{12345}=+1$ .

The supersymmetry-preserving breaking of the SU(5) GUT gauge group to the SM SU(3)$\times$SU(2)$\times$U(1) gauge group occurs via a vev of the adjoint Higgs field $\Sigma$ in the
direction
\begin{equation}
 \langle \Sigma \rangle =V\cdot {\rm diag}(2,2,2,-3,-3)~,
\end{equation}
with $V = 4 \mu_\Sigma/\lambda^\prime$.
The GUT gauge bosons then acquire masses $M_X = 5 g_5 V$,
where $g_5$ is the SU(5) gauge coupling. Since the coloured triplets in the $H$ and $\overline{H}$ representations
are required to be heavy (as they mediate proton
decay), while the corresponding doublets need
to be light (they correspond to the MSSM
Higgs doublets), we must split the masses of the 
doublet and triplet, which requires a fine-tuning condition $\mu_H -3\lambda V \ll V$. In this case, the color-triplet Higgs
states have masses $M_{H_C} = 5\lambda V$.
In addition, the masses of the
color and weak adjoint components of $\Sigma$ are equal to $M_\Sigma =
5\lambda^\prime V/2$, while the singlet component of $\Sigma$ acquires a
mass $M_{\Sigma_{24}} = \lambda^\prime V/2$. 

The Yukawa couplings $\left(h_{\bf 10}\right)_{ij}$ and $\left(h_{\overline{\bf 5}}\right)_{ij}$ in
Eq.~(\ref{W5}) have redundant degrees of freedom, some
of which can be eliminated by field re-definitions of $\Psi_i$ and $\Phi_i$. The coupling $\left(h_{\bf 10}\right)_{ij}$ is a symmetric matrix, and thus has six independent complex components, while $\left(h_{\overline{\bf 5}}\right)_{ij}$ has nine. The field
redefinitions form an ${\rm U}(3)\otimes {\rm U}(3)$ transformation group, and thus the number of
physical degrees of freedom is $12+18- (9\times 2) =12$. Among
them, six correspond to quark mass eigenvalues and four to the CKM matrix
elements, so there are two additional GUT phases \cite{Ellis:1979hy}. In this
paper, we follow Ref.~\cite{Hisano:1992jj} by adopting the basis in which
\begin{align}
 \left(h_{\bf 10}\right)_{ij}&= e^{i\varphi_i} \delta_{ij}h_{{\bf 10}, i}  ~, \label{GUTphase} \\
 \left(h_{\overline{\bf 5}}\right)_{ij}&= V^*_{ij} h_{\overline{\bf 5}, j}  ~,
\end{align}
where $h_{{\bf 10}, i}$ and $h_{\overline{\bf 5}, j}$ are the eigenvalues of $\left(h_{\bf 10}\right)_{ij}$ and $\left(h_{\overline{\bf 5}}\right)_{ij}$, respectively. 
The phase factors $\varphi_i$ satisfy the condition 
\begin{equation}
    \sum_{i=1}^3 \varphi_i=0 ~,
\end{equation}
and thus two of them are independent. We discuss the effect of these GUT phases on the nucleon decay rates in Section~\ref{sec:GUTphi}.

The eigenvalues $h_{{\bf 10}, i}$ and $h_{\overline{\bf 5}, i}$ are obtained from the quark Yukawa couplings at the GUT scale. We use the same matching conditions as used in Ref.~\cite{eemno}, namely
\begin{equation}
    h_{{\bf 10}, 3} = \frac{1}{4} f_{u_3} (M_{\text{GUT}})~, \qquad
    h_{\overline{\bf 5}, 3} = \frac{1}{\sqrt{2}} 
    \left[f_{d_3}(M_{\text{GUT}}) + f_{e_3} (M_{\text{GUT}})\right]
    ~,
\end{equation}
for the third-generation Yukawa couplings,~\footnote{These conditions are also used in Ref.~\cite{emo}. } while for the first- and second-generations we use 
\begin{equation}
    h_{{\bf 10}, i} = \frac{1}{4} f_{u_i}(M_{\text{GUT}})~, \qquad
    h_{\overline{\bf 5}, i} = \sqrt{2} f_{d_i} (M_{\text{GUT}})~,
\end{equation}
where $f_{u_i}(M_{\text{GUT}})$, $f_{d_i}(M_{\text{GUT}})$, and $f_{e_i}(M_{\text{GUT}})$ are the up-type, down-type, and charged lepton Yukawa couplings at the GUT scale $M_{\text{GUT}}$, respectively, and $V_{ij}$ are the CKM matrix elements. We note that there is an ambiguity in the determination of $h_{\overline{\bf 5}, i}$, and discuss the implications of this ambiguity in Section~\ref{sec:yukawauni}.\\

The MSSM matter superfields are embedded into the SU(5) matter
multiplets as
\begin{align}
 \Psi_i&\ni \{
Q_i,~e^{-i\varphi_i}\overline{U}_i,~V_{ij}\overline{E}_j
\}~,~~~~~~
\Phi_i\ni\{
\overline{D}_i,~L_i
\}~,
\end{align}
so Eq.~\eqref{W5} leads to:~\footnote{To simplify the following expression, we replace $(f_{d_3}+ f_{e_3} )/2$ by $f_{d_3}$ in this equation, but we use the full expression in our calculations.}
\begin{align}
 W_{\rm Yukawa}&=f_{u_i}
(Q^{a}_i\cdot H_u)\overline{U}_{ia}-V^*_{ij}f_{d_j} (Q^{a}_i\cdot H_d)
\overline{D}_{ja}-f_{d_i}
\overline{E}_i (L_i\cdot H_d)\nonumber \\[2pt]
&-\frac{1}{2}e^{i\varphi_i}\epsilon_{abc}f_{u_i}
(Q^{a}_i \cdot Q^{b}_i) H^c_C
+V^*_{ij}f_{d_j}(Q^{a}_i\cdot L_j)\overline{H}_{Ca}
\nonumber \\[2pt]
&+f_{u_i}V_{ij}\overline{U}_{ia}
\overline{E}_jH^a_C
-V^*_{ij}f_{d_j}e^{-i\varphi_i}\epsilon^{abc}
\overline{U}_{ia}\overline{D}_{jb}\overline{H}_{Cc}~,
\label{eq:wyukawa}
\end{align}
where $a,b,c$ denote the colour indices, and the dots indicate contractions of SU(2)$_L$ indices with the anti-symmetric tensor. We note that 
the new GUT phase factors (\ref{GUTphase}) appear only in the couplings of the colour-triplet
Higgs multiplets, as expected since they are unobservable at the electroweak scale. 

In the super-GUT models we discuss below, the 
supersymmetry-breaking soft mass terms for the MSSM
fields are determined at the GUT scale (defined as the energy scale when the electroweak gauge couplings are unified) by a set of matching conditions. 
The soft supersymmetry-breaking terms in the minimal supersymmetric SU(5) GUT are
\begin{align}
 {\cal L}_{\rm soft} = &- \left(m_{\bf 10}^2\right)_{ij}
 \widetilde{\psi}_i^* \widetilde{\psi}_j
- \left(m_{\overline{\bf 5}}^2\right)_{ij} \widetilde{\phi}^*_i
 \widetilde{\phi}_j
- m_H^2 |H|^2 -m_{\overline{H}}^2 |\overline{H}|^2 - m_\Sigma^2 {\rm Tr}
\left(\Sigma^\dagger \Sigma\right)
\nonumber \\
&-\biggl[
\frac{1}{2}M_5 \widetilde{\lambda}^{A} \widetilde{\lambda}^A
+ A_{\bf 10} \left(h_{\bf 10}\right)_{ij}
 \epsilon_{\alpha\beta\gamma\delta\zeta} \widetilde{\psi}_i^{\alpha\beta}
 \widetilde{\psi}^{\gamma\delta}_j H^\zeta
+ A_{\overline{\bf 5}}\left(h_{\overline{\bf 5}}\right)_{ij}
 \widetilde{\psi}_i^{\alpha\beta} \widetilde{\phi}_{j \alpha}  \overline{H}_\beta
\nonumber \\
&+ B_\Sigma \mu_\Sigma {\rm Tr} \Sigma^2 +\frac{1}{6} A_{\lambda^\prime
 } \lambda^\prime  {\rm Tr} \Sigma^3 +B_H \mu_H \overline{H} H+
 A_\lambda \lambda \overline{H} \Sigma H +{\rm h.c.}
 \biggr]~,
\end{align}
where $\widetilde{\psi}_i$ and $\widetilde{\phi}_i$ are the scalar
components of $\Psi_i$ and $\Phi_i$, respectively,
the $\widetilde{\lambda}^A$ are the SU(5) gauginos, and we
use the same symbols for the scalar components of the Higgs fields as for the
corresponding superfields.

In the CMSSM and its generalizations considered here we impose universality
conditions for the soft-mass parameters at a soft supersymmetry-breaking mass input scale
$M_{\rm in}$:
\begin{align}
 \left(m_{\bf 10}^2\right)_{ij} =
\left(m_{\overline{\bf 5}}^2\right)_{ij}
&\equiv m_0^2 \, \delta_{ij} ~,
\nonumber \\[3pt]
m_H = m_{\overline{H}} = m_\Sigma &\equiv m_0 ~,
\nonumber \\[3pt]
A_{\bf 10} = A_{\overline{\bf 5}} = A_\lambda = A_{\lambda^\prime}
&\equiv A_0 ~,
\nonumber \\[3pt]
 M_5 &\equiv m_{1/2} ~.
\label{eq:inputcond}
\end{align}
The bilinear soft SUSY-breaking therms $B_\Sigma$ and $B_H$ are discussed below. 
When $M_{\rm in} =
M_{\rm GUT}$, the above conditions are equivalent to those in the CMSSM and the renormalization group equations (RGEs) are run between the weak and the GUT scale. When $M_{\rm in} <
M_{\rm GUT}$, as in sub-GUT models \cite{sub-GUT},
the RGEs are run only up to $M_{\rm in}$. In contrast, 
in the super-GUT scenario where $M_{\rm in} > M_{\rm GUT}$~\cite{super-GUT,emo,eemno},
RGE running occurs for MSSM parameters between
the weak and GUT scale, where they are matched into
SU(5) GUT parameters which are then run up to $M_{\rm in}$. 
Boundary conditions are set at both
the electroweak scale (e.g., for the gauge and Yukawa couplings) and at $M_{\rm in}$ (e.g., for the soft supersymmetry-breaking parameters).
For a more complete discussion of the RGE evolution in super-GUT models, see Ref.~\cite{eemno}. 

In the super-GUT scenario with $M_{\rm in} > M_{\rm GUT}$, after the parameters listed in Eq.~\eqref{eq:inputcond} are run down to $M_{\rm GUT}$ they are matched onto the corresponding MSSM parameters. 
Of particular importance are the matching conditions for the gauge couplings and gaugino masses. We use the $\overline{\rm DR}$ scheme for the gauge couplings, and include the one-loop threshold corrections due to the GUT-scale fields. We can then take linear combinations of the matching conditions to obtain the following convenient expressions \cite{eemno, Hisano:1992jj, Hisano:1992mh, Hisano:2013cqa}: 
\begin{align}
 \frac{3}{g_2^2(Q)} - \frac{2}{g_3^2(Q)} -\frac{1}{g_1^2(Q)}
&=-\frac{3}{10\pi^2} \ln \left(\frac{Q}{M_{H_C}}\right)
-\frac{96cV}{M_P}
~,\label{eq:matchmhc} \\[3pt]
 \frac{5}{g_1^2(Q)} -\frac{3}{g_2^2(Q)} -\frac{2}{g_3^2(Q)}
&= -\frac{3}{2\pi^2}\ln\left(\frac{Q^3}{M_X^2 M_\Sigma}\right) ~,
\label{eq:matchmgut}
\\[3pt]
 \frac{5}{g_1^2(Q)} +\frac{3}{g_2^2(Q)} -\frac{2}{g_3^2(Q)}&= -\frac{15}{2\pi^2} \ln\left(\frac{Q}{M_X}\right) + \frac{6}{g_5^2(Q)} -\frac{144cV}{M_P} ~,\label{eq:matchg5}
\end{align}
where $g_1$, $g_2$, $g_3$, and $g_5$ are the U(1), SU(2), SU(3), and SU(5) gauge
couplings, respectively, and $Q$ is a renormalization scale taken in our analysis to be
the unification scale: $Q = M_{\rm GUT}$.
\footnote{Other implications of gauge coupling unification and proton decay on supersymmetric models were considered in \cite{pok}.}
The last terms in
equations (\ref{eq:matchmhc}) and (\ref{eq:matchg5}) represent the contribution from the dimension-five
operator 
\begin{equation}
 W_{\rm eff}^{\Delta g} = \frac{c}{M_P} {\rm Tr}\left[
\Sigma {\cal W} {\cal W}
\right] ~,
\label{eq:SigmaWW}
\end{equation}
where ${\cal W}\equiv  {\cal W}^A T^A$ denotes the
superfields corresponding to the field strengths of the SU(5) gauge vector bosons
${\cal V} \equiv {\cal V}^A T^A$. Since $V/M_P \simeq 10^{-2}$, these terms
can be comparable to the one-loop threshold corrections, and thus should
be taken into account when discussing gauge-coupling unification~\cite{Tobe:2003yj}.
In what follows, we use the notation
\begin{equation}
\epsilon \equiv 8cV/M_P\, .
\label{defepsilon}
\end{equation}
The matching conditions for the gaugino
masses are given by \cite{Tobe:2003yj, Hisano:1993zu, Evans:2019oyw}:
\begin{align}
 M_1 &= \frac{g_1^2}{g_5^2} M_5
-\frac{g_1^2}{16\pi^2}\left[10 M_5 -10(A_{\lambda^\prime} -B_\Sigma)
 -\frac{2}{5}B_H\right]
- \frac{\epsilon g_1^2 (A_{\lambda^\prime} -B_\Sigma)}{2} ~,
\label{eq:m1match}
\\[3pt]
M_2 &= \frac{g_2^2}{g_5^2} M_5
-\frac{g_2^2}{16\pi^2}\left[6 M_5 -6A_{\lambda^\prime} +4B_\Sigma
 \right]
-\frac{3 g_2^2 \epsilon (A_{\lambda^\prime} -B_\Sigma)}{2} ~,
\label{eq:m2match}
\\[3pt]
M_3 &= \frac{g_3^2}{g_5^2} M_5
-\frac{g_3^2}{16\pi^2}\left[4 M_5 -4A_{\lambda^\prime} +B_\Sigma
-B_H \right]
+\epsilon g_3^2(A_{\lambda^\prime} -B_\Sigma)
~.
\label{eq:m3match}
\end{align}
Again we see that the contributions of the dimension-five operator
can be comparable to those of the one-loop threshold corrections.

In the absence of the dimension-five operator (i.e., when $\epsilon = 0$), Eqs.~(\ref{eq:matchmhc}--\ref{eq:matchg5}) provide three conditions on the masses of $M_{H_C}$, $M_\Sigma$, and $M_X$,
as well as $g_5$. In addition, 
we can relate the three masses to the GUT Higgs vev $V$ through the couplings $\lambda$, $\lambda^\prime$, and $g_5$ respectively. This gives us six 
constraints on seven quantities:
the three masses, the three couplings, and $V$. Thus only 
one of the two GUT couplings, 
$\lambda$ or $\lambda^\prime$ 
can be chosen as a free parameter.
On the other hand, if $\epsilon \ne 0$, $\lambda$ and $\lambda^\prime$ can be chosen independently with the following condition on the 
dimension-five coupling:
\begin{equation}
\epsilon = \frac{1}{6 g_3^2 (M_{\rm GUT})}- \frac{1}{6 g_1^2 (M_{\rm GUT})}- \frac{1}{40\pi^2}\ln \biggl(\frac{M_{\rm GUT}}{M_{H_C}}\biggr)~,
\end{equation}
which can be obtained from Eq.~\eqref{eq:matchmhc} with $g_1 (M_{\rm GUT}) = g_2 (M_{\rm GUT})$. We note that this relation for $\epsilon$ can also be used in the CMSSM (in which $M_{\rm in} = M_{\rm GUT}$)
if $\lambda$ and $\lambda^\prime$
are specified, even if no running
above the GUT scale is considered. As we will see, `turning on' $\epsilon$ enables the coloured Higgs mass to be increased, and thus increases the proton lifetime.

The remaining MSSM soft supersymmetry-breaking mass terms and trilinear couplings are related at $M_{\rm GUT}$ by
\begin{align}
 m^2_{Q} = m_{U}^2 = m^2_{E} = m^2_{{\bf 10}} ~,&
~~~~~~ m_{D}^2 = m_{L}^2 = m_{\overline{\bf 5}}^2 ~, \nonumber \\
 m_{H_u}^2 = m_H^2 ~,& ~~~~~~ m_{H_d}^2 = m_{\overline{H}}^2 ~,
\nonumber \\
 A_t = A_{\bf 10} ~,& ~~~~~~
 A_b = A_\tau = A_{\overline{\bf 5}} ~.
 \label{gutmatch}
\end{align}
The MSSM $\mu$ and $B$ terms are \cite{Borzumati:2009hu}
\begin{align}
 \mu &= \mu_H - 3 \lambda V\left[
1+ \frac{A_{\lambda^\prime} -B_\Sigma}{2 \mu_\Sigma}
\right] ~,
\label{eq:matchingmu}
 \\[3pt]
 B &= B_H + \frac{3\lambda V \Delta}{\mu}
+ \frac{6 \lambda}{\lambda^\prime \mu} \left[
(A_{\lambda^\prime} -B_\Sigma) (2 B_\Sigma -A_{\lambda^\prime}
 +\Delta) -m_\Sigma^2
\right]~,
\label{eq:matchingb}
\end{align}
with
\begin{equation}
 \Delta \equiv A_{\lambda^\prime} - B_\Sigma - A_\lambda +B_H ~.
\label{eq:deltadef}
\end{equation}
In the absence of a more elegant solution for the separation of the GUT and weak scales, we must tune $|\mu_H -3\lambda V|$
to be ${\cal O}(M_{\rm SUSY})$. 
In practice, $\mu$ and $B$
are determined at the electroweak scale
by the minimization of the Higgs potential as in the CMSSM. These are then run up to the scale where Eqs.~\eqref{eq:matchingmu}
and \eqref{eq:matchingb} are applied. Then, using $\Delta = 0$ (which is stable against radiative corrections
as shown in Ref.~\cite{Kawamura:1994ys}),
we can solve for $B_H$ and $B_\Sigma$ at the GUT scale, which are needed in the matching conditions for the gaugino masses.

As was pointed out in \cite{eemno}, Eqs. (\ref{eq:matchingb}) and (\ref{eq:deltadef})  have no solution unless we take $A_{\lambda'}\gtrsim 8m_\Sigma^2$. For super-GUT theories with $A_0=0$, like in the focus-point case we consider below, this conditions is generally not satisfied. However, these types of models can satisfy this condition if the following Giudice-Masiero terms
\cite{GM} are added to the K\"ahler potential:
\begin{eqnarray}
\Delta K= c_\Sigma {\rm Tr}(\Sigma^2)+c_HH\bar H +h.c.
\end{eqnarray}
Although the above terms only give a small correction to the $B$ terms,
\begin{align}
\Delta B_\Sigma&=\frac{2c_\Sigma m_{3/2}^2}{\mu_\Sigma}\, , \\
\Delta B_H&=\frac{2c_H m_{3/2}^2}{\mu_H}~,
\end{align}
they give an important contribution to \eqref{eq:matchingb}, since 
\begin{eqnarray}
\frac{3\lambda V \Delta}{\mu}\to \left(c_H-\frac{12\lambda}{\lambda'}c_\Sigma\right)\frac{2m_{3/2}^2}{\mu}~,
\end{eqnarray}
where $m_{3/2}$ is the gravitino mass which sets the scale for $m_0$ and $m_{1/2}$.
The additional contributions in the above equations make it trivial to satisfy \eqref{eq:matchingb} even with $A_0=0$, see \cite{eenno} for more discussion. 

To summarize, the well-studied CMSSM is characterized by four parameters and one sign:
\begin{equation}
 m_0,\ m_{1/2},\ A_0,\  \tan \beta,\ {\rm
  sign}(\mu) \, .
  \label{cmssmpara}
\end{equation}
In sub-GUT models we must in addition specify
the input universality scale:
\begin{equation}
  M_{\rm in} ,
\end{equation}
and in super-GUT models we also need to specify the Higgs couplings:
\begin{equation}
  \lambda,\ \lambda' .
\end{equation}
which
are fixed at $Q=M_{\rm GUT}$.
As noted above, CMSSM models with
non-zero values of $\epsilon$
also require fixing these GUT Higgs couplings.

\subsection{Nucleon Decay in Minimal Supersymmetric SU(5)}
\label{sec:decaybasics}

We now review the calculation of the proton decay lifetime in the minimal supersymmetric SU(5) model \cite{Hisano:1992jj, Goto:1998qg, mp, Hisano:2013exa, Nagata:2013sba, Nagata:2013ive, evno, eelnos, Evans:2019oyw}. As we have seen in Section~\ref{sec:BDKexp}, the forthcoming experiments are expected to offer great sensitivities to nucleon decay. To make the best of these experiments, therefore, it is desirable to formulate a precise and systematic method for the computation of nucleon decay lifetimes. To that end, we adopt the method of effective field theories. In this method, the fundamental theory is matched onto a low-energy effective theory at a high-energy scale (the GUT scale in our case), where the effect of heavy states (the GUT-scale fields) is included into the Wilson coefficients of higher-dimensional effective operators. We then run the Wilson coefficients down to the hadronic scale by using RGEs, which allows us to resum large logarithmic radiative corrections due to the large hierarchy in energy scales. The long-distance QCD effect is taken into account through the calculation of hadronic matrix elements. With this procedure, we can separate the short- and long-range contributions to the decay amplitude in a consistent manner. Note that this prescription is the same as those used for the calculation of precision physics observables, such as flavour observables \cite{Buchalla:1995vs} and the dark matter-nucleon scattering cross section \cite{Hisano:2015bma, Ellis:2018dmb}.

As we discussed above, the most important decay mode is $p\to K^+ \bar{\nu}$, which is 
induced by the exchange of the colour-triplet Higgs multiplets \cite{Sakai:1981pk}. 
The effective Lagrangian for this contribution is
\begin{equation}
{\cal L}_5^{\rm eff}= C^{ijkl}_{5L}{\cal O}^{5L}_{ijkl}
+C^{ijkl}_{5R}{\cal O}^{5R}_{ijkl}
~~+~~{\rm h.c.}~,
\label{eq:leff5}
\end{equation}
where the effective operators ${\cal O}^{5L}_{ijkl}$ and ${\cal
O}^{5R}_{ijkl}$ are defined by
\begin{align}
 {\cal O}^{5L}_{ijkl}&\equiv\int d^2\theta~ \frac{1}{2}\epsilon_{abc}
(Q^a_i\cdot Q^b_j)(Q_k^c\cdot L_l)~,\nonumber \\
{\cal O}^{5R}_{ijkl}&\equiv\int d^2\theta~
\epsilon^{abc}\overline{U}_{ia}\overline{E}_j\overline{U}_{kb}
\overline{D}_{lc}~,
\end{align}
and the Wilson coefficients $C^{ijkl}_{5L}$ and $C^{ijkl}_{5R}$ are
given by
\begin{align}
 C^{ijkl}_{5L}(M_{\rm GUT})&
=\frac{2 \sqrt{2}}{M_{H_C}}h_{{\bf 10}, i}
e^{i\varphi_i}\delta^{ij}V^*_{kl}h_{\overline{\bf 5}, l}~,\nonumber \\
C^{ijkl}_{5R}(M_{\rm GUT})
&=\frac{2 \sqrt{2}}{M_{H_C}}h_{{\bf 10}, i}V_{ij}V^*_{kl}h_{\overline{\bf 5}, l}
e^{-i\varphi_k}
~.
\label{eq:wilson5}
\end{align}
We note that antisymmetry with respect to the colour indices requires
that the operators include at least two generations of quarks. For this
reason, the dominant decay modes generally contain a strange quark in the final
state, such as $p\to K^+\bar{\nu}$~\cite{strangeBDK}.

The Wilson coefficients $C^{ijkl}_{5L}$ and $C^{ijkl}_{5R}$ are run down to the supersymmetric scale $M_{\rm SUSY}$ using the RGEs 
\begin{align}
 \frac{d}{d\ln Q} C^{ijkl}_{5L} &=
\frac{1}{16\pi^2}\biggl[-\frac{2}{5}g_1^2 -6g_2^2 -8g_3^2
+f_{u_i}^2 +f_{d_i}^2 + f_{u_j}^2 +f_{d_j}^2 + f_{u_k}^2 + f_{d_k}^2 +
 f_{e_l}^2
\biggr] C^{ijkl}_{5L}~, \nonumber \\
 \frac{d}{d\ln Q} C^{ijkl}_{5R} &=
\frac{1}{16\pi^2}\biggl[-\frac{12}{5}g_1^2 -8g_3^2
+2f_{u_i}^2 + 2f_{e_j}^2 + 2f_{u_k}^2 + 2f_{d_l}^2
\biggr] C^{ijkl}_{5R}~,
\end{align}
where $Q$ denotes the renormalization scale.
At the scale $M_{\rm SUSY}$, sfermions are integrated out via the wino- or Higgsino-exchange one-loop diagrams. The low-energy effective Lagrangian below the supersymmetric scale is then given by
\begin{align}
 {\cal L}^{\text{eff}}_{\rm SM}&=C^{\widetilde{H}}_i {\cal O}_{1i33}
+ C^{\widetilde{W}}_{jk}\widetilde{\cal O}_{1jjk}
+ C^{\widetilde{W}}_{jk}\widetilde{\cal O}_{j1jk}
+ \overline{C}^{\widetilde{W}}_{jk}\widetilde{\cal O}_{jj1k}
~,
\end{align}
where the effective operators have the form
\begin{align}
 {\cal O}_{ijkl} &\equiv \epsilon_{abc}(u^a_{Ri}d^b_{Rj})
(Q_{Lk}^c \cdot L_{Ll}) ~, \nonumber \\
 \widetilde{\cal O}_{ijkl} &\equiv \epsilon_{abc} \epsilon^{\alpha\beta}
\epsilon^{\gamma\delta} (Q^a_{Li\alpha}Q^b_{Lj\gamma})
(Q_{Lk\delta}^c L_{Ll\beta}) ~,
\label{eq:effopPD}
\end{align}
with $i =1,2$, $j=2,3$, and $k=1,2,3$, and their Wilson coefficients are evaluated as 
\begin{align}
 C_i^{\widetilde{H}} 
(M_{\text{SUSY}})&=\frac{f_tf_\tau}{(4\pi)^2}C^{*331i}_{5R}(M_{\text{SUSY}})
F(\mu, m_{\widetilde{t}_R}^2,m_{\tau_R}^2)~, \nonumber \\
 C^{\widetilde{W}}_{jk}(M_{\text{SUSY}}) &=
\frac{\alpha_2}{4\pi}C^{jj1k}_{5L}(M_{\text{SUSY}})
[F(M_2, m_{\widetilde{Q}_1}^2,
 m_{\widetilde{Q}_j}^2) +F(M_2, m_{\widetilde{Q}_j}^2,
 m_{\widetilde{L}_k}^2)] ~, \nonumber \\
 \overline{C}^{\widetilde{W}}_{jk}(M_{\text{SUSY}}) &=
-\frac{3}{2}
\frac{\alpha_2}{4\pi}C^{jj1k}_{5L}(M_{\text{SUSY}})
[F(M_2, m_{\widetilde{Q}_j}^2,
 m_{\widetilde{Q}_j}^2) +F(M_2, m_{\widetilde{Q}_1}^2,
 m_{\widetilde{L}_k}^2)] ~,
\label{eq:effWCPD}
\end{align}
with
 \begin{align}
F(M, m_1^2, m_2^2) &\equiv
\frac{M}{m_1^2-m_2^2}
\biggl[
\frac{m_1^2}{m_1^2-M^2}\ln \biggl(\frac{m_1^2}{M^2}\biggr)
-\frac{m_2^2}{m_2^2-M^2}\ln \biggl(\frac{m_2^2}{M^2}\biggr)
\biggr]~.
\label{eq:funceq}
\end{align}
From the supersymmetric breaking scale to the electroweak scale, we use the RGEs \cite{Alonso:2014zka}
\begin{align}
 \mu \frac{d}{d\mu} C^{\widetilde{H}}_{i}
&= \biggl[\frac{\alpha_1}{4\pi}\biggl(-\frac{11}{10}\biggr)
+\frac{\alpha_2}{4\pi}\biggl(-\frac{9}{2}\biggr)
+\frac{\alpha_3}{4\pi}(-4) +\frac{1}{2}\frac{y_{t}^2}{16\pi^2}
\biggr]C^{\widetilde{H}}_{i}
 ~,\nonumber \\
 \mu \frac{d}{d\mu} C^{\widetilde{W}}_{jk}
&= \biggl[\frac{\alpha_1}{4\pi}\biggl(-\frac{1}{5}\biggr)
+\frac{\alpha_2}{4\pi}(-3)
+\frac{\alpha_3}{4\pi}(-4) +\frac{y_{u_j}^2}{16\pi^2}
\biggr]C^{\widetilde{W}}_{jk}
+\frac{\alpha_2}{4\pi}(-4)[
2C^{\widetilde{W}}_{jk} + \overline{C}^{\widetilde{W}}_{jk}]
 ~,\nonumber \\
 \mu \frac{d}{d\mu} \overline{C}^{\widetilde{W}}_{jk}
&= \biggl[\frac{\alpha_1}{4\pi}\biggl(-\frac{1}{5}\biggr)
+\frac{\alpha_2}{4\pi}(-3)
+\frac{\alpha_3}{4\pi}(-4) +\frac{y_{u_j}^2}{16\pi^2}
\biggr]\overline{C}^{\widetilde{W}}_{jk}
+\frac{\alpha_2}{4\pi}(-4)[
2C^{\widetilde{W}}_{jk} + \overline{C}^{\widetilde{W}}_{jk}] ~,
\end{align}
where $y_{u_j}$ denotes the SM up-type Yukawa couplings.

Below the electroweak scale, the effective interactions that give rise to the $p\to K^+\bar{\nu}_k$ decay mode are described by 
\begin{align}
 {\cal L}(p\to K^+\bar{\nu}_i^{})
=&C_{RL}(usd\nu_i)\bigl[\epsilon_{abc}(u_R^as_R^b)(d_L^c\nu_i^{})\bigr]
+C_{RL}(uds\nu_i)\bigl[\epsilon_{abc}(u_R^ad_R^b)(s_L^c\nu_i^{})\bigr]
\nonumber \\
+&C_{LL}(usd\nu_i)\bigl[\epsilon_{abc}(u_L^as_L^b)(d_L^c\nu_i^{})\bigr]
+C_{LL}(uds\nu_i)\bigl[\epsilon_{abc}(u_L^ad_L^b)(s_L^c\nu_i^{})\bigr]
~,
\end{align}
where the coefficients of these interactions are obtained at the electroweak scale as 
\begin{align}
 C_{RL}(usd\nu_\tau)&=-V_{td}C^{\widetilde{H}}_{2}(M_Z)~,\nonumber \\
 C_{RL}(uds\nu_\tau)&=-V_{ts}C^{\widetilde{H}}_{1}(M_Z)~,\nonumber \\
 C_{LL}(usd\nu_k)&=\sum_{j=2,3}V_{j1}V_{j2}
C^{\widetilde{W}}_{jk}(M_Z)~,\nonumber \\
 C_{LL}(uds\nu_k)&=\sum_{j=2,3}V_{j1}V_{j2}
C^{\widetilde{W}}_{jk}(M_Z)~.
 \label{eq:effWCPD_weak}
\end{align}
We then use the two-loop RGE given in Ref.~\cite{Nihei:1994tx} to evolve these coefficients down to the hadronic scale $\mu_{\text{had}}= 2$~GeV, and finally obtain the partial decay width of the $p\to K^+ \bar{\nu}_i$
mode as
\begin{equation}
 \Gamma(p\to K^+\bar{\nu}_i)
=\frac{m_p}{32\pi}\biggl(1-\frac{m_K^2}{m_p^2}\biggr)^2
\vert {\cal A}(p\to K^+\bar{\nu}_i)\vert^2~,
\end{equation}
where $m_p$ and $m_K$ are the masses of proton and kaon,
respectively. Since the experiments cannot determine the flavor of the neutrino, below, we will use the notation that $p\to K^{+}{\bar \nu}$ represents the sum of the decays to all neutrino flavors. The decay amplitude ${\cal A}(p\to K^+\bar{\nu}_i)$ is the sum of the products of Wilson coefficients with hadronic matrix elements:
\begin{align}
\label{eq:ApKpnm}
     {\cal A}(p\to K^+\bar{\nu}_i)&=
C_{RL}(usd\nu_i)\langle K^+\vert (us)_Rd_L\vert p\rangle
+
C_{RL}(uds\nu_i)\langle K^+\vert (ud)_Rs_L\vert p\rangle 
\nonumber \\
&+
C_{LL}(usd\nu_i)\langle K^+\vert (us)_Ld_L\vert p\rangle
+C_{LL}(uds\nu_i)\langle K^+\vert (ud)_Ls_L\vert p\rangle
~.
\end{align}
The hadronic matrix elements are evaluated at the scale $\mu_{\text{had}}= 2$~GeV using QCD lattice simulations, which we discuss in detail in Section~\ref{sec:ME}. 

The dimension-five effective interactions in Eq.~\eqref{eq:leff5} also induce other nucleon decay modes, such as $p \to \pi^+ \overline{\nu}$ and $n \to \pi^0 \overline{\nu}$. The calculation of the decay rates of these mode is the same as for $p \to K^+ \overline{\nu}$ above the electroweak scale. The effective interactions for these decay modes below the electroweak scale are 
\begin{align}
 {\cal L}(N\to \pi \bar{\nu}_i)
&=C_{RL}(udd\nu_i)\bigl[\epsilon_{abc}(u_R^ad_R^b)(d_L^c\nu_{Li}^{})\bigr]
+C_{LL}(udd\nu_i)\bigl[\epsilon_{abc}
(u_L^ad_L^b)(d_L^c\nu_{Li}^{})\bigr]~,
\label{eq:leffppinu}
\end{align}
where the matching conditions for the Wilson coefficients are 
\begin{align}
 C_{RL}(udd\nu_\tau)&=-V_{td}C^{\widetilde{H}}_1 (M_Z)~,\nonumber \\
 C_{LL}(udd\nu_k)&=\sum_{j=2,3} V_{j1} V_{j1} C_{jk}^{\widetilde{W}}
(M_Z)~.
\label{eq:CRLCLLMZ}
\end{align}
Using these interactions, we compute the partial decay widths of $p \to \pi^+ \overline{\nu}$ and $n \to \pi^0 \overline{\nu}$ as
\begin{align}
     \Gamma (p\to  \pi^+ \bar{\nu}_i) &=
\frac{m_p}{32\pi}\biggl(1-\frac{m_\pi^2}{m_p^2}\biggr)^2
\vert {\cal A}(p\to \pi^+ \bar{\nu}_i) \vert^2~, \\
\Gamma (n\to  \pi^0 \bar{\nu}_i) &=
\frac{m_n}{32\pi}\biggl(1-\frac{m_\pi^2}{m_n^2}\biggr)^2
\vert {\cal A}(n\to \pi^0 \bar{\nu}_i) \vert^2~,
\end{align}
where $m_n$ and $m_\pi$ are the masses of neutron and pion, respectively, and
\begin{align}
 {\cal A}_L(p\to \pi^+ \bar{\nu}_i)&=
C_{RL}(udd\nu_i)\langle \pi^+\vert (ud)_Rd_L\vert p\rangle
+C_{LL}(udd\nu_i)\langle \pi^+\vert (ud)_Ld_L\vert p\rangle~, \\
 {\cal A}_L(n\to \pi^0 \bar{\nu}_i)&=
C_{RL}(udd\nu_i)\langle \pi^0\vert (ud)_Rd_L\vert n\rangle
+C_{LL}(udd\nu_i)\langle \pi^0\vert (ud)_Ld_L\vert n\rangle~.
\end{align}
Again, since the experiments are unable to determine the flavor of the neutrino, we will use $p\to \pi^{+}{\bar \nu}$ and $n\to \pi^{+}{\bar \nu}$ to represent the sum of decays to all neutrino flavors. The dimension-six nucleon decay can also be computed in a similar way, which we show in the Appendix, for completeness.

\section{Uncertainties in Nucleon Decay Calculations}
\label{sec:uncertainties}

In this Section we discuss various uncertainties in the calculation of the proton decay rate for fixed values of the supersymmetric GUT model parameters. We start with the uncertainties in the hadronic matrix elements and the experimental input value of $\alpha_s$, which are generally the most important.~\footnote{The effect of the uncertainties in $\alpha_s$ is greatly diminished when
the operator (\ref{eq:SigmaWW}) is considered, as we discuss below.} We also consider the effects of uncertainties in the weak mixing angle, quark masses, loop corrections to quark Yukawa couplings, and quark mixing parameters. The experimental inputs for $\alpha_s$ and the quark masses that we use are listed in Table~\ref{tab:expinputs}~\cite{PDG}. 
\footnote{We note that there the PDG~\cite{PDG} lists an updated value of $\alpha_s = 0.1179 \pm 0.0010$, a change that is a small fraction of the uncertainty and does not affect our results significantly.} As many of the contributions to the proton lifetime uncertainty are small compared to those arising from the matrix elements and the value of $\alpha_s$, when we compute the ``total" uncertainty in $\tau_p \equiv 1/\sum_i \Gamma (p \to K^+ \bar{\nu}_i)$, we propagate only the effects of the matrix elements and the direct effect of $\alpha_s$ on $M_{H_C}$, which is the most important source of sensitivity to the value of $\alpha_s$.~\footnote{For example, changes in the renormalization-group (RG) running below $M_Z$ shift the Yukawa couplings of the light quarks at the GUT Scale by about 2 percent, a little more for charm. The shifts in the Yukawa couplings and the Wilson coefficients from the RG running above $M_Z$ are estimated to be smaller, as are the effects on the soft masses.}

\begin{table}[!ht]
\begin{center}
\caption{\it{Experimental Inputs~\cite{PDG}.} 
\label{tab:expinputs}}
\vspace{-0.7cm}
\begin{tabular}{|c|l|}
\multicolumn{2}{c}{}\\
\hline
\hline
$\alpha_s$& $0.1181\pm 0.0011 $ \\
$m_t$& $172.9\pm 0.4$ GeV \\
$m_b$& $4.18^{+ 0.03}_{-0.02}$ GeV\\
$m_c$& $1.27\pm 0.02$ GeV\\
$m_s$& $0.093^{+0.011}_{- 0.005}$ GeV\\
$m_d$& $0.00467^{+0.00048}_{-0.00017}$ GeV\\
$m_u$& $0.00216^{+0.00049}_{-0.00026}$ GeV\\
\hline
\hline
\end{tabular}
\end{center}
\end{table}

\subsection{Hadronic Matrix-Element Uncertainties}
\label{sec:ME}

The hadronic matrix elements that determine directly the most relevant
proton partial decay rates have been updated recently~\cite{AISS}. The improvement in the update provides a total accuracy of 10 to 15 \% of the matrix elements.  Table~\ref{tab:w0_error} lists the values of the most relevant baryonic decay matrix elements calculated previously in~\cite{Aoki:2013yxa} (second column) and recently in~\cite{AISS} (fourth column), including both the statistical and systematic uncertainties, which are indicated by (...)(...). Also shown in the third and fifth columns are the corresponding total errors after combining these uncertainties in quadrature.
We see that in some cases the calculated matrix elements have changed significantly between \cite{Aoki:2013yxa} and \cite{AISS}, and that the uncertainties have been reduced substantially in every case.
Both of these two simulations utilize the same gauge ensemble of $N_f = 2+ 1$ domain-wall fermions with the same lattice spacing $a = 0.11$~fm, lattice volume $(2.65)^3$~fm$^3$, and the pion mass 0.34--0.69~GeV, and directly compute the three-point function of the nucleon-to-pseudoscalar transition with an insertion of the baryon-number violating operators (``direct method'').~\footnote{These papers also show the results obtained with the ``indirect'' method, where the hadron matrix elements are evaluated through the low-energy constants in the baryon chiral perturbation theory. It is found in Ref.~\cite{AISS} that the  matrix elements obtained with the indirect method tend to be larger in magnitude than those obtained with the direct method. As discussed in Ref.~\cite{AISS}, the direct method is expected to be more reliable since in the nucleon decay processes the pion in the final state has a sizable momentum, which spoils the validity of the chiral perturbation theory. For this reason, in our analysis, we only use the matrix elements obtained with the direct method. Notice that with this choice the resultant proton lifetimes tend to be longer, and thus we obtain a conservative limit.  } In Ref.~\cite{AISS}, they use an algorithm called all-mode-averaging (AMA) \cite{Blum:2012uh}, with which the statistical error has significantly been reduced. In addition, an error in the renormalization-scheme matching factors was corrected in Ref.~\cite{AISS}; this leads to a 6--7\% change in the matrix elements (see Footnote~2 in Ref.~\cite{AISS}). Notice that these simulations are still performed with an unphysical pion mass. A simulation at the physical point is on-going \cite{Yoo:2018fyn}, which is expected to reduce the systematic uncertainty associated with the chiral extrapolation. All in all, a precision with $< 10$\% uncertainty is expected to be achieved in 5 years \cite{Cirigliano:2019jig}. 

\begin{table}[!ht]
\begin{center}
\caption{\it Comparison of lattice hadronic matrix element calculations. 
}
\vspace{-0.7cm}
\label{tab:w0_error}
\begin{tabular}{|c|c|c|c|c|}
\multicolumn{5}{c}{}\\
\hline
\hline
Matrix element & Previous value &  Total error &  New value & Total error \\
 & \cite{Aoki:2013yxa} & & \cite{AISS} &  \\
 \hline
\footnotesize{$\langle \pi^0|(ud)_Ru_L|p\rangle$} & $-$0.103(23)(34) &  0.041&$-0.131(4)(13)$ &0.013\\
\footnotesize{$\langle \pi^0|(ud)_Lu_L|p\rangle$} & 0.133(29)(28)      &  0.040  & 0.134(5)(16)  & 0.016\\
\footnotesize{$\langle \pi^+|(du)_Rd_L|p\rangle$} & $-$0.146(33)(48)  &  0.058  &$-0.186 (6)(18)$ & 0.019\\
\footnotesize{$\langle \pi^+|(du)_Ld_L|p\rangle$} &0.188(41)(40)        &  0.057  & 0.189(6)(22) & 0.023\\
\hline
\footnotesize{$\langle K^0|(us)_Ru_L|p\rangle$} & 0.098(15)(12)       &  0.019  & 0.103(3)(11)  &0.011 \\
\footnotesize{$\langle K^0|(us)_Lu_L|p\rangle$} & 0.042(13)(8)        & 0.015   & 0.057(2)(6)   & 0.006\\
\footnotesize{$\langle K^+|(us)_Rd_L|p\rangle$} & $-$0.054(11)(9)     &  0.014   &$-0.049(2)(5)$  & 0.006\\
\footnotesize{$\langle K^+|(us)_Ld_L|p\rangle$} & 0.036(12)(7)          &  0.014   &0.041(2)(5)  &0.006 \\
\footnotesize{$\langle K^+|(ud)_Rs_L|p\rangle$} &$-$0.093(24)(18)   &  0.030 & $-0.134$(4)(14) & 0.014\\
\footnotesize{$\langle K^+|(ud)_Ls_L|p\rangle$} &  0.111(22)(16)       &  0.027  &  0.139(4)(15) & 0.016\\
\footnotesize{$\langle K^+|(ds)_Ru_L|p\rangle$} & $-$0.044(12)(5)    &  0.013  & $-0.054(2)(6) $ &0.006\\
\footnotesize{$\langle K^+|(ds)_Lu_L|p\rangle$} & $-$0.076(14)(9)     &  0.017  & $-0.098(3)(10)$ & 0.010\\
\hline
\footnotesize{$\langle \eta|(ud)_Ru_L|p\rangle$} & 0.015(14)(17)     & 0.022,   & 0.006(2)(3) & 0.003\\
\footnotesize{$\langle \eta|(ud)_Lu_L|p\rangle$} & 0.088(21)(16)       & 0.026  & 0.113(3)(12) & 0.012\\
\hline\hline
\end{tabular}
\end{center}
\end{table}

\begin{figure}[htb!]
\begin{minipage}{6in}
\hspace*{-1.2in}
\includegraphics[height=2.9in]{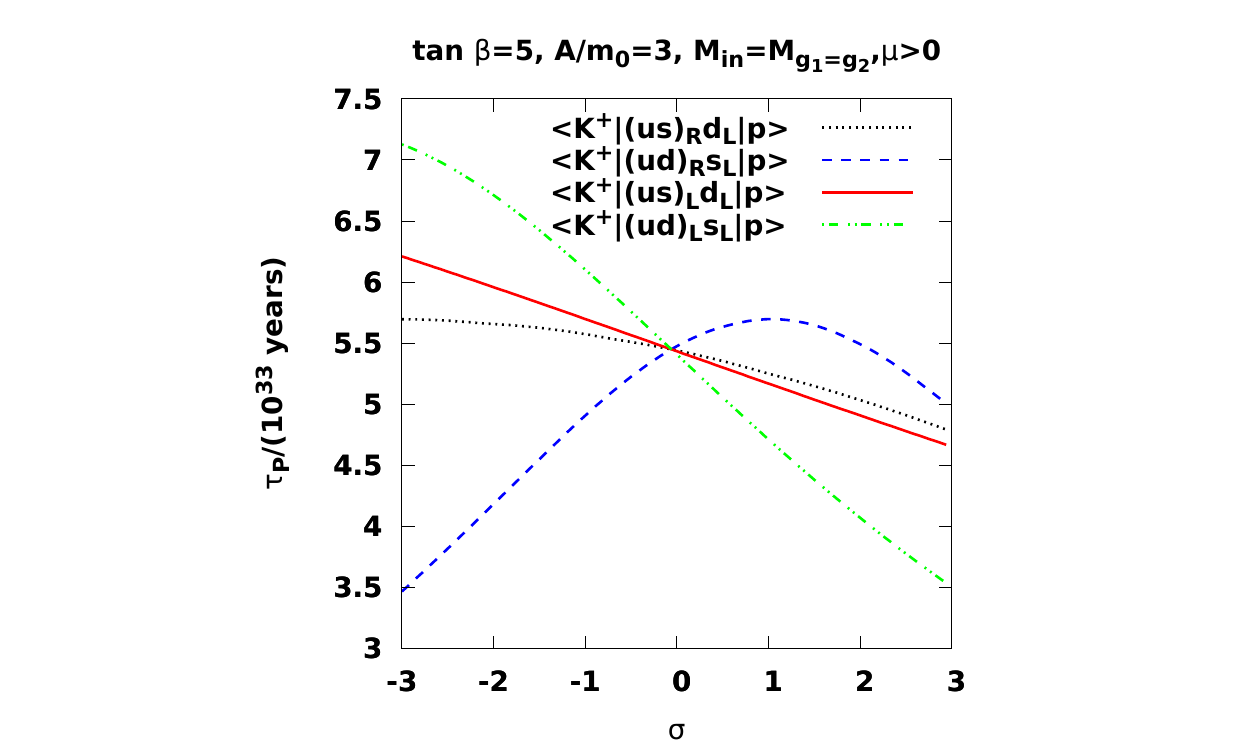}
\hspace*{-1.8in}
\includegraphics[height=2.9in]{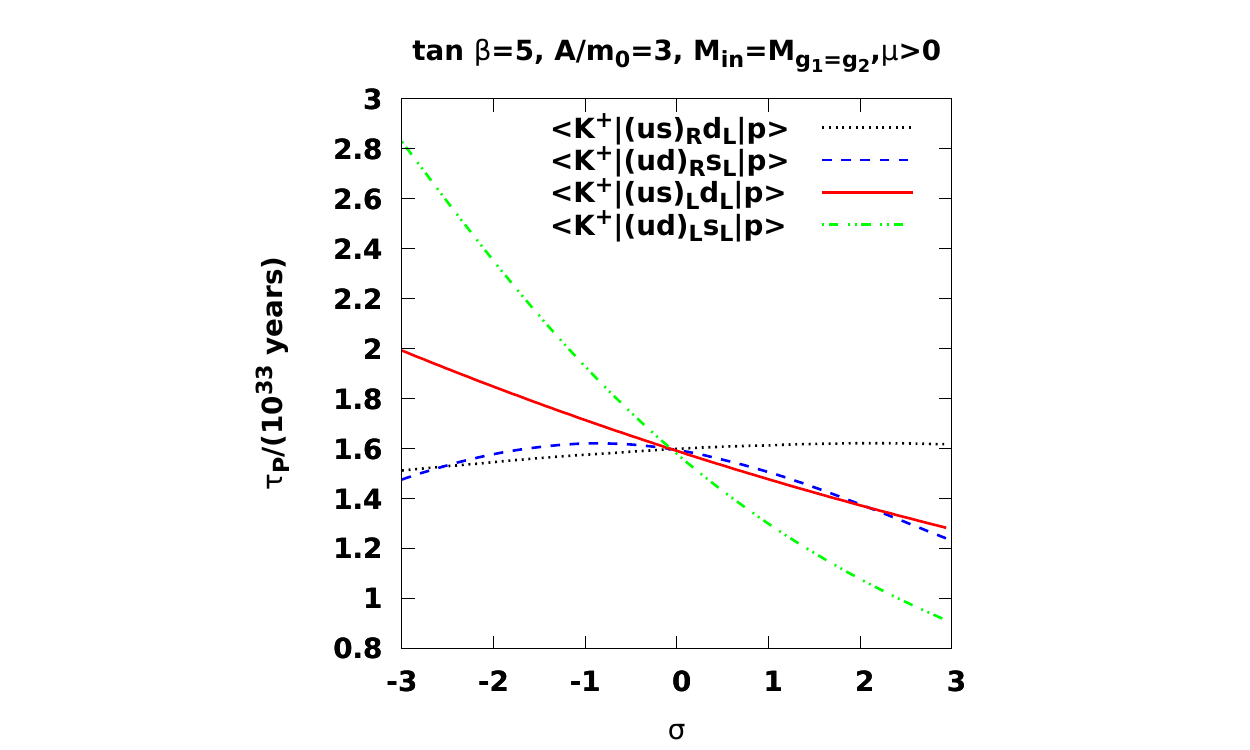}
\hfill
\end{minipage}
\begin{minipage}{6in}
\hspace*{-1.2in}
\includegraphics[height=2.9in]{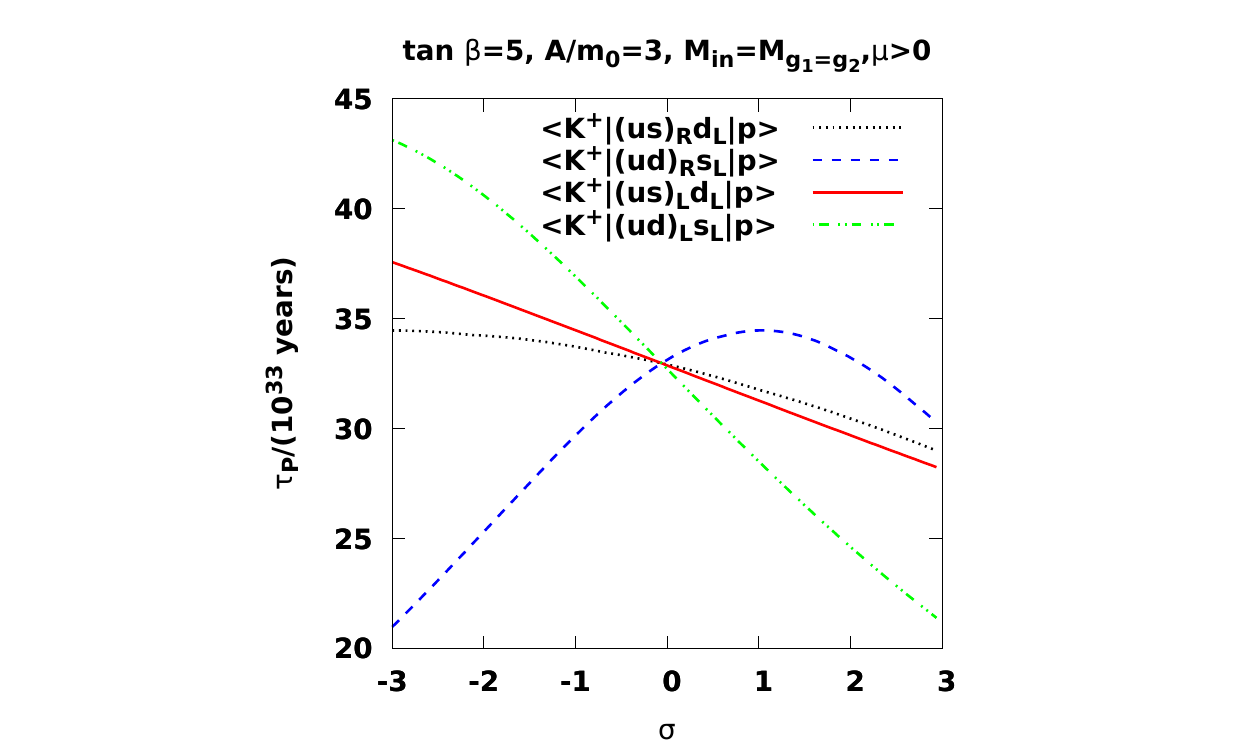}
\hspace*{-1.8in}
\includegraphics[height=2.9in]{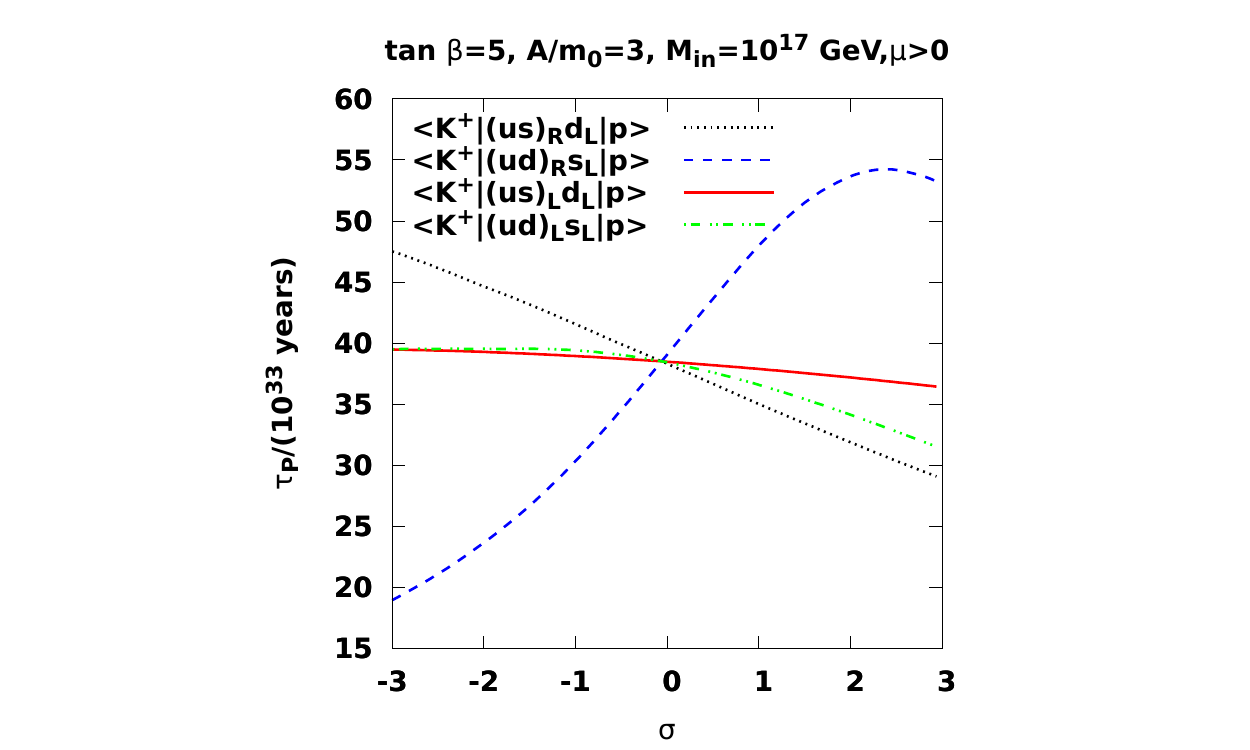}
\hfill
\end{minipage}
\vspace{-5mm}
\caption{
{\it Sensitivities of the proton decay rate to variations in units of the standard deviations of the matrix elements of the indicated 3-quark operators.
In all cases, we chose reference points with $\tan \beta = 5$, $A_0/m_0 = 3$, $m_{1/2} = 9.79$ TeV and $\mu >0$. We also chose (in TeV) ($m_0,M_{\rm in}$) = (14.13, $M_{\rm GUT}$) (upper left); (10, $M_{\rm GUT}$) (upper right); (14.13, $M_{\rm GUT}$), with $c\ne 0$ (lower left); (15.45, $10^{17}$~GeV) (lower right). In each case, $\sigma = 0$ corresponds to the current central value of the matrix element from~\cite{AISS}.}}
\label{fig:hadME}
\end{figure}

We focus in this paper primarily on the $p \to K^+ \bar{\nu}$ decay mode, which is determined by the four matrix elements that are featured in 
Fig.~\ref{fig:hadME}. This figure illustrates the sensitivities of the $p \to K^+ \bar{\nu}$ decay rate to variations of these four hadronic matrix elements, which are each given in units of the total uncertainties of the new elements given in column 5 of the Table. Thus, $\sigma = 0$ corresponds to the current central value of each of the four matrix elements and $\sigma = \pm 1$ corresponds to adding (subtracting) the 1-$\sigma$ total uncertainty to (from) the corresponding hadronic matrix element. The previous values of these matrix elements would lie at $\sigma = -0.83$ for $\langle K^+\vert (us)_R ^{}d_L^{}\vert p\rangle$ and $\langle K^+\vert (us)_L ^{}d_L^{}\vert p\rangle$, $\sigma = 2.93$ for $\langle K^+\vert (ud)_R ^{}s_L^{}\vert p\rangle$ and $\sigma = -1.75$ for $\langle K^+\vert (ud)_L ^{}s_L^{}\vert p\rangle$, which are  significant changes, especially for the last two.  These large changes in the matrix elements are responsible for the bulk of the reduction in the $p$ lifetime relative to values found in previous work \cite{eemno}.

The sensitivities to the four matrix elements shown in Fig.~\ref{fig:hadME} are for specific points in the CMSSM and super-GUT parameter spaces. In all cases, the reference points were chosen with $\tan \beta = 5$, $A_0/m_0 = 3$, $m_{1/2} = 9.79$~TeV and $\mu >0$. We also chose ($m_0,M_{\rm in}$) = (14.13 TeV, $M_{\rm GUT}$) (upper left); (10~TeV, $M_{\rm GUT}$) (upper right); (14.13 TeV, $M_{\rm GUT}$) with $c\ne 0$ (lower left); (15.45 TeV, $10^{17}$~GeV) (lower right).

The upper two panels of Fig.~\ref{fig:hadME} are based on a CMSSM input spectrum. In the upper left panel, the point lies on the stop coannihilation strip with $m_h = 125$ GeV (see below for more information on the relevance of this choice), whereas in the upper right panel the 
point is at lower $m_0$ where the stop mass is significantly larger. In the lower left panel, we
allow the dimension-five coupling $c$ in Eq.~\eqref{eq:SigmaWW} to be non-zero for the same stop-coannihilation point. In this case, we see that while the relative sensitivity to the matrix elements is similar, the lifetime is significantly increased. Finally, in the lower right
panel we show a super-GUT example with $M_{\rm in} = 10^{17}$~GeV. Because the super-GUT running tends to reduce the stop mass relative to the other sfermion masses, the proton lifetime becomes much more sensitive to the Wilson coefficient arising from Higgsino exchange. This alters the sensitivity to the hadron matrix elements. 

\begin{figure}[!htb]
\begin{minipage}{6in}
\begin{center}
\includegraphics[height=3.5in]{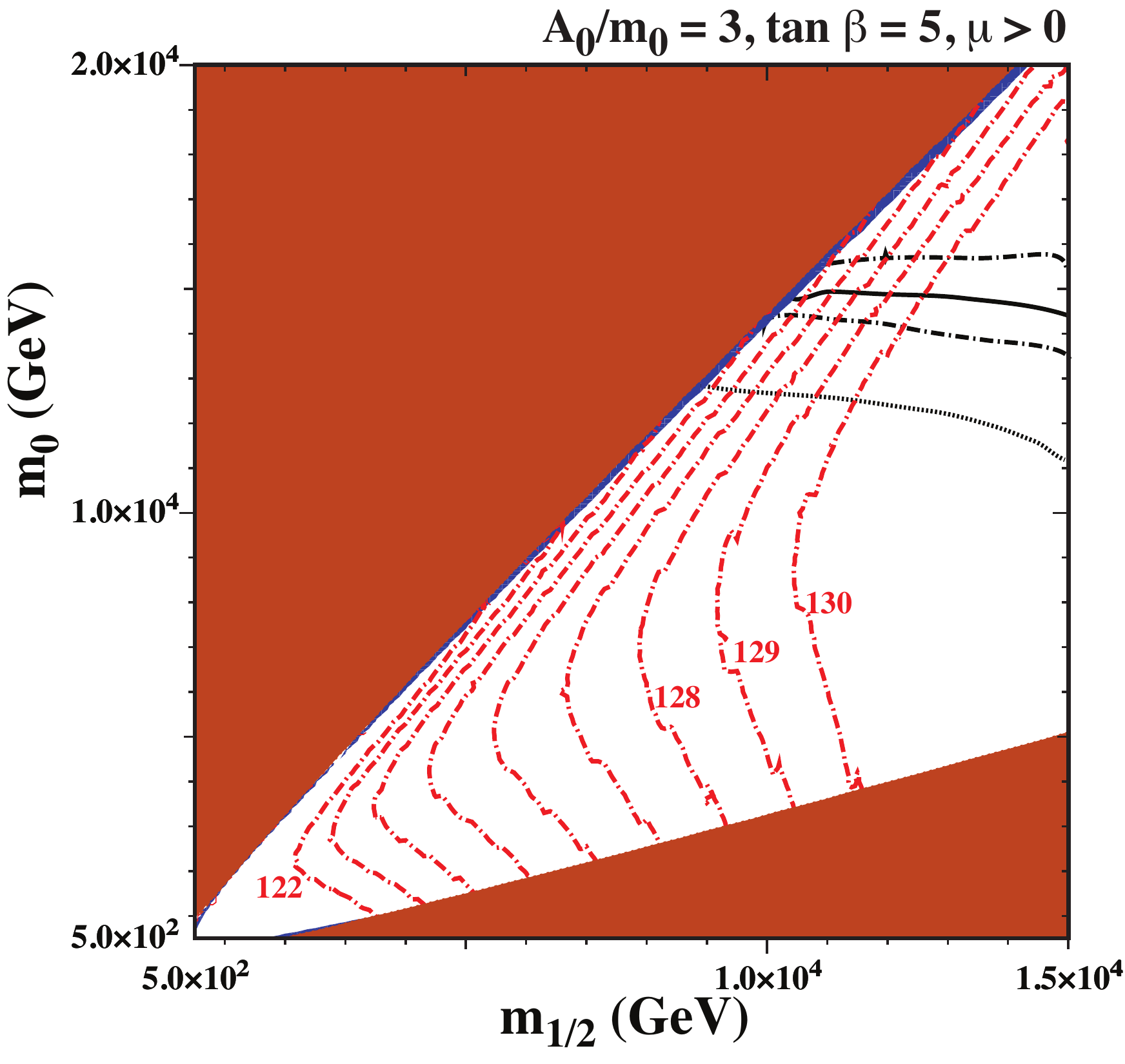}
\end{center}
\hfill
\end{minipage}
\vspace{-5mm}
\caption{
{\it The $(m_{1/2}, m_0)$ plane for $\tan \beta = 5$, $A_0/m_0 = 3$ and $\mu > 0$ in the CMSSM with GUT-scale universality, exhibiting in black the constraint on the lifetime of $p \to K^+ \bar{\nu}$ for the central value of all parameters in solid, for a one-standard-deviation variations in the hadronic matrix elements of the relevant 3-quark operators indicated in Table~\ref{tab:w0_error} (dot-dashed lines), and for a one-standard-deviation variations in both the hadronic matrix elements and $\alpha_s$ (dotted). The red dot-dashed contours show the lightest Higgs mass as calculated using {\tt FeynHiggs}~\cite{FH,FeynHiggs}. In the upper red shaded region, the lighter stop is the LSP, and in the lower red shaded region, the lighter stau is the LSP. In the blue
strip along the stop LSP region, the LSP has an enlarged relic density range, $0.01 < \Omega_\chi h^2 < 2.0$, to enhance its visibility. }}
\label{fig:had}
\end{figure}

We illustrate in  Fig.~\ref{fig:had} the effect of the uncertainties in the hadronic matrix elements
on the allowed ranges of the input parameters in a representative $(m_{1/2}, m_0)$ plane of the CMSSM. Here and in the remaining figures in this Section,
we consider the $(m_{1/2}, m_0)$ plane for the fixed values $\tan \beta = 5$,
$A_0/m_0 = 3$ and $\mu > 0$. The input universality scale is taken to be the GUT scale, defined as the renormalization scale for which $g_1 = g_2$. There are in general two dark red shaded regions in each figure.
The lower region where $m_{1/2} \gg m_0$ is excluded because there the lighter stau
is the lightest supersymmetric particle (LSP), whereas the upper region with $m_0 \gg m_{1/2}$ is excluded because there the lighter stop is the LSP. 
Along the boundary of  the stop LSP region, there is a very thin
blue strip where stop coannihilation \cite{stopco,eds,eoz,interplay,raza} is effective in reducing the LSP relic density to match the cold dark matter density determined by Planck \cite{Planck}.
We allow the relic density to vary between $0.01 < \Omega_\chi h^2 < 2.0$ to enhance the visibility of the strip in the figures. The points chosen in
the left panels of Fig.~\ref{fig:hadME} lie on this strip at $m_{1/2} = 9.79$ TeV. We note that regions of the $(m_{1/2}, m_0)$ plane that would correspond in a conventional cosmological scenario with adiabatic expansion to a cold dark matter density larger than that determined by Planck \cite{Planck} would, however, be allowed in scenarios with late entropy generation.~\footnote{For a recent analysis in such a scenario, see~\cite{EGNNO5}.} The bulk regions of the displayed planes could therefore be allowed in such a case.
The red dot-dashed lines are contours of constant Higgs masses between $m_h = 122$ and 130 GeV in increments of 1 GeV as calculated using {\tt FeynHiggs}~\cite{FH,FeynHiggs}.

The solid black curve in Fig.~\ref{fig:had} shows the contour of constant $p \to K^+ \bar{\nu}$ lifetime set at its
current lower limit of $0.066 \times 10^{35}$ yrs for central values of the matrix elements and other parameters. Its location is significantly higher than in previous work \cite{eemno}, due to the numerous updates incorporated here. 
These include updates to the hadronic matrix elements, the value of $\alpha_s$ and the value of $\sin^2 \theta_W$, as well as the one-loop correction to the charm quark Yukawa coupling. The dotted black curve shows the shift in the lower limit induced by a 1-$\sigma_{\rm tot}$ increase in the $p \to K^+ \bar{\nu}$ lifetime where, as described above, $\sigma_{\rm  tot}$ includes only
the contributions from the variations in the matrix elements, $\sigma_{\rm{had}}$,  and the effect of the variation in $\alpha_s$ on $M_{H_C}$, $\sigma_{\tau_p}^2$ (discussed further in the following Section) added in quadrature: 
\begin{equation}
\sigma_{\rm tot}=\sqrt{\sigma_{\rm{had}}^2+\sigma_{\tau_p}^2}\, .
\end{equation}
Here $\sigma_{\rm{had}}$ is the combined uncertainty due to all the hadronic matrix elements entering into $\tau_p$, which can be read from the amplitude of \eq{eq:ApKpnm}. The contour corresponding to a 1-$\sigma$ decrease in the proton lifetime is off the scale of the plot. 
The pair of dot-dashed curves show the $\pm 1 \sigma$ uncertainty stemming from the hadronic matrix elements alone. That is, we are plotting
the contours where $\tau_p \pm \sigma_{\rm had} = 0.066 \times 10^{35}$~years. 


\subsection{Dependence on the Strong Coupling}
\label{sec:alphas}

Despite the impressive reduction of the uncertainty in the experimental value of $\alpha_s$ \cite{PDG}, the proton lifetime still varies drastically for a 1-$\sigma$ change in $\alpha_s$. 
This can be understood from Eqs.~(\ref{eq:matchmhc}--\ref{eq:matchg5}).
In particular, Eq.~\eqref{eq:matchmhc}
indicates that the mass of the coloured Higgs field, $M_{H_C}$ is exponentially sensitive to changes in $\alpha_s = g_3^2/4 \pi$ \cite{Hisano:2013cqa}.~\footnote{However, as we discuss in more detail below, this sensitivity is substantially reduced when the dimension-five coupling $c$ in Eq.~\eqref{eq:SigmaWW} is allowed to be non-zero 
and vary while $\lambda$ and $\lambda^\prime$ are kept fixed.} As the relevant Wilson coefficients are inversely proportional to the coloured Higgs mass, the proton decay rate is proportional to $M_{H_C}^{-2}$, and so is quite sensitive to variations in $\alpha_s$.

To estimate numerically the sensitivity of the proton decay width to variation in $\alpha_s$, we solve Eq.~\eqref{eq:matchmhc} for $M_{H_C}$ and assume $\Gamma_p=K/M^2_{H_C}$ where $K$ is independent of $\alpha_s$. We then compute 
\begin{align}
\sigma_{\tau_p} \equiv \frac{d\tau_p}{d g^2_3 (M_Z)}(4\pi\Delta_{\alpha_s}) =-\frac{1}{\Gamma^2_p}\frac{d\Gamma_p}{d g^2_3(M_Z)}(4\pi\Delta_{\alpha_s})~,    
\end{align}
where $\Delta_{\alpha_s}$ is the 1-$\sigma$ uncertainty in $\alpha_s$ and we have assumed that the variation of $M_{H_C}$ is dominantly determined by $g_3$. This then gives
\begin{align}
\sigma_{\tau_p} &\simeq \tau_p \left(\frac{10\pi}{3}\right)\left(\frac{\Delta_{\alpha_s}}{\alpha_s(M_Z)^2} \right)
\nonumber\\
&= 0.83 \left(\frac{\Delta_{\alpha_s}}{0.0011}\right)\left(\frac{0.1181}{\alpha_s(M_Z)} \right)^2 \tau_p.
\label{withoute}
\end{align}
However, a variation in $\alpha_s$
also leads to changes in the light quark masses of roughly 2\% for a 1$\sigma$ change in $\alpha_s$, which
leads in turn to an 8\% change in the proton decay rate. As indicated above, when
we consider the `total' proton lifetime 
uncertainty, the variation connected to the quark masses is not included, though it is included when we show the isolated effect using $\alpha_s \pm \Delta_{\alpha_s}$. 
For this reason, the 1-$\sigma$ spread due to $\alpha_s$ shown below appears larger than the `total' uncertainty.

\begin{figure}[htb!]
\begin{minipage}{6in}
\begin{center}
\includegraphics[height=3.5in]{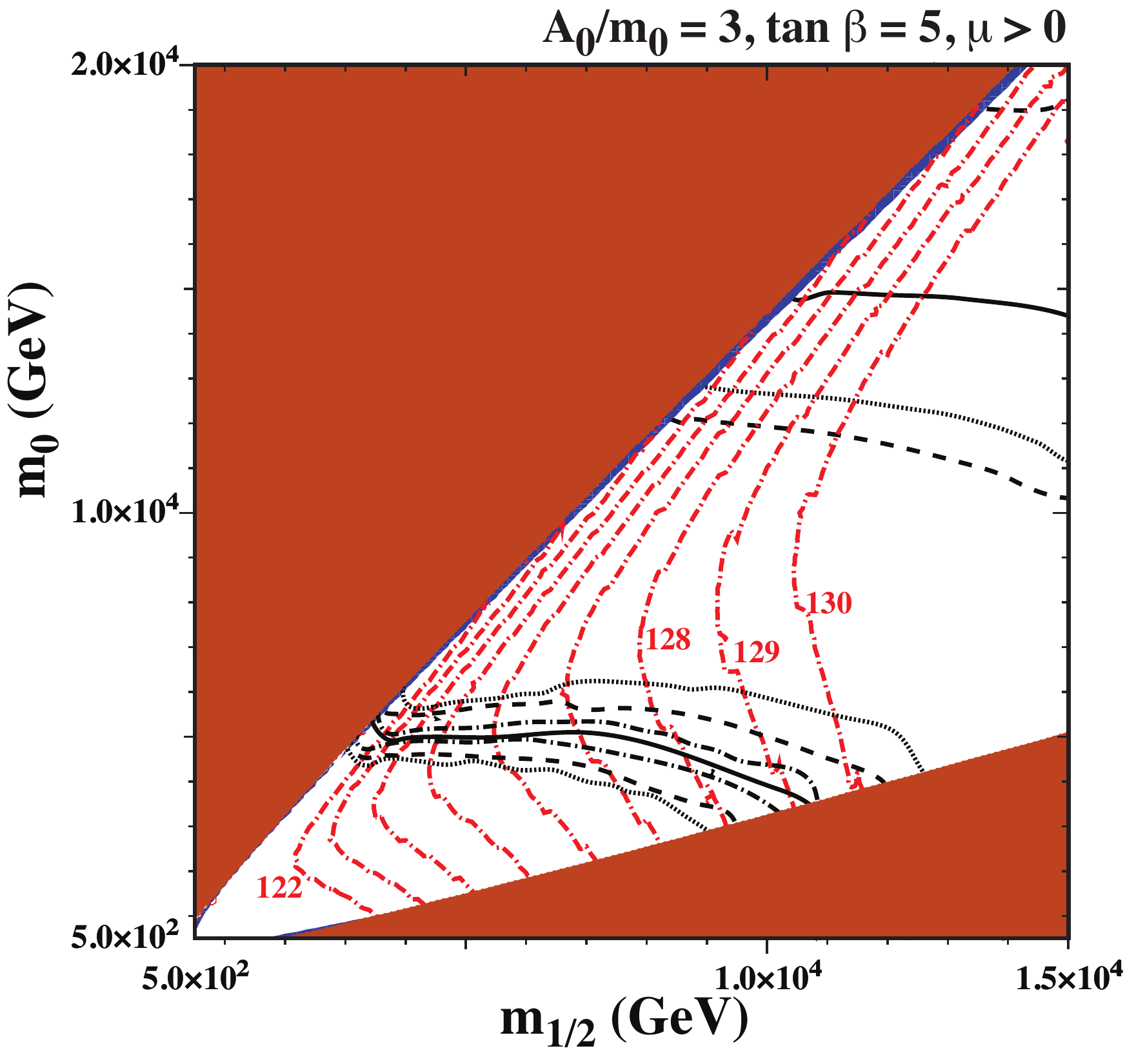}
\end{center}
\hfill
\end{minipage}
\vspace{-5mm}
\caption{
{\it
Sensitivity of the $p \to K^+ \bar{\nu}$ lifetime to $\alpha_s$, assuming $\tan \beta = 5, A_0/m_0 = 3, \mu > 0$ in the CMSSM with GUT-scale universality. In the upper set of curves (with $c = 0$), the solid black curve corresponds to the current lower limit of $0.066 \times 10^{35}$ yrs on $\tau ( p \to K^+ \bar{\nu})$ for central values of the matrix elements and model parameters, and the black dashed lines correspond to the variation of $\alpha_s$ within one standard deviation: $\alpha_s = 0.1181 \pm .0011$. The dotted curve corresponds to the shift in the solid curve when the decay rate is decreased by $\sigma_{\rm tot}$. In the lower set of curves, the coupling associated with the dimension 5 operator is non-zero, and the meanings of the curves are similar. Also shown is the propagated uncertainty in $\alpha_s$ alone, shown by the dot-dashed curves, which are now clearly distinct from the dotted curves using $\sigma_{\rm tot}$.
}}
\label{fig:alphas}
\end{figure}

The upper set of curves in Fig.~\ref{fig:alphas} have been produced assuming that the coupling of the dimension-five operator $c$ vanishes. As in 
Fig.~\ref{fig:had}, 
the solid black curve in Fig.~\ref{fig:alphas} shows the contour of constant $p \to K^+ \bar{\nu}$ lifetime set at its
current lower limit of $0.066 \times 10^{35}$~yrs for central values of the matrix elements and model parameters.
Similarly, the dotted black curve shows the shift in the lower limit induced by a 1-$\sigma_{\rm tot}$ decrease in the proton decay rate.

We also show as dashed black lines in Fig.~\ref{fig:alphas} the result of varying $\alpha_s$ within its uncertainty $\pm 0.0011$ when determining all the supersymmetric spectrum and other observables, including $\tau(p \to K^+ \bar{\nu})$. The lower black dashed line in Fig.~\ref{fig:alphas} corresponds to $\tau(p \to K^+ \bar{\nu}) = 0.066 \times 10^{35}$ yrs when calculated with $\alpha_s=0.1192$, whereas the upper black dashed line corresponds to calculations with $\alpha_s=0.117$, i.e., $\pm 1-\sigma$ excursions in $\alpha_s$. 
As noted above, the strong dependence on $\alpha_s$ can be understood largely from the fact that varying $\alpha_s$ affects $M_{H_C}$, but also from the variations in the values of the light quark masses when evolved both between $M_Z$ and $M_{\rm GUT}$  and between $2$ GeV or $m_c$ and $M_Z$. 

 In particular, there are two aspects that affect the evolution between $M_Z$ and $M_{\rm GUT}$. First, the running of the light quark masses is completely controlled by $\alpha_s$ due to the relatively small values of the couplings $f_d$, $f_s$ and $f_c$: 
 \begin{align}
\beta_{f^2_d} = & \frac{f^2_d}{8\pi^2}\left(6 f^2_d +3 (f^2_s+f^2_b) -\frac{16}{3}g^2_3 -3 g^2_2 -\frac{7}{9} g^{\prime 2}\right) ~,\\
\beta_{f^2_s} = & \frac{f^2_s}{8\pi^2} \left( 6 f^2_s+ 3(f^2_d + f^2_b) + f^2_c -\frac{16}{3}g^2_3 -3 g^2_2 -\frac{7}{9}g^{\prime 2}\right) ~,\\
\beta_{f^2_c}= &\frac{f^2_c}{8\pi^2}\left(6 f^2_c + 3 f^2_t + f^2_s -\frac{16}{3}g^2_3 -3 g^2_2 -\frac{13}{9} g^{\prime 2} \right)   ~,
\end{align}
where $g^\prime = g_1 \sqrt{3/5}$.
This contrasts with the running of the $b$ and $t$ Yukawa couplings, e.g.,
\begin{equation}
\beta_{f^2_b}= \frac{f_b^2}{8\pi^2}\ \left( f_t^2 + 6 f_b^2 + f^2_\tau - \frac{16}{3} g_3^2 -3 g_2^2 - \frac{7}{9} g^{\prime 2} \right) \, .
\end{equation}
Secondly, the one-loop corrections, $\Delta m_{q}$, to light-quark masses due to gluino loops are also controlled by $\alpha_s$, e.g., for $m_s$ \cite{Pierce:1996zz}:
\begin{align}
    \Delta m_s \supset &- \frac{g_3^2}{12\pi^2}
    \biggl\{
     B_1(M_3,m_{\tilde s_1} )+ B_1(M_3,m_{\tilde s_2})  
     \nonumber \\
     &-\sin(2 \theta_{m_s}) \biggl(\frac{M_3}{m_s}\biggr)
     \biggl[B_0(M_3,m_{\tilde s_1})-B_0(M_3, m_{\tilde s_2})\biggr] 
    \biggr\} ~,
\end{align}
where $B_0$ and $B_1$ are the Passarino-Veltman functions given in Ref.~\cite{Pierce:1996zz}, and $\theta_{m_s}$ represents the mixing angle between the strange squarks $\tilde s_1$ and $\tilde s_2$ with mass eigenvalues of $m_{\tilde{s}_1}$ and $m_{\tilde{s}_2}$, respectively, which is determined by $m_s(A_s+\mu\tan \beta)/(M^2_{{\tilde Q}_2}- M^2_{{\tilde D}_2})$. 
Therefore,
for larger $\alpha_s$ there is a bigger change in $m_q$, and hence a bigger change in the value of the decay amplitude for the proton lifetime. 

We note that the sensitivity to $\alpha_s$ is more pronounced for $\mu < 0$ than for $\mu > 0$, simply because the angle $\theta_{m_s}$ above changes sign due to the change of sign in $\mu$, giving a bigger contribution to $\Delta m_s$ for $\mu<0$. For example, in the region of $m_0\sim 10$ TeV and $M_{1/2}\sim 10$ TeV, the contribution to $\Delta m_s$ from gluinos is of the order of 26\% for $\mu$ negative, while for $\mu$ positive it is of order $13\%$.

 As for the values of the light quark masses at $M_Z$, it is well known that $\alpha_s$ increases considerably between $M_Z$ and $m_c$. Hence a 1-$\sigma$ change 
 in $\alpha_s$ affects by about $2\%$ the estimation of the light quark masses at $M_Z$ that we use as inputs to our calculations at $M_Z$, as shown in Table~\ref{tab:mqMz}.

\begin{table}[!ht]
\begin{center}
\caption{{\it Quark Masses at $M_Z$ in the
$\overline{\rm{DR}}$ Prescription.}
\label{tbl:qkmassesDRMZ}}
\label{tab:mqMz}
\vspace{-7mm}
\begin{tabular}{|c|c|c|c|}
\multicolumn{3}{c}{}\\
\hline
\hline
 & $m_q(M_Z)_{\alpha_s=0.117}$  & $m_q(M_Z)_{\alpha_s=0.1181}$  & $m_q(M_Z)_{\alpha_s=0.1192}$\\
\hline
$m_d$ &  $2.70\times 10^{-3}$ & $2.67\times 10^{-3}$ & $2.64\times 10^{-3}$  \\
$m_s$ &  $5.37\times 10^{-2}$ &   $5.31\times 10^{-2}$   & $5.25\times 10^{-2}$   \\
$m_c$ &  $0.633$                     & $0.622$ & $0.610$ \\
\hline
\hline
 \end{tabular}
  \end{center}
 \end{table}

In contrast to the above analysis, we show in the lower set of curves in Fig.~\ref{fig:alphas} the contours of $\tau(p \to K^+ \bar{\nu}) = 0.066 \times 10^{35}$ yrs when the dimension-five operator coupling, $c$, is allowed to be non-zero with $\lambda$ and $\lambda^\prime$ fixed, e.g., here we set $\lambda = 0.6$ and $\lambda^\prime = 0.0001$. In this case, Eq.~\eqref{eq:matchmgut} fixes the combination $M_X^2 M_\Sigma$ that in turn fixes the the coloured Higgs mass:
\begin{equation}
    M_{H_C} = \lambda \left( \frac{2 M_X^2 M_\Sigma }{\lambda^\prime g_5^2} \right)^{1/3}\, .
    \label{mhc}
\end{equation}
Then Eq. (\ref{withoute}) becomes 
\begin{align}
\sigma_{\tau_p} &\simeq \tau_p \left(\frac{2\pi}{9}\right)\left(\frac{\Delta_{\alpha_s}}{\alpha_s(M_Z)^2} \right)    \nonumber \\
&=0.055 \left(\frac{\Delta_{\alpha_s}}{0.0011}\right)\left(\frac{0.1181}{\alpha_s(M_Z)} \right)^2 \tau_p \, .
\label{withe}
\end{align}
Thus we expect the uncertainty in the proton lifetime to be significantly less sensitive to the uncertainty in $\alpha_s$, by a factor $\sim 1/15$. 

This is seen in the lower set of curves in Fig.~\ref{fig:alphas}.
Since $M_{H_C}$ is now essentially fixed,
the proton lifetime is substantially larger, and the limit contour (shown again as the solid curve) appears at lower $m_{1/2}$ and $m_0$.
On either side of the central curve, we show 3 sets of curves displaying the
uncertainty due to $\alpha_s$.
Nearest the centre, the dot-dashed curves correspond to the propagated variation due to $\alpha_s$ alone. In the upper set of curves, we did not show this, as it would have been indistinguishable from the dotted curve showing the total sensitivity, which was dominated by $\alpha_s$. Here we see clearly that, with $c\ne0$, the 
uncertainty due to $\alpha_s$ is greatly diminished. The dashed curves show the shift in the limit contour when the values
$\alpha_s = 0.1170$ and 0.1192
are used for the supersymmetric spectrum and all other observables, as was done in computing the dashed curves in the upper part of the figure with $c = 0$.
Finally, the dotted curves show the total propagated uncertainty,
which is now dominated by the 
uncertainty in the matrix elements. 


\subsection{Dependence on the Weak Mixing Angle}
\label{sec:sin2thetaW}

It was assumed in previous work~\cite{eemno} that $\sin^2 \theta_W = 0.2325$ in the $\overline{\rm MS}$ prescription, and this value was taken as an input condition at $M_Z$. However, the precision of measurements of electroweak symmetry breaking (EWSB) observables warrants paying careful attention to the precise input value of $\sin^2 \theta_W$, and in our calculations here we specify $\sin^2 \theta_W$ in the $\overline{\text{DR}}$ scheme, which can be extracted from
\begin{align}
\label{eq:DRWeinbergA}
\sin^2 \theta_W\vert_{\overline{\rm DR}} &
=1-
    \left(\frac{M_{W,\text{susy}}^{\text{\DRbar}}(M_Z)}{M_{Z,\text{susy}}^{\text{\DRbar}}(M_Z)}\right)^2,
\end{align}
where the quantities ${M_{(W,Z),\text{susy}}^{\text{\DRbar}}(M_Z)}$ contain one-loop corrections to the $W$ and $Z$ boson masses calculated in the $\DRbar$ scheme.
Alternatively, $\sin^2 \theta_W\vert_{\overline{\rm DR}}$
can be extracted from  
\begin{align}
\sin^2 \theta_W\vert_{\text{\DRbar}}= \frac{\sin^2  \theta_{W,\text{eff}}(M_Z)}{\rm{Re} \ \hat \kappa_{\ell}} \, ,
\end{align}
where the current measurement of $\sin^2 \theta_{W,\text{eff}}(M_Z)$ is $0.23155(4)$ \cite{PDG}, and 
\begin{align}
\label{eq:kdeff}
\hat \kappa_{\ell}=1 +\frac{\hat c}{\hat s}\frac{\Pi_{Z\gamma}(M^2_Z)-\Pi_{Z\gamma}(0)}{M_Z^2}+ \frac{\hat\alpha}{\pi}\frac{\hat{c}^2}{\hat{s}^2} \log c^2 - \frac{\hat \alpha}{4\pi \hat{s}^2} V_{\ell} (M^2_Z) \, ,
\end{align}
where $\Pi_{Z\gamma}(p^2)$ is the mixed self-energy of $Z$ and $\gamma$ at the momentum scale $p^2$, and $V_{\ell}$ is a function of the
$\overline{\text{DR}}$ quantities, $\hat{s} \equiv \sin \theta_{W,\text{susy}}^{\text{\DRbar}}(M_Z)$ and $\hat{c}^2=1-\hat{s}^2$ \cite{Degrassi:1990ec,Pierce:1996zz}. In the expression Eq.~\eqref{eq:kdeff} above, as a first approximation $\hat{c}$ on the right-hand side of the equation   can be taken as  $c=\cos \theta_{W,\text{eff}}(M_Z)$, rather than $\hat{c}=\cos\theta_{W,\text{susy}}^{\text{\DRbar}}(M_Z)$.
We choose to use $\sin^2 \theta_{W,\text{eff}}$ as our starting-point, since the MSSM corrections to $\sin^2 \theta_{W,\text{eff}}(M_Z)$ have been  studied in some depth. In particular, for supersymmetric masses bigger than 1 TeV these corrections are known to be ${\cal O}(10^{-4})$ and always negative \cite{Pierce:1996zz}. 

The factor $\hat \kappa_{\ell}$ in \eq{eq:kdeff} can be interpreted as the conversion factor to the $\overline{\text{DR}}$ scheme, where typically  $1/{{\rm{Re} \ \hat \kappa_{\ell}}}$ represents a decrease by another amount of ${\cal O}(10^{-4})$. Therefore we expect $\sin^2 \theta_{W,\text{susy}}^{\text{\DRbar}}(M_Z)$ to vary in the range $(0.2312, 0.2315)$, depending on the supersymmetric contribution to $\sin^2 \theta_{W,\text{eff}}(M_Z)$. 
Since the proton lifetime is relatively insensitive to variations of ${\cal O}(10^{-4})$ in $\sin^2 \theta_W\vert_{\overline{\rm DR}}$, as we see below, we consider a precise computation for each point of the parameter space to lie beyond the scope of this work. 

\begin{figure}[htb!]
\begin{minipage}{6in}
\begin{center}
\includegraphics[height=3.5in]{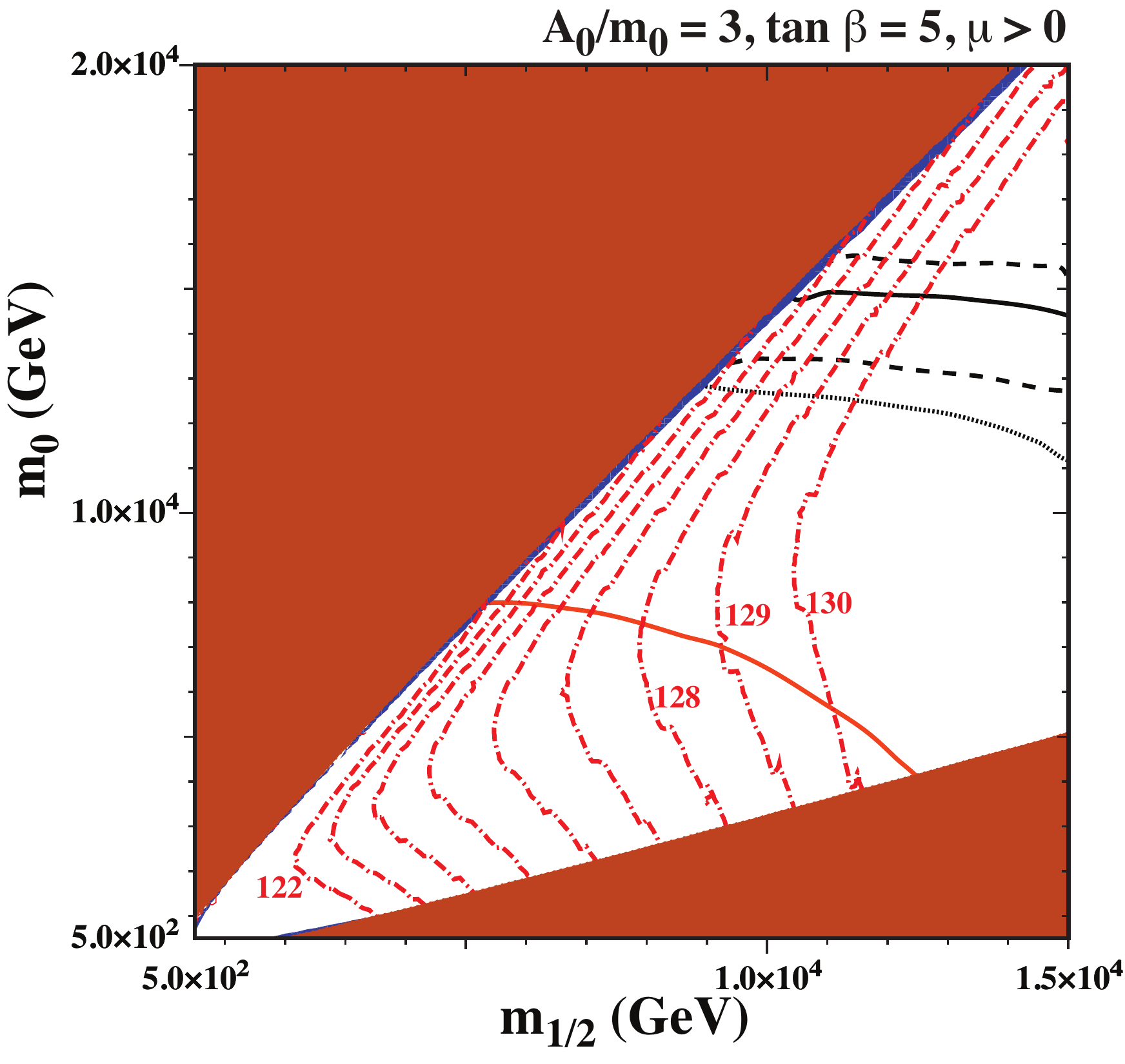}
\end{center}
\hfill
\end{minipage}
\vspace{-5mm}
\caption{
{\it Sensitivity of the proton decay rate to $\sin^2 \theta_W\vert_{\overline{DR}}$ in the range 0.2312 to 0.2315 (upper and lower black dashed lines). As in the previous Figure, the solid black contour shows the contour $\tau(p \to K^+ \bar{\nu}) = 0.066 \times 10^{35}$ yrs for central values of the inputs, and the dotted black contour shows the shift in this limit due to the $\sigma_{\rm tot}$ uncertainty. The solid brown curve shows the corresponding contour calculated using $\sin^2 \theta_W = 0.2325$ in the $\overline{MS}$ prescription.}}
\label{fig:sin2theta}
\end{figure}

We illustrate in Fig.~\ref{fig:sin2theta} the sensitivity
of the $p \to K^+ \bar{\nu}$ lifetime calculation to varying $\sin^2 \theta_W\vert_{\overline{\rm DR}}$ over the range 0.2312 to 0.2315, corresponding to the upper and lower black dashed lines.
As in Fig.~\ref{fig:alphas}, the solid black contour shows the position of the limit $\tau(p \to K^+ \bar{\nu}) = 0.066 \times 10^{35}$ yrs for central values of the inputs, and the dotted
black contour shows the shift in this limit due to the $\sigma_{\rm tot}$ uncertainty. 
We see that the induced uncertainty associated with $\sin^2 \theta_W$ is significantly smaller than that due to the hadronic matrix elements and $\alpha_s$. On the other hand, using the previous value of 0.2325 would have given quite different results,
as illustrated by the solid brown curve in Fig.~\ref{fig:sin2theta}. 

\subsection{Sensitivities to Quark Masses}
\label{sec:mq}

The amplitudes ${\cal A}(p\to K^+\bar{\nu}_i)$ for $i = e, \mu, \tau$ are the following sums of products of the Wilson
coefficients with hadronic matrix elements:
\begin{align}
 {\cal A}(p\to K^+\bar{\nu}_e)&=
C_{LL}(usd\nu_e)\langle K^+\vert (us)_Ld_L\vert p\rangle
+C_{LL}(uds\nu_e)\langle K^+\vert (ud)_Ls_L\vert p\rangle ~,
\nonumber \\
 {\cal A}(p\to K^+\bar{\nu}_\mu)&=
C_{LL}(usd\nu_\mu)\langle K^+\vert (us)_Ld_L\vert p\rangle
+C_{LL}(uds\nu_\mu)\langle K^+\vert (ud)_Ls_L\vert p\rangle ~,
\nonumber \\
 {\cal A}(p\to K^+\bar{\nu}_\tau)&=
C_{RL}(usd\nu_\tau)\langle K^+\vert (us)_Rd_L\vert p\rangle
+
C_{RL}(uds\nu_\tau)\langle K^+\vert (ud)_Rs_L\vert p\rangle 
\nonumber \\
&+
C_{LL}(usd\nu_\tau)\langle K^+\vert (us)_Ld_L\vert p\rangle
+C_{LL}(uds\nu_\tau)\langle K^+\vert (ud)_Ls_L\vert p\rangle
~.
\label{eq:ApKnui}
\end{align}
Wino exchange contributes to the Wilson coefficients $C_{LL} (usd\nu_i)$ and $C_{LL} (uds\nu_i)$,
which may be approximated by
\begin{align}
 C_{LL} (usd\nu_i) &= C_{LL} (uds\nu_i) \nonumber \\
&\simeq
\frac{2\alpha_2^2}{\sin 2\beta}\frac{m_t m_{d_i} M_2}{ m_W^2M_{H_C}M_{\rm
 SUSY}^2} V_{ui}^*V_{td}V_{ts}e^{i\phi_3}\left(1 +
 e^{i(\phi_2-\phi_3)}\frac{m_c V_{cd}V_{cs}}{m_tV_{td}V_{ts}}\right) ~,
\label{eq:cllaprox}
\end{align}
where $m_{d_i}$ are the masses of the down-type quarks. 
On the other hand, as can be seen from Eq.~\eqref{eq:effWCPD_weak}, Higgsino exchange contributes only to 
${\cal A}(p\to K^+\bar{\nu}_\tau)$, via $C_{RL}(usd\nu_\tau)$ and $C_{RL}(uds\nu_\tau)$, which 
 are given approximately by 
 \begin{align}
 C_{RL} (usd\nu_\tau) &\simeq -
 \frac{\alpha_2^2}{\sin^2 2\beta}\frac{m_t^2 m_s m_\tau \mu}{ m_W^4 M_{H_C}M_{\rm SUSY}^2}V_{tb}^*
 V_{us} V_{td}e^{-i(\phi_2+\phi_3)} ~, \label{eq:crlaprox1} \\
 C_{RL} (uds\nu_\tau) &\simeq -
 \frac{\alpha_2^2}{\sin^2 2\beta}\frac{m_t^2 m_d m_\tau \mu}{m_W^4 M_{H_C} M_{\rm SUSY}^2}V_{tb}^*
 V_{ud} V_{ts}e^{-i(\phi_2+\phi_3)} ~,
\label{eq:crlaprox}
\end{align}
We find that the total decay width $\Gamma(p\rightarrow K^+ \overline{\nu})=$ $\sum_{i=e,\mu,\tau} \Gamma(p\rightarrow K^+ \overline{\nu}_i)$
is dominated throughout the plane by the contributions 
\begin{align}
C_{RL}(uds\nu_\tau)\langle K^+|(us)_Rd_L|p\rangle,\quad
C_{LL}(uds\nu_\mu)\langle K^+|(ud)_Ls_L|p\rangle \, ,
\end{align}
as a result of the dependences on quark masses, CKM elements and phases that we describe in this and the following Sections.

We first discuss the sensitivity to $m_s$ in the range $m_s = 93^{+11}_{-5}$~MeV when $\tan \beta = 5$ and $A_0/m_0 = 3$, assuming GUT-scale universality. 
When the GUT phases are zero, all the contributions in Eq.~(\ref{eq:ApKnui}) are of the same order and the contribution of the second term in $C_{LL}(usd\nu_2)$ is maximized, see Eq.~(\ref{eq:cllaprox}), rendering this contribution of the same size or, in most of the parameter space, even larger than that proportional to $C_{RL}(uds\nu_\tau)$.

\begin{figure}[htb!]
\begin{minipage}{6in}
\begin{center}
\includegraphics[height=3.5in]{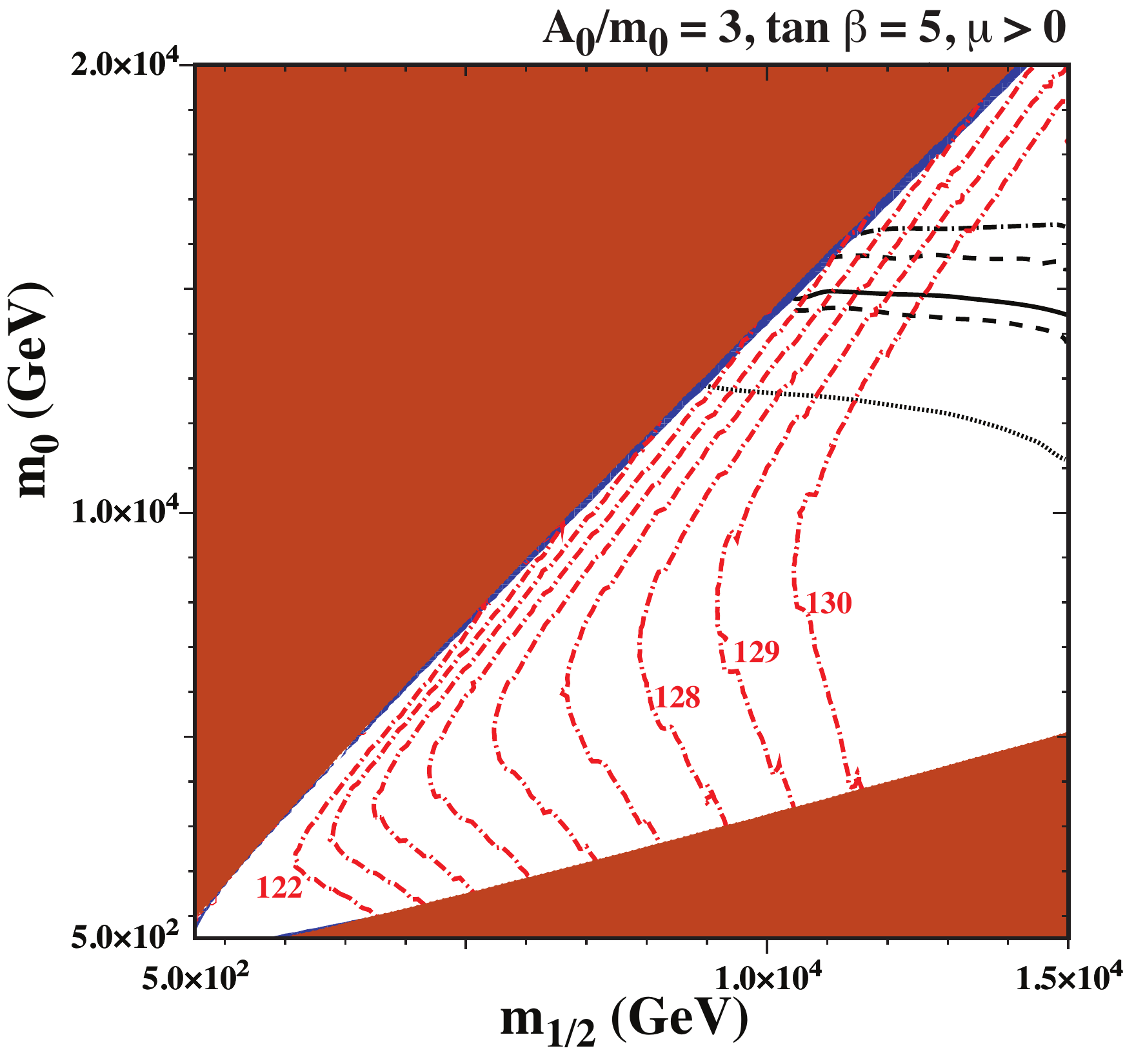}
\end{center}
\hfill
\end{minipage}
\vspace{-5mm}
\caption{
{\it Sensitivity of the $p\rightarrow K^+ \overline{\nu}$ lifetime calculation to $m_s$ for $\tan \beta = 5, A_0/m_0 = 3, M_{\rm in} = M_{\rm GUT}$ and $\mu > 0$. The dashed lines correspond to the variation of $m_s$ within one standard deviation ($m_s = 93^{+11}_{-5}$~MeV) and the dotted line corresponds to the combined uncertainty from the hadronic matrix elements and $\alpha_s$. The dot-dashed curve shows the position of the lifetime limit when the 1-loop correction to $m_s$ is removed.}}
\label{fig:ms}
\end{figure}

Since $C_{LL}(usd\nu_2)$ is proportional to $m_s$ and the uncertainty in $m_s$ is between $- 5\%$ and $+ 12\%$, any change in $m_s$ affects the $p\rightarrow K^+ \overline{\nu}$ lifetime more than the other quark masses. 
We show in Fig.~\ref{fig:ms} the $p\rightarrow K^+ \overline{\nu}$ lifetime calculated with the central value of
$m_s$ (solid black line), while the black dashed lines correspond to $114$ MeV (upper line) and  $88$ MeV (lower line). We see that this uncertainty is much smaller than that corresponding to the combined uncertainty from the hadronic matrix elements and $\alpha_s$ (lower dotted line).

\begin{figure}[!htb]
\begin{minipage}{6in}
\begin{center}
\includegraphics[height=3.5in]{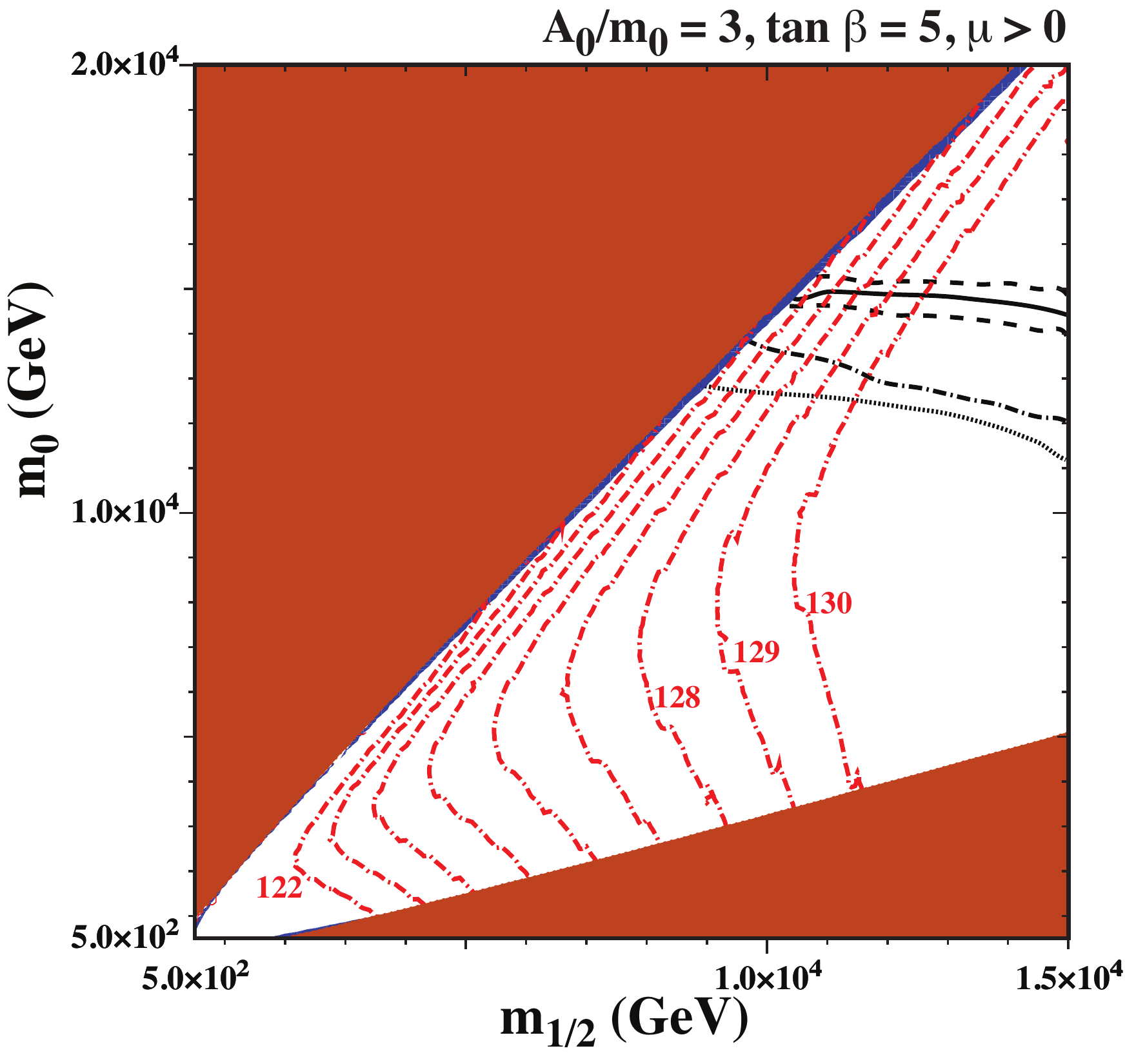}
\end{center}
\hfill
\end{minipage}
\vspace{-5mm}
\caption{
{\it
Sensitivity to $m_c$ for $\mu > 0$, assuming $\tan \beta = 5$, $A_0/m_0 = 3$ and $M_{\rm in} = M_{\rm GUT}$. The dashed lines correspond to the variation of $m_c$ within one standard deviation ($m_c = 1.27 \pm 0.02$~GeV) and the dotted line corresponds to the combined uncertainty from the hadronic matrix elements and $\alpha_s$. The dot-dashed curve shows the position of the lifetime limit when the 1-loop correction to $m_c$ is removed.
}}
\label{fig:mc}
\end{figure}

The second most important quark-mass sensitivity is that to $m_c$, which contributes to the second term in Eq.~(\ref{eq:cllaprox}). We vary $m_c$ in the range $m_c = 1.27 \pm 0.02$~GeV, and find that, when $\tan \beta = 5, A_0/m_0 = 3$ and $M_{\rm in} = M_{\rm GUT}$ the sensitivity to $m_c$ is less than that due to the uncertainty in $m_s$ as seen in Fig.~\ref{fig:mc}. 

The sensitivity to the  masses $m_d$, $m_b$ and $m_t$ is very mild.  We can see from Eq.~(\ref{eq:crlaprox})
that $C_{RL}(uds\nu_\tau)\propto m_d$, but the contribution to $\tau(p\rightarrow K^+ \overline{\nu})$ from this Wilson coefficient is suppressed when the GUT phases are chosen so as to maximize
the contribution to ${\cal A}(p\to K^+\bar{\nu}_\mu)$ proportional to
$C_{LL}(usd\nu_\mu)\langle K^+\vert (us)_Ld_L\vert p\rangle$.
Given the precision in the measurement of the top mass, $m_t = 172.9 \pm 0.4$~GeV \cite{PDG}, its effect on the p-lifetime is negligible, even though all the dominant Wilson coefficients, Eqs.~(\ref{eq:cllaprox}) and (\ref{eq:crlaprox}), are proportional to $f_t$. The sensitivity to $m_b = 4.18^{+0.03}_{-0.02}$~GeV is also
negligible, because it does not enter in either of the leading contributions, $C_{RL}(uds\nu_\tau)\langle K^+|(us)_Rd_L|p\rangle$ and
$C_{LL}(uds\nu_\mu)\langle K^+|(ud)_Ls_L|p\rangle$, to the total decay amplitude.

\subsection{Sensitivities to One-Loop Mass Renormalization Effects}
\label{sec:charm}

In Section \ref{sec:alphas} we detailed how $\alpha_s$ enters, and controls, the 1-loop corrections. In Table~\ref{tab:mqMz1loop} we illustrate the effects of varying $\alpha_s$ in the 1-loop corrections to the quark masses with examples for two choices of ($m_{1/2}, m_0$). By comparing Tables~\ref{tbl:qkmassesDRMZ} and \ref{tab:mqMz1loop} we see that for the case of $m_d$ and $m_s$ the effects range from $16\%$ up to $20\%$, while for $m_c$ the effect is no more than $10\%$.

In the previous Section, we discussed how $m_d$, $m_s$, and $m_c$ enter the total decay width,  as they contribute to  Eqs.~(\ref{eq:cllaprox}--\ref{eq:crlaprox}). In Figs.~\ref{fig:ms} and \ref{fig:mc} we show the effects of the  1-loop corrections to $m_s$
and $m_c$. In Fig.~\ref{fig:ms}, the dot-dashed line shows the position of the lifetime limit
when the 1-loop corrections to $m_s$ are ignored. As one can see, the curve lies above the nominal central contour (where the correction is included), indicating that the correction to $m_s$ increases the lifetime and weakens the limit (allowing lower sparticle masses). The effect of the correction to $m_d$ is qualitatively similar but less important and is not shown. In contrast, 
the dot-dashed line in Fig.~\ref{fig:mc} shows the
p-lifetime limit calculating $m_c(M_Z)$ without loop corrections. In this case, we see that the one-loop correction significantly decreases the proton lifetime, making the constraint stronger so that the central limit lies at higher sparticle masses. This effect has a bigger impact than the $1.6\%$ variation due to the uncertainty in $m_c$.

\begin{table}[!htb]
\begin{center}
\caption{\it One-Loop-Corrected Quark Masses at $M_Z$. 
}
\label{tab:mqMz1loop}
\vspace{-7mm}
\begin{tabular}{|c|c|c|c|}
\multicolumn{4}{c}{}\\
\hline
\hline
\multicolumn{4}{|c|}{$m_q^{1\ell}(M_Z)$ \small{for $m_{1/2}=7$ TeV, $m_0=5$ TeV}} \\
\hline
& $m_q(M_Z)_{\alpha_s=0.117}$  & $m_q(M_Z)_{\alpha_s=0.1181}$  & $m_q(M_Z)_{\alpha_s=0.1192}$\\
\hline
$m_d$ & $2.27\times 10^{-3}$ & $2.25\times 10^{-3}$ & $2.22\times 10^{-3}$ \\
$m_s$ & $4.53 \times 10^{-2}$ & $4.47\times 10^{-2}$ & $4.41\times 10^{-2}$\\
$m_c$ &  $0.700$& $0.689$ & $0.677$\\
\hline
\multicolumn{4}{|c|}{$m_q^{1\ell}(M_Z)_B$ \small{for $m_{1/2}=10$ TeV, $m_0=15$ TeV}} \\
\hline
& $m_q(M_Z)_{\alpha_s=0.117}$  & $m_q(M_Z)_{\alpha_s=0.1181}$  & $m_q(M_Z)_{\alpha_s=0.1192}$\\
\hline
$m_d$ &  $2.22\times 10^{-3}$  & $2.19\times 10^{-3} $ & $2.16\times 10^{-3} $ \\
$m_s$ &  $4.42 \times 10^{-2}$ & $4.36\times 10^{-2}$ &  $4.30\times 10^{-2}$\\
$m_c$ &  $0.706$ & $0.694$ & $0.682$ \\
\hline
\hline
 \end{tabular}
  \end{center}
 \end{table}

\subsection{Quark Mixing Uncertainties}
\label{sec:mixing}

The minimal SU(5) GUT does not contain a way to describe fermion mixing, but we know that any additional part of the theory which can describe it must reproduce at low energy the CKM elements within their experimental error. We therefore explore the sensitivity to this uncertainty through the fitted values of the Wolfenstein parameterisation of the CKM matrix~\cite{PDG}.

\begin{figure}[htb!]
\begin{minipage}{6in}
\begin{center}
\includegraphics[height=3.5in]{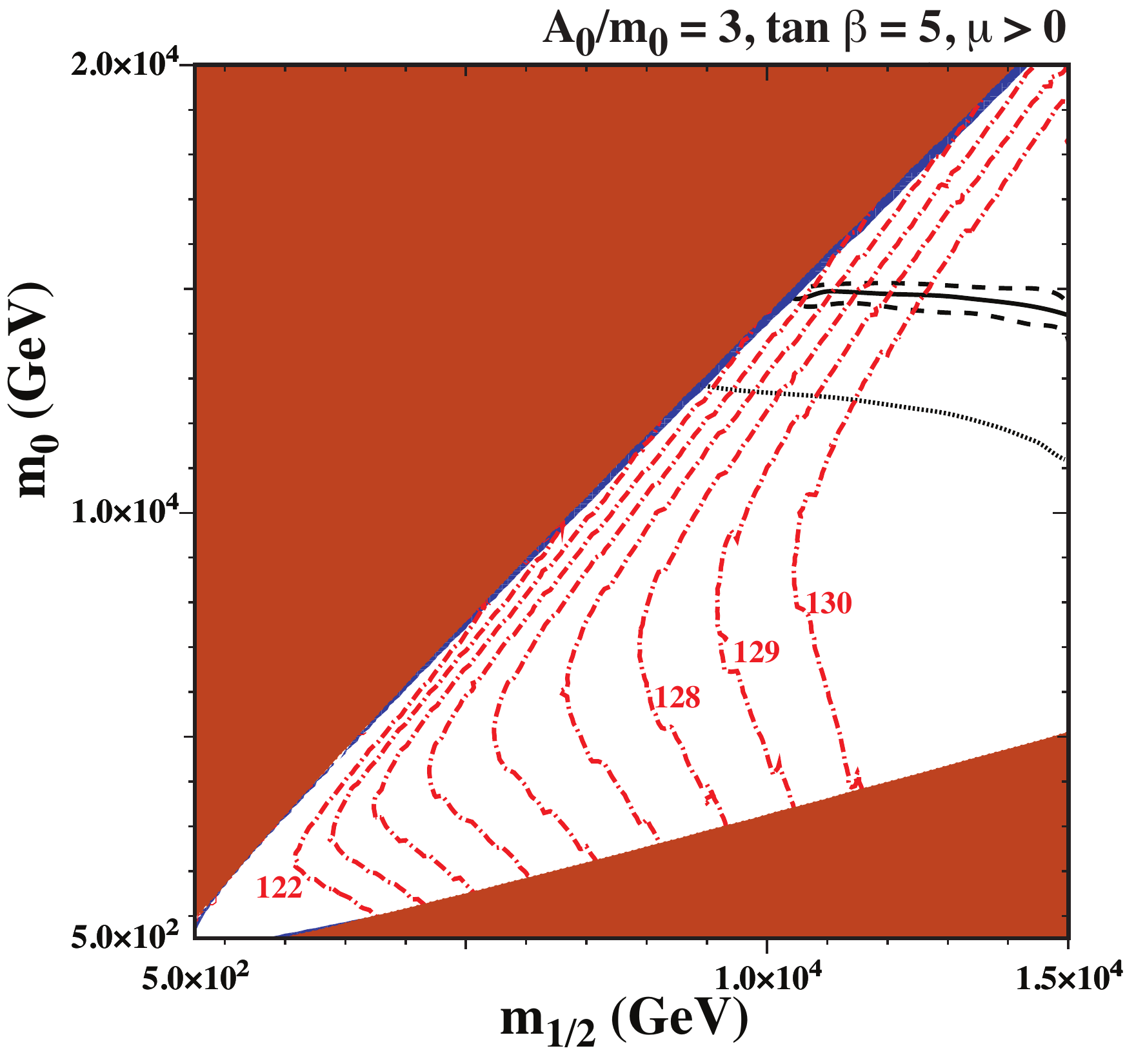}
\end{center}
\hfill
\end{minipage}
\vspace{-7mm}
\caption{
{\it Sensitivity to the Wolfenstein $A$ parameter for $\mu > 0$  assuming $\tan \beta = 5, A_0/m_0 = 3$ and $M_{\rm in} = M_{\rm GUT}$. The dashed lines correspond to the variation of $A$ within one standard deviation ($A = 0.836 \pm 0.015$).
}}
\label{fig:WolfA}
\end{figure}

The CKM phase $\delta$ plays no role, so the 3 relevant parameters are $A, \rho$ and $\lambda$. Of these, by far the greatest sensitivity is to $A$, as we illustrate in Fig.~\ref{fig:WolfA} assuming $A = 0.836 \pm 0.015$, $\tan \beta = 5, A_0/m_0 = 3$, $M_{\rm in} = M_{\rm GUT}$ and $\mu > 0$.

To understand the sensitivity to the uncertainty on $A$, we see from Eq.~\eqref{eq:cllaprox} that $C_{LL}(uds\nu_mu)\propto  V_{us}^* V_{td}V_{ts}$.~\footnote{The second term in $C_{LL}(uds\nu_\mu)$ is in principle proportional to $V_{us}^* V_{cd}V_{cs}$. However, this term is proportional to the phase factor $e^{i(\phi_2-\phi_3)}$,
and we scan over phases so as to minimize the rate, bringing this term as close as possible to $-1$, thus concealing the dependence on $V_{cd}V_{cs}$. } In terms of the Wolfenstein parametrization this can be written as $C_{LL}(uds\nu_mu)\propto  A^2\lambda^7$. We can then see from Eq.~\eqref{eq:crlaprox} that $C_{RL}(uds\nu_\tau)\propto  V^*_{tb}V_{ud} V_{ts}\approx -A\lambda^3$. Hence the dependence on the quark mixing matrix reduces to those on the $A$ and $\lambda$ parameters, which have uncertainties of $1.8\%$ and $0.2\%$, respectively.  It is not a surprise, then, that the sensitivity to $A$ is comparable to that of that of $m_c$, whose uncertainty is $1.6\%$, see Fig.~\ref{fig:mc}.

\subsection{GUT Phases}
\label{sec:GUTphi}

\begin{figure}[htb!]
\centering
\begin{minipage}{4.5in}
\hspace*{0.5in}
\includegraphics[height=3in]{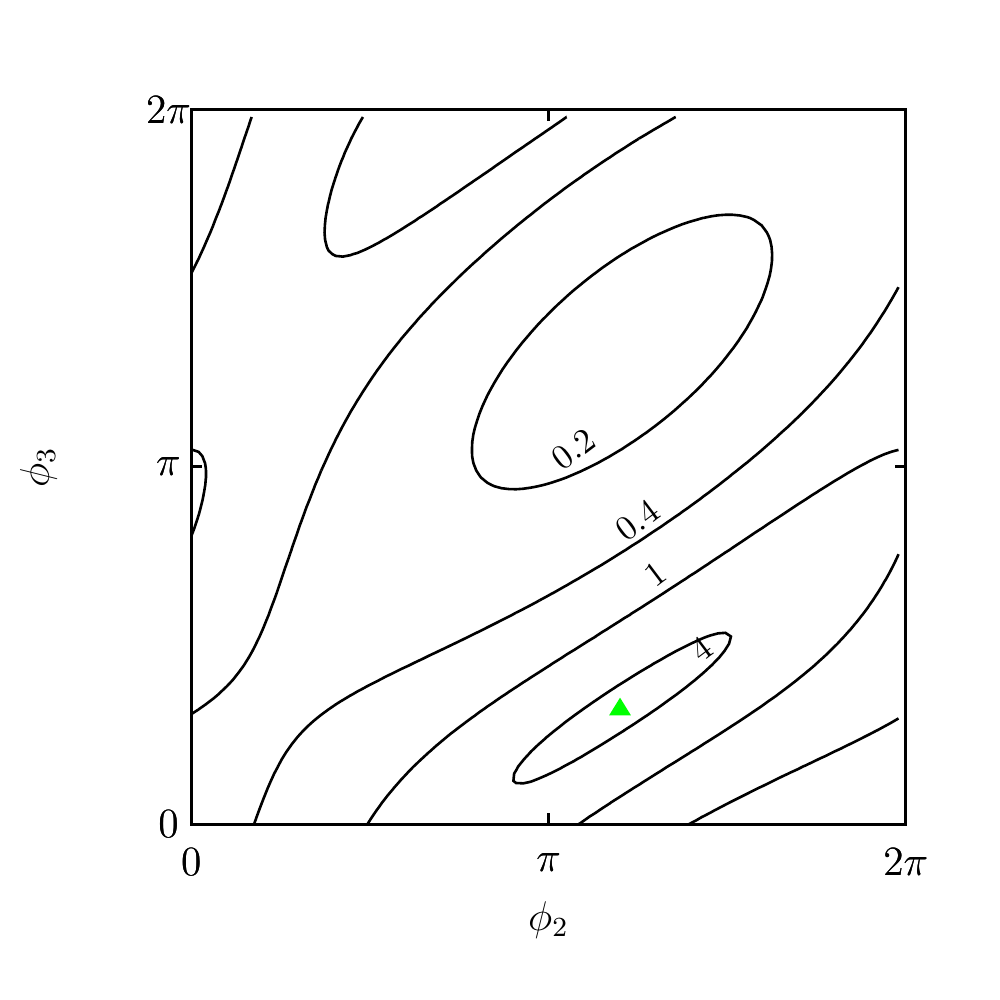}
\hfill
\end{minipage}
\vspace*{-0.5cm}
\caption{
{\it Sensitivity of the $p \to K^+ \bar{\nu}$ lifetime to the GUT phases $(\phi_2, \phi_3)$ for the CMSSM parameter choices $\tan \beta=5, A_0/m_0=3, m_0= 14.13$~TeV and $m_{1/2} = 9.79$~TeV in units of $10^{33}$ years. The maximum value of the lifetime is indicated by a green triangle.}}
\label{fig:GUTphi3D}
\end{figure}

We now discuss the uncertainties associated with the GUT phases (\ref{GUTphase}). Since the  two terms in Eq.~\eqref{eq:cllaprox} have comparable magnitudes, the Wilson coefficients 
$C_{LL} (usd\nu_i)$ and $C_{LL} (uds\nu_i)$ may be suppressed in certain ranges of the GUT phases. A general overview of the dependence of the lifetime for $p \to \pi^+ \bar{\nu}$ in the plane of the two GUT phases $(\phi_2, \phi_3)$ for the CMSSM parameter choices $\tan \beta=5, A_0/m_0=3, m_0= 14.13$~TeV and $m_{1/2} = 9.79$~TeV is shown in Fig.~\ref{fig:GUTphi3D}.
The maximum value of the lifetime is indicated by a green triangle.

\begin{figure}[htb!]
\begin{minipage}{6in}
\includegraphics[height=3in]{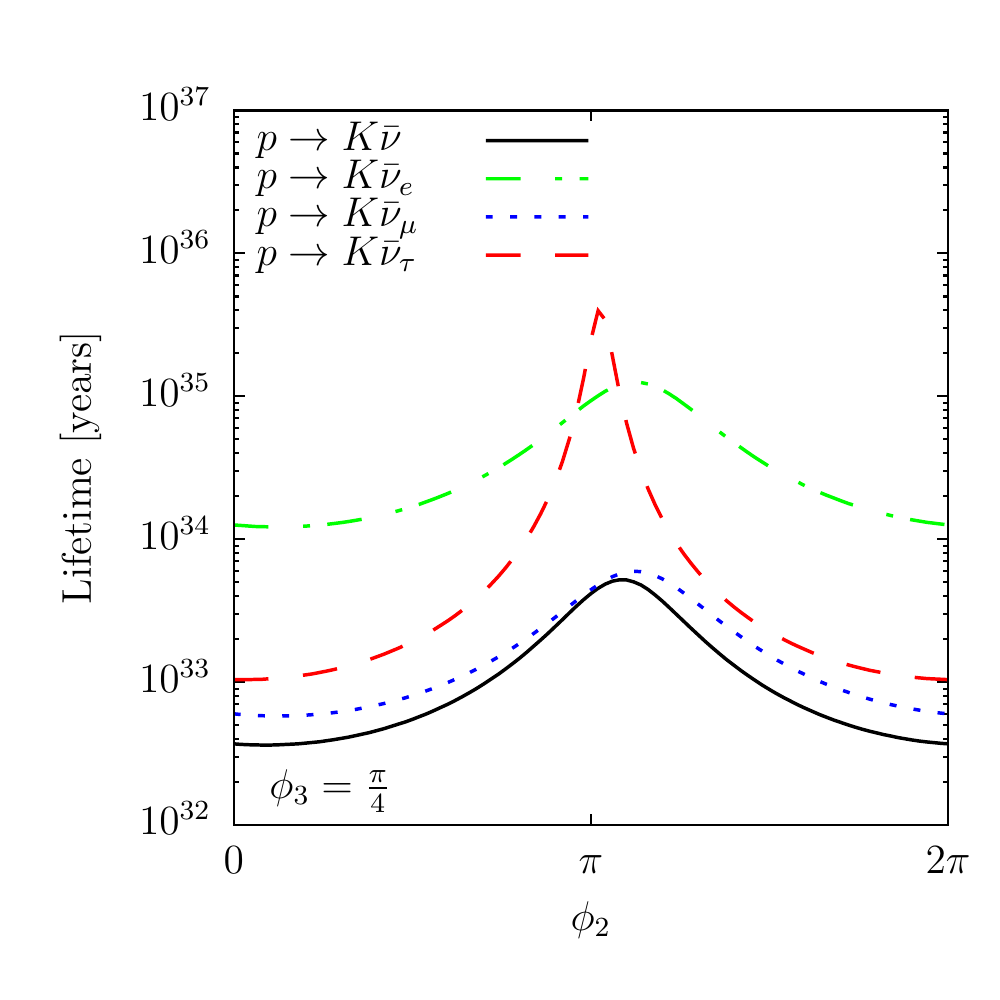}
\includegraphics[height=3in]{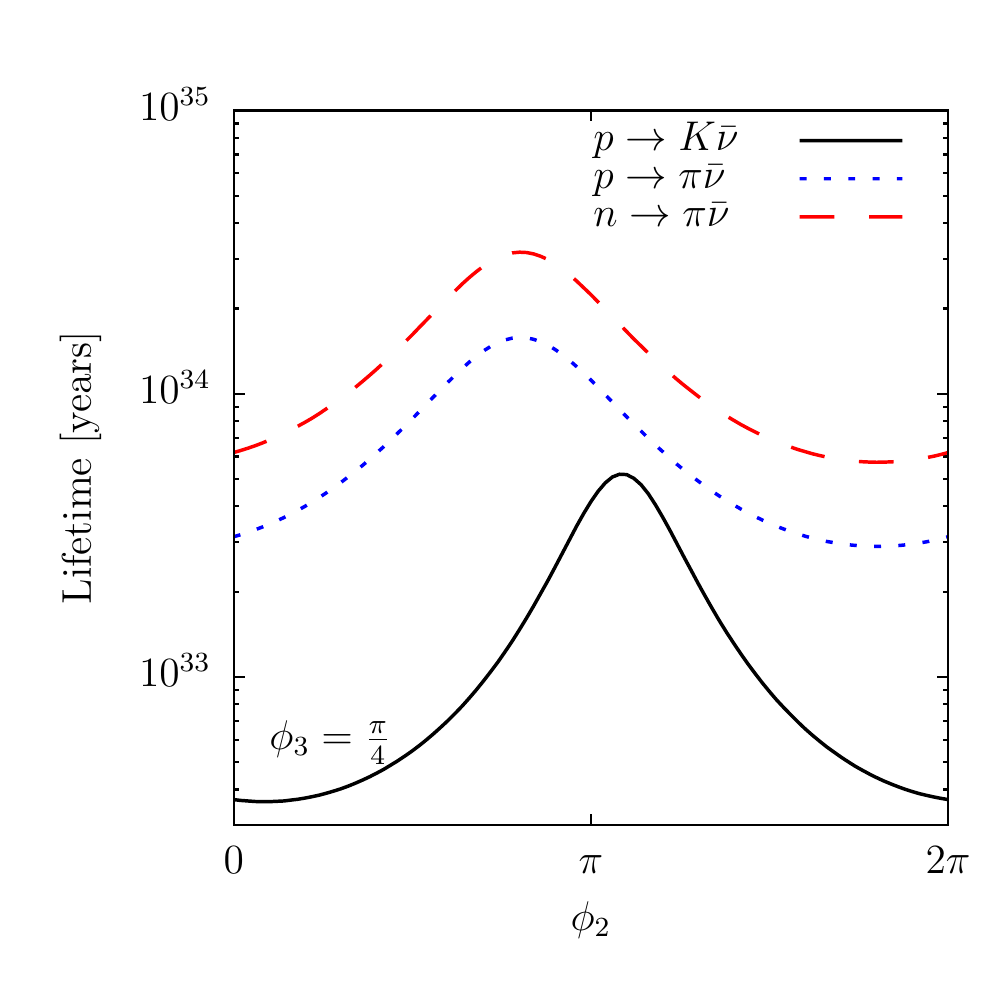}
\hfill
\end{minipage}
\vspace*{-0.5cm}
\caption{
{\it Left panel: Sensitivity of the $p \to K^+ \bar{\nu}$ partial lifetimes to the GUT phase $\phi_2$ for the same point considered in Fig.~\ref{fig:GUTphi3D}. Right panel: Comparison of partial lifetimes for $p \to K^+ \bar{\nu}$ (black solid line), $p \to \pi^+\bar{\nu}$ (blue dotted line) and $n \to \pi^0 \bar{\nu}$ (red dashed line) as functions of $\phi_2$.}}
\label{fig:GUTphi}
\end{figure}

We mentioned in Section \ref{sec:mq} that Higgsino exchange contributes only to 
${\cal A}(p\to K^+\bar{\nu}_\tau)$, via
$C_{RL}(usd\nu_\tau)$ and $C_{RL}(uds\nu_\tau)$. These coefficients are approximately given by Eqs.~\eqref{eq:crlaprox1} and \eqref{eq:crlaprox}, respectively, where we see that, unlike the coefficients in Eq.~\eqref{eq:cllaprox}, their absolute
values do not change when the phases vary. However,
the difference in the phase structure from that in Eq.~\eqref{eq:cllaprox} contributes to the GUT phase dependence of ${\cal A}(p\to K^+\bar{\nu}_\tau)$, which is different from that of ${\cal A}(p\to K^+\bar{\nu}_{e, \mu})$.
This feature is seen in the left panel of Fig.~\ref{fig:GUTphi}, where we choose $\tan \beta=5, A_0/m_0=3, m_0= 15.75$~TeV and $m_{1/2} = 11$~TeV as in Fig.~\ref{fig:GUTphi3D}, and $\phi_3$ is chosen to maximize approximately the $p \to K^+ \bar{\nu}$ lifetime. We see that the ratio between the rates for $p\to K^+\bar{\nu}_{e, \mu}$ (green dot-dashed line and blue dotted line, respectively) is independent of the GUT phase $\phi_2$, whereas the rate for $p\to K^+\bar{\nu}_{\tau}$ (red dashed line) has a quite different dependence on $\phi_2$. The solid black line is the total $p\to K^+\bar{\nu}$ decay rate.

Another potentially important proton decay mode is $p \to \pi^+ \bar{\nu_i}$, whose decay amplitude is
\begin{align}
 {\cal A}(p\to \pi^+ \bar{\nu}_i)&=
C_{RL}(udd\nu_i)\langle \pi^+\vert (ud)_Rd_L\vert p\rangle
+C_{LL}(udd\nu_i)\langle \pi^+\vert (ud)_Ld_L\vert p\rangle~,
\end{align}
where $C_{RL}(udd\nu_i)$ is non-vanishing only for $i = 3$. 
The neutron decay mode $n \to \pi^0 \bar{\nu}$ is also potentially important, and is given by the same Wilson operators:
\begin{align}
 {\cal A}(n\to \pi^0 \bar{\nu}_i)&=
C_{RL}(udd\nu_i)\langle \pi^0\vert (ud)_Rd_L\vert n\rangle
+C_{LL}(udd\nu_i)\langle \pi^0\vert (ud)_Ld_L\vert n\rangle~.
\end{align}
As seen in the right panel of Fig.~\ref{fig:GUTphi}, the rate for
$p \to \pi^+ \bar{\nu}$ (blue dotted line) is smaller than that for $p \to K^+ \bar{\nu}$ for all values of $\phi_2$, though it becomes comparable for $\phi_2 \sim 200^o$.~\footnote{In~\cite{eemno}, which used older hadronic matrix elements, there were phase values where the rate to $\pi^+ \bar{\nu}$ was dominant.} 
The rate for $n \to \pi^0 \bar{\nu}$ (red dashed line) is always smaller than that for $p \to \pi^+ \bar{\nu}$ for the central values of the hadronic matrix elements that we use. As mentioned earlier, we do not consider the experimental searches for $p \to \pi^+ \bar{\nu}$ and $n \to \pi^0 \bar{\nu}$, as the current limits on these decays are significantly weaker than those for $p \to K^+ \bar{\nu}$, and no detailed studies are yet available for the next-generation detectors.

\subsection{Yukawa Non-Unification}
\label{sec:yukawauni}

As can be seen in Eq.~\eqref{eq:wyukawa}, in minimal SU(5) the lepton
and down-type quark Yukawa couplings should be equal at the GUT scale.  However, when
running the physical values up from the electroweak scale, 
one finds that they are quite different
for the first two generations, whereas Yukawa unification is a good approximation for the $b$ and $\tau$. The differences for the lighter generations can,
however, easily be compensated by effects from physics above the GUT scale. In particular, operators of higher mass dimension induced at the Planck scale
that contribute to the Yukawa couplings may account
for this difference~\cite{Ellis:1979fg, Panagiotakopoulos:1984wf, Nath:1996qs, Nath:1996ft,
Bajc:2002pg}.~\footnote{Another approach is to utilize higher-dimensional Higgs representations \cite{Georgi:1979df}, but we do not consider this possibility here, focusing instead on the minimal field content.}
Among such operators, those of lowest dimension are
\begin{align}
  W_{\rm eff}^{\Delta h} &= \frac{c_{\Delta h,1 }^{ij}}{M_P} \Phi_{i \alpha}
  \Sigma^\alpha_{~\beta} \Psi^{\beta\gamma}_j \overline{H}_\gamma 
+\frac{c_{\Delta h, 2}^{ij}}{M_P} \Psi_i^{\alpha\beta} \Phi_{j\alpha}
\Sigma^\gamma_{~\beta} \overline{H}_\gamma
~,
\label{eq:weffdelh}
\end{align}
which yield Yukawa interaction terms when $\Sigma$ acquires a vev. In particular, the first operator in Eq.~\eqref{eq:weffdelh} splits the lepton and down-type quark Yukawa couplings by the product of the superpotential coupling with $V/M_P \sim 10^{-2}$, which is sufficient to explain the differences in the lepton and down-type quark Yukawa couplings for all of the three generations.

The operators in Eq.~\eqref{eq:weffdelh} also modify the couplings of the colour-triplet Higgs fields to the quark and lepton fields, and thus directly affect the proton decay amplitude. Our ignorance of the coefficients $c_{\Delta h}$ in Eq.~\eqref{eq:weffdelh} leads to ambiguity in these couplings, which then results in the uncertainty in the Wilson coefficients in Eq.~\eqref{eq:wilson5}.

The range of this uncertainty is indicated by the differences between the quark and lepton Yukawa couplings. In the previous Sections we 
have chosen the quark Yukawa couplings, i.e., $f_s$ and $f_d$. Since $f_s < f_\mu$, we would expect that in general using the strange-quark Yukawa coupling may yield a longer lifetime than using the muon coupling whereas, since $f_d > f_e$, using the down-quark Yukawa coupling may give a shorter lifetime than using the electron coupling. These expectations are borne out in tests we have made using a CMSSM GUT point with $\tan \beta = 5, A_0/m_0 = 3, m_{1/2} = 9.8$~TeV, $m_0 = 14.1$~TeV and $\mu > 0$. Our default choice of Yukawa couplings, $f_{s,d}$, yields a proton lifetime $\simeq 5.4  \times 10^{33}$~y, whereas using $f_{s,e}$ yields a lifetime $\simeq 5.6 \times 10^{33}$~y, a 4\% difference. On the other hand, replacing $f_s$ by $f_\mu$ yields a lifetime that is 23 times smaller. Thus, our choice $f_{s,d}$ is quite conservative, and the most conservative choice $f_{s,e}$ would have resulted in an insignificant difference.

In principle, couplings of the type (\ref{eq:weffdelh}) could also modify the pattern of quark mixing in GUT Higgs triplet interactions. However, we would not expect this to modify the generic prediction that the dominant proton decay mode should be into $K^+ \overline \nu$, which results from the combination of colour and flavour antisymmetry in the effective dimension-five interaction~\cite{strangeBDK}. Nevertheless, 
it is clear that more detailed studies of this ambiguity in specific GUT models are warranted, though they lie beyond the scope of this paper.

\section{Results}
\label{sec:results}

In this Section we display  $(m_{1/2}, m_0)$ planes for various choices of the supersymmetric model parameters. For CMSSM models with GUT scale universality, we show 2 sets of proton decay limit contours. Those in black are for the minimal supersymmetric SU(5) GUT, and those in green are calculated assuming that the dimension-five operator in Eq.~\eqref{eq:SigmaWW} is present with $c \ne 0$.  In both cases, the solid lines correspond to the proton decay lifetime limit of $0.066 \times 10^{35}$ yrs using the Standard Model inputs described in the previous Section. 
We also show dot-dashed lines corresponding a lifetime $\tau(p \to K^+ \bar{\nu}) = 5 \times 10^{34}$~yr, corresponding to the estimated 3-$\sigma$ discovery sensitivity of the DUNE experiment after 20~yrs of operation (see Table~\ref{tab:baryondecaylimits}).
The dashed contours surrounding the solid contour correspond to the 1$\sigma_{\rm tot}$ uncertainty in the position of the limit. As in the previous Section, $\sigma_{\rm tot}$ takes into account the propagated uncertainties from the hadronic matrix elements and the strong coupling as it affects $M_{H_C}$. For super-GUT CMSSM models, the dimension-five operator is needed to satisfy the boundary conditions, and only one set of contours are shown and coloured black. 
At each point in the supersymmetric space, we choose the unknown GUT phases so as to minimize the $p \to K^+ \bar{\nu}$ decay rate. 

In each $(m_{1/2}, m_0)$ plane, we show contours of $m_h$ 
calculated using {\tt FeynHiggs~2.14.1} \cite{FeynHiggs} that are consistent with the measured Higgs mass within the estimated calculational uncertainties. These are shown as red dot-dashed contours. Regions of the planes that are shaded brick red are excluded because there the LSP would be charged. Typically, in such regions at large $m_0$,
the LSP is the lighter stop, and when present, brick red regions at lower $m_0$ contain a stau LSP. Regions shaded pink are excluded because there is no consistent electroweak symmetry-breaking vacuum. In addition, there are very narrow strips shaded blue where the LSP density calculated in standard Big Bang cosmology falls within the range allowed by Planck and other measurements. Here, to make these good relic density regions visible on the scale plotted, we allow the relic density to vary between $0.01 < \Omega_\chi h^2 < 2.0$.  In other regions of the $(m_{1/2}, m_0)$ planes the LSP would generally be overdense in the absence of some scenario for modified cosmological evolution with entropy generation (see, e.g., Ref.~\cite{EGNNO5}).

\subsection{The CMSSM}
\label{sec:CMSSM}

\begin{figure}[!htb]
\begin{minipage}{8in}
\includegraphics[height=3.in]{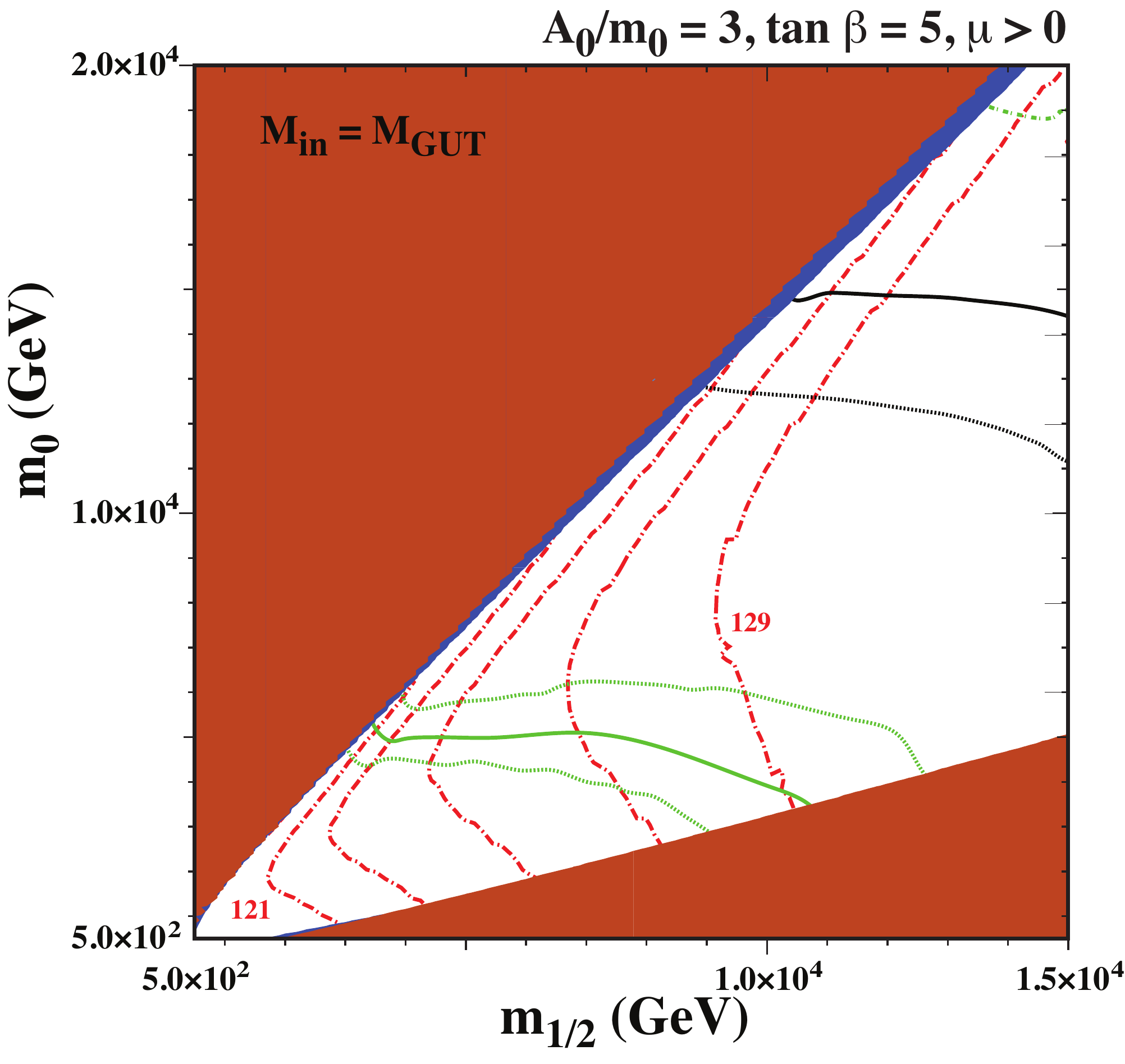}
\includegraphics[height=3.in]{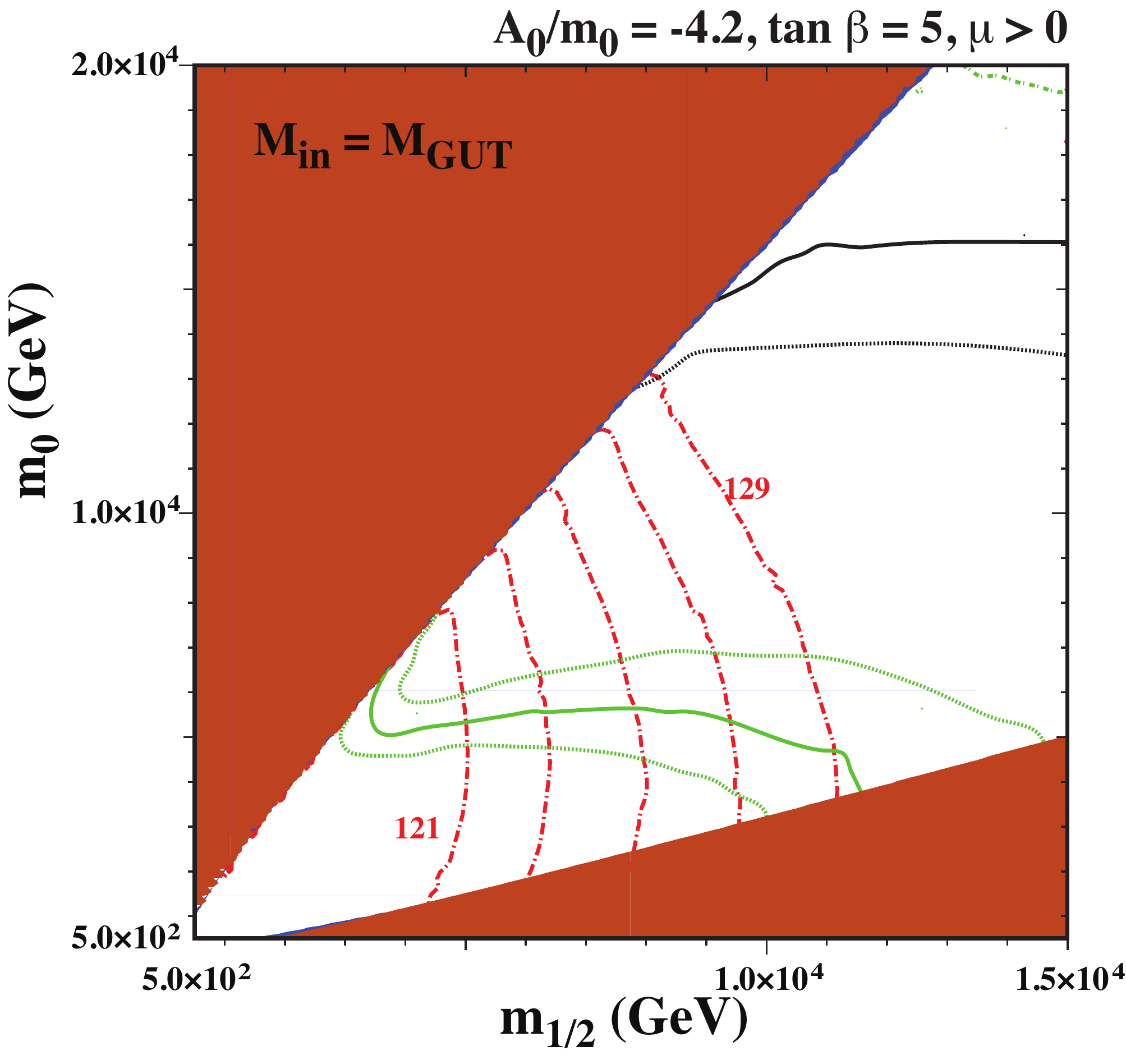}
\end{minipage}
\begin{minipage}{8in}
\includegraphics[height=3.in]{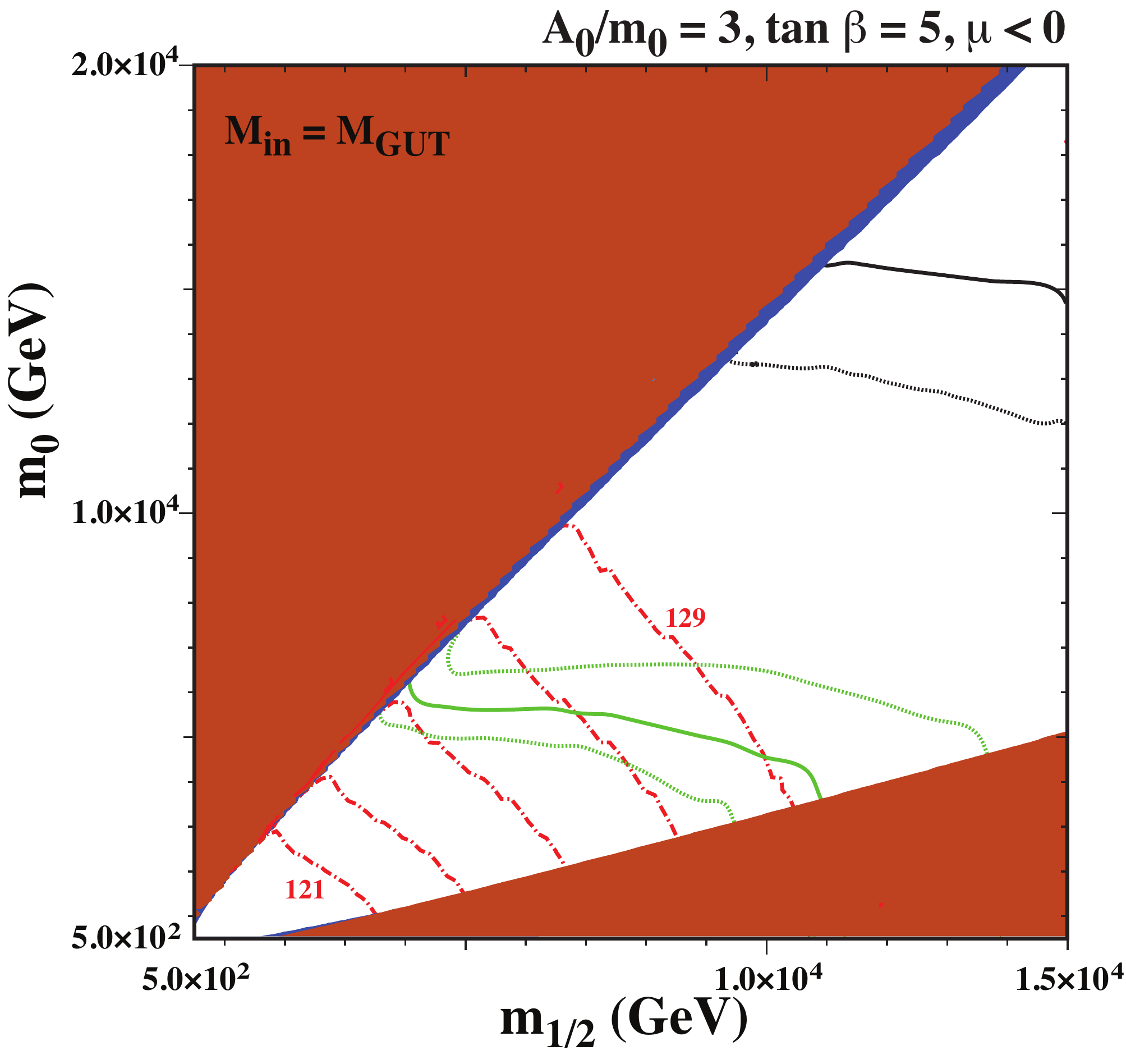}
\includegraphics[height=3.in]{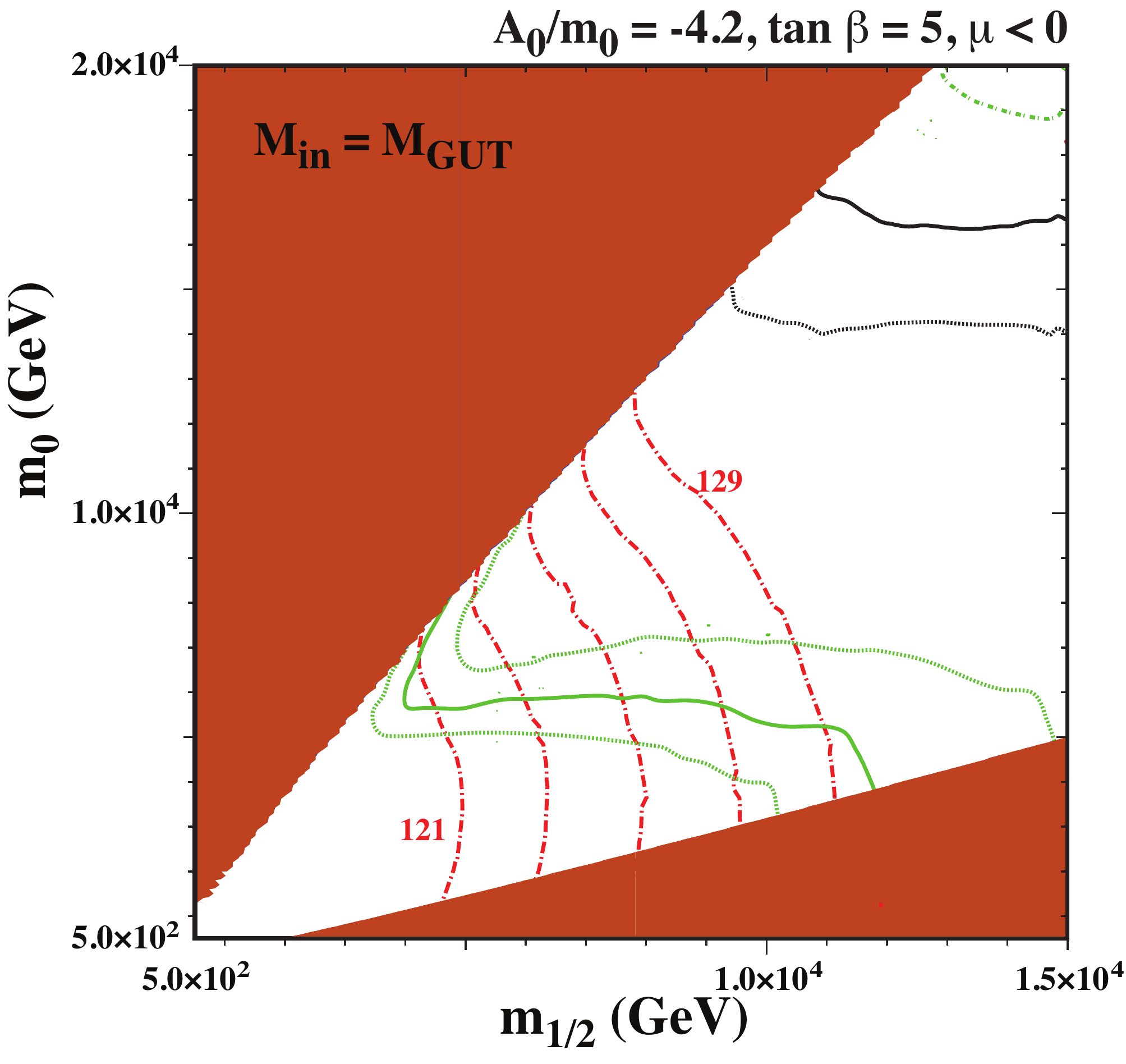}
\hfill
\end{minipage}
\caption{
{\it Some $(m_{1/2}, m_0)$ planes in the CMSSM for $\tan \beta = 5$, $\mu > 0$ (upper panels), $\mu < 0$ (lower panels), $A_0/m_0 = 3$ (left panels), $A_0/m_0 = - 4.2$ (right panels). The black lines are contours of the $p \to K^+ \bar{\nu}$ lifetime, as calculated varying the GUT phases to minimize this decay rate, using the central parameter values and their combined 1-$\sigma$ variations.
The green lines are corresponding results including the dimension-5 contribution discussed in the text. When present, the green dot-dashed curves correspond to the DUNE discovery sensitivity. The red lines are the indicated contours of $m_h$. 
}}
\label{fig:CMSSM}
\end{figure}

We begin the discussion of our main results with the CMSSM.
We recall from Eq.~\eqref{cmssmpara} that the CMSSM is defined by four parameters given at the GUT scale, defined to be where the two electroweak gauge couplings are equal.  
Because we are primarily interested in calculating proton decay rates, we need to determine
the mass of the coloured Higgs triplet, $M_{H_C}$, which we obtain from the matching conditions in Eqs.~(\ref{eq:matchmhc}--\ref{eq:matchg5}). 
As discussed earlier, in the CMSSM with $M_{\rm in} = M_{\rm GUT}$,
we do not run the RGEs above the GUT scale, and no additional matching to GUT scale parameters is needed. As a result, we can define CMSSM models with
the dimension-five operator  turned off, i.e., $c = 0$.  When this operator is turned on,  fixing $M_{H_C}$ requires specifying the SU(5) Higgs couplings $\lambda$ and $\lambda^\prime$. In all figures below with $A_0\ne 0$, we have fixed $\lambda = 0.6$ and $\lambda^\prime = 0.0001$. For $A_0=0$, we take $\lambda=0.1$ instead, since otherwise the focus-point region would be pushed to very large values of $m_0$ where the RGE running becomes unstable. For more on the dependence of $\tau_p$ on these two GUT couplings, see \cite{eemno}. 
We show in Fig.~\ref{fig:CMSSM} four examples of CMSSM planes. 
In the two left panels, we assume $\tan \beta = 5$ with $A_0/m_0 = 3$, whereas in the right panels we take $A_0/m_0 = -4.2$ for the same value of $\tan \beta$. In the upper two panels we take $\mu > 0$,
whereas $\mu < 0$ in the lower panels.  These values are chosen so as to bring the relic density strip (shaded blue) in a position to intersect with experimentally viable values of the Higgs mass (allowing for uncertainties in the theoretical calculation of the Higgs mass). 

The upper left panel of Fig. \ref{fig:CMSSM} corresponds to the same choice of parameters as used in the previous Section.  Indeed, this panel is essentially a simplified version of that shown in Fig.~\ref{fig:alphas},
keeping only the central contour limits (for both $c = 0$ (black) and $c \ne 0$ (green)) along with the variation of these contours by $\pm 1\sigma_{\rm tot}$. (We recall that the stronger limit lies off the scale shown in the plot when $c = 0$.) Here and in all the other panels, the limit on the parameter space is greatly weakened when $c \ne 0$, as the coloured Higgs mass is much larger, being determined by Eq. (\ref{mhc}) with our choices of $\lambda$ and $\lambda^\prime$. In this panel, as in subsequent panels, we also show the location of the
DUNE sensitivity $\tau_p = 5 \times 10^{34}$~yrs by the dot-dashed curve with $c\ne 0$.  When $c = 0$, the contour often lies beyond the range shown. It is found in the upper right corner of the panels of Fig.~\ref{fig:CMSSM}, shown by the dot-dashed green curves, except for the case of $A_0/m_0=3$ and $\mu<0$ (lower left panel), where it is outside the parameter ranges shown.

As an example, consider a point near ($m_{1/2}, m_0$) = (9.8,14.1) TeV. It corresponds to a bino LSP
with mass of roughly 4.8 TeV that is nearly degenerate with the lighter stop.  The Higgs mass is close to 125 GeV. With $c = 0$, this point has a lifetime which is slightly less than the experimental limit, but is well within  one $\sigma$ of the limit. However, when $c \ne 0$ it lies safely above the 
lifetime limit. More specifically, with $c = 0$, we find $\tau_p = 5.4\pm 4.6 \times 10^{33}$ yrs, whereas for the same choice of parameters when $c \ne 0$, we find $\epsilon = 8cV/M_P = .0024$, and $\tau_p = 3.3 \pm 0.6 \times 10^{34}$ yrs, a factor of over 6 times larger.
This and other examples discussed in this section are summarized in Table~\ref{tab:re}.

\begin{table}[!ht]
\begin{center}
\captionsetup{justification=centering}
\caption{\it{Lifetimes at points with $\Omega_\chi h^2 \approx 0.12$, and $m_h \approx 125$ GeV. Masses are in TeV and lifetimes in units of $10^{33}$ years.\label{tab:re} }}
\vspace{-7mm}
\begin{tabular}{|l|c|c|c|c|c|c|c|}
\multicolumn{7}{c}{}\\
\hline
\hline
 & $m_{1/2}$ & $m_0$ & $m_\chi$ & $\tau_p (c = 0)$ & $\epsilon$ & $\tau_p (c \ne 0)$  \\
\hline
Fig. \ref{fig:CMSSM}a & 9.8 & 14.1 & 4.8 & $5.4 \pm 4.6$ & 0.0024  &  $33 \pm 6$  \\
\hline
Fig. \ref{fig:CMSSM}b & 6.4 & 10.6 & 3.0 & $1.5\pm 1.3$ & 0.0030 &  $16\pm 4$  \\
\hline
Fig. \ref{fig:CMSSM}c & 3.7 & 5.7 & 1.7 & $0.10 \pm 0.09$ & 0.0052  &  $6.0 \pm 1.4$  \\
\hline
Fig. \ref{fig:CMSSM}d & 6.1 & 10.2 & 2.9 & $0.65 \pm 0.58$ & 0.0036  &  $11 \pm 4$  \\
\hline
Fig. \ref{fig:CMSSMFP}a & 14.9 & 48.9 & 1.1 & $2.0 \pm 1.7$ & 0.0024  &  $12 \pm 2$  \\
\hline
Fig. \ref{fig:CMSSMFP}b & 9.5 & 27.9 & 1.1 & $0.24 \pm 0.21$ & 0.0032  &  $3.0 \pm 0.6$  \\
\hline
Fig. \ref{fig:super-GUT}a & -- & -- & -- & -- & --  &  --  \\
\hline
Fig. \ref{fig:super-GUT}b & 7.2 & 13.0 & 2.5 & -- & 0.0032  &  $14\pm 5$  \\
\hline
Fig. \ref{fig:super-GUT}c & 3.6 & 6.1 & 1.4 & -- & 0.0054  &  $5.7\pm 1.8$  \\
\hline
Fig. \ref{fig:super-GUT}d & 6.4 & 12.1 & 2.2 & -- & 0.0033  &  $6.8\pm 2.2$  \\
\hline
Fig. \ref{fig:super-GUTFP} & 18.9 & 59.4 & 1.1 & -- & 0.0021  &  $22\pm 4$  \\
\hline
Fig. \ref{fig:sub-GUT} & 6.2 & 12.3 & 4.4 & $2.2 \pm 1.9$ & 0.0029  &  $20\pm 4$  \\
\hline
\hline
\end{tabular}
\end{center}
\end{table}

When $\mu < 0$ as in the lower left panel of Fig.~\ref{fig:CMSSM}, we see that the most important change is in the Higgs mass, which now requires significantly lower values of 
($m_{1/2}, m_0$) to obtain $m_h = 125$ GeV with the correct relic density.  In this case, unless $c \ne 0$, the proton decay lifetime limit is badly violated. The DUNE sensitivity lies beyond the range shown, implying that this experiment should be able to explore fully this parameter range. 
In the right panels of Fig.~\ref{fig:CMSSM} where $A_0/m_0 = -4.2$, the 125 GeV Higgs mass contours intersect the relic density strip at intermediate values of ($m_{1/2}, m_0$). However, both examples require $c \ne 0$ to be compatible with proton lifetime limit. 

\begin{figure}[!htb]
\begin{minipage}{8in}
\includegraphics[height=3.in]{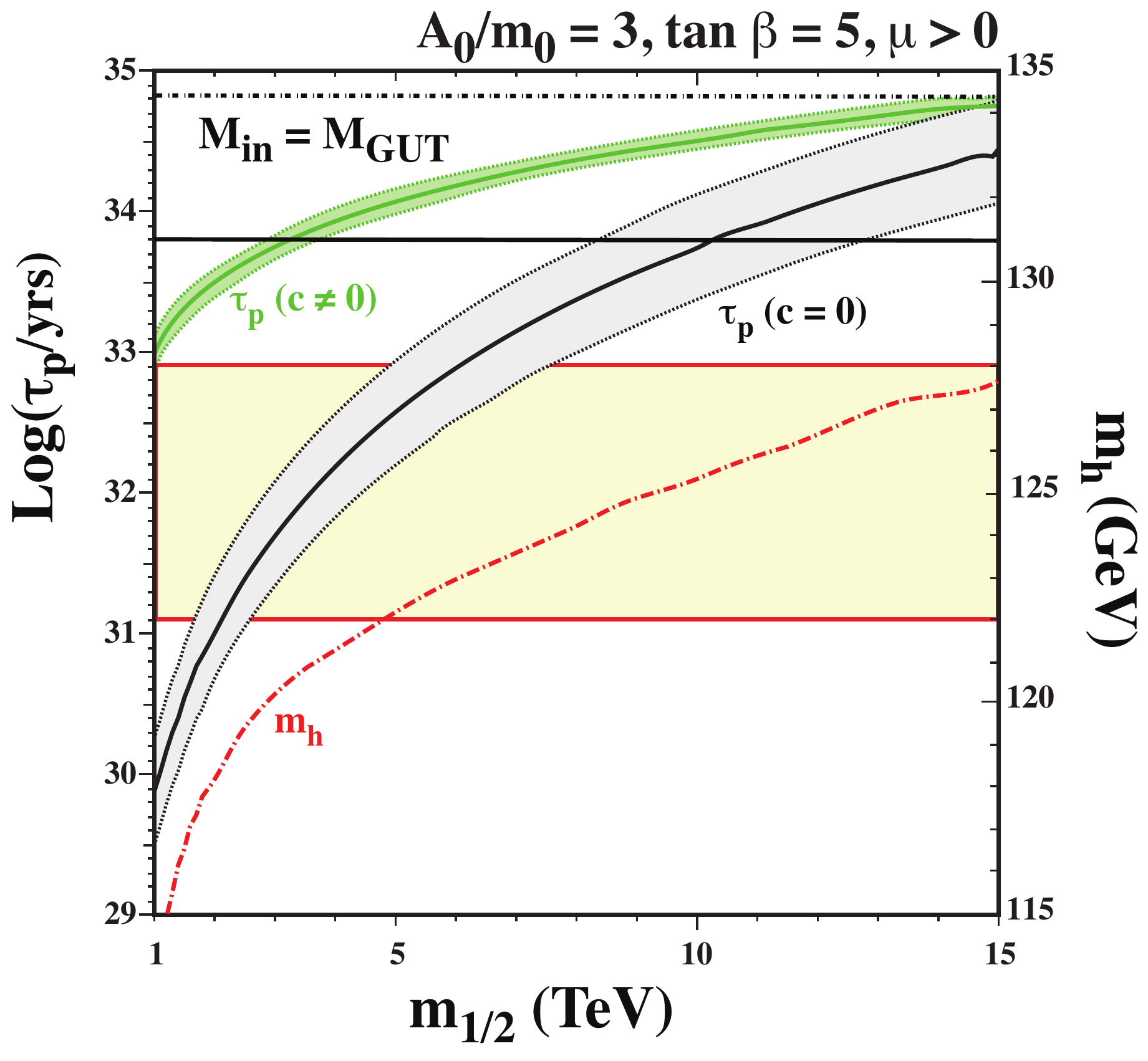}
\includegraphics[height=3.in]{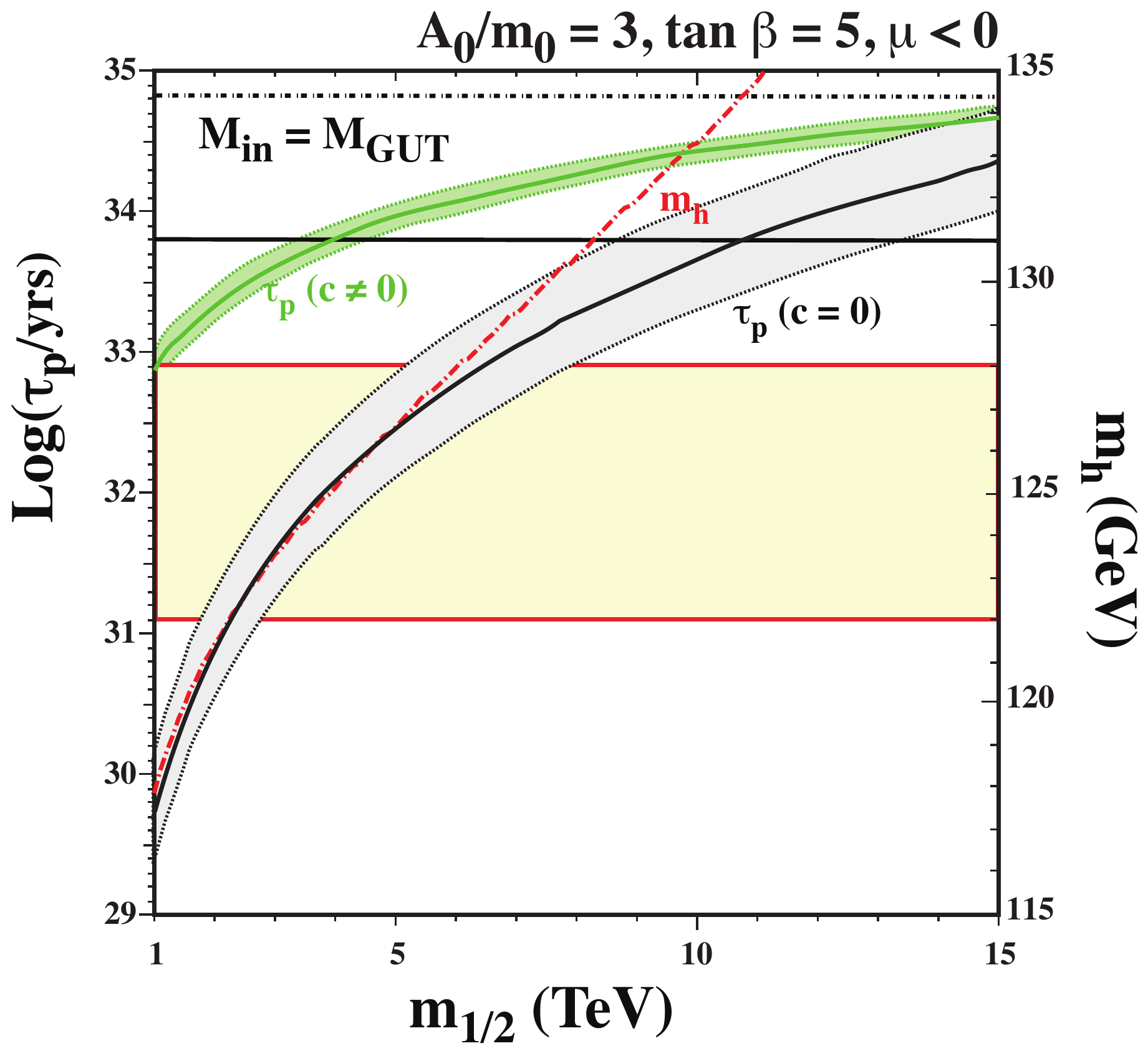}
\end{minipage}
\begin{minipage}{8in}
\includegraphics[height=3.in]{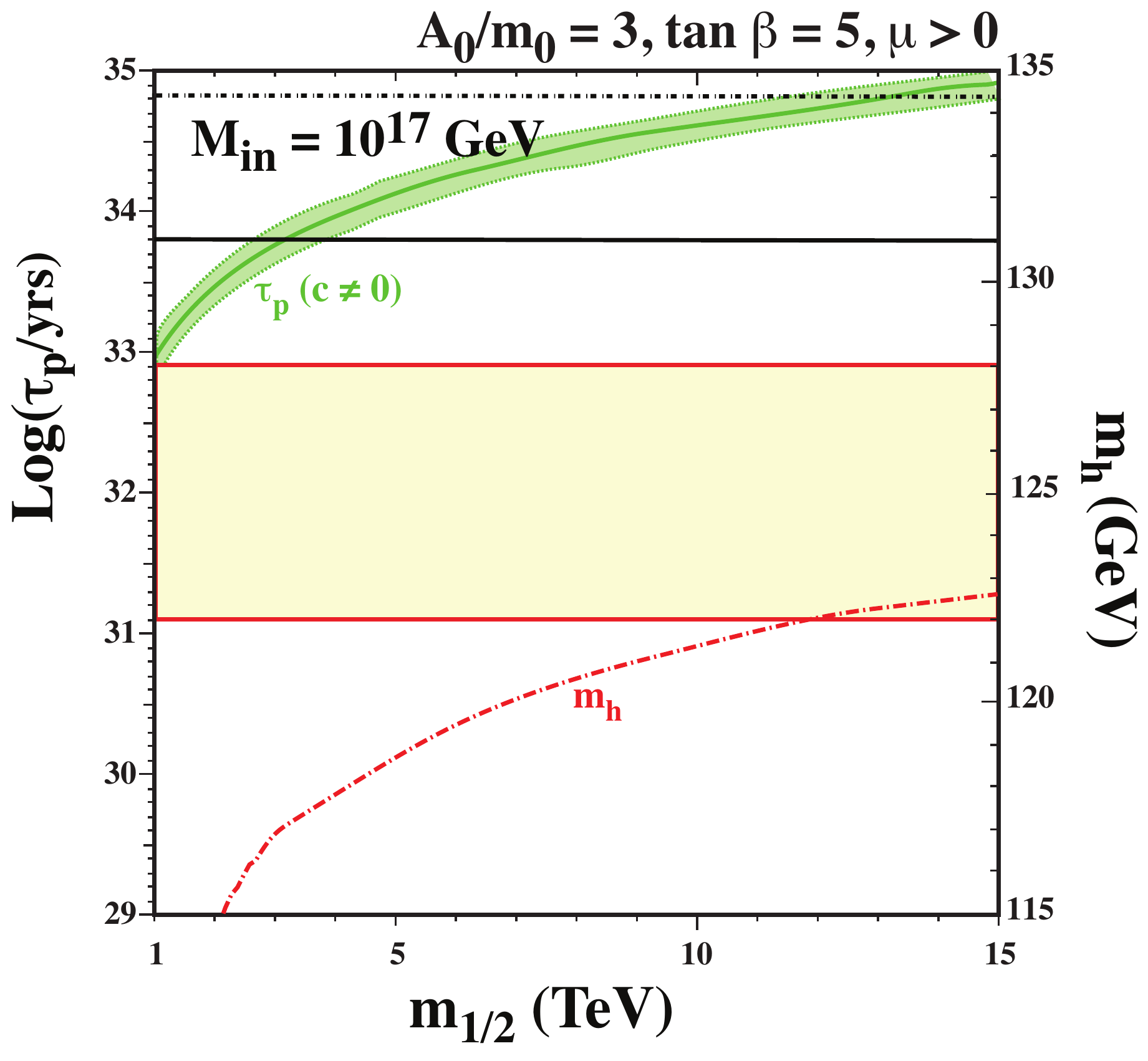}
\includegraphics[height=3.in]{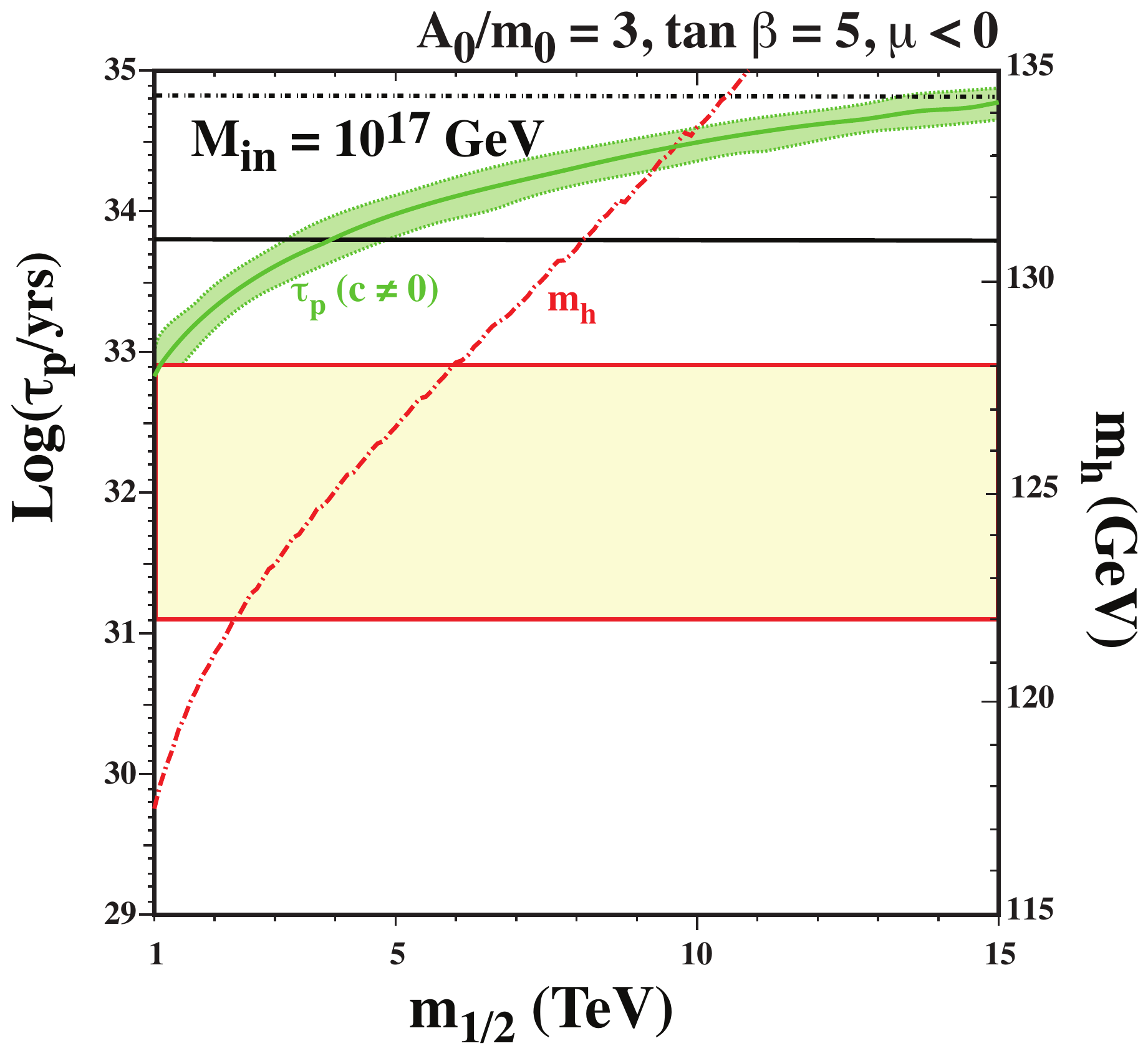}
\hfill
\end{minipage}
\caption{
{\it Some profiles of stop coannihilation strips in the CMSSM (upper panels) and super-GUT models with $M_{\rm in} = 10^{17}$~GeV (lower panels), for $\tan \beta = 5, A_0/m_0 = 3$, $\mu > 0$ (left panels) and $\mu < 0$ (right panels). The rising solid black lines are contours of the proton lifetime evaluated with $c = 0$, the solid green lines are contours of the proton lifetime evaluated with $c \ne 0$ (legends on the left axes). The bands surrounding these lines represent $\pm 1 \sigma_{\rm tot}$ uncertainties. The horizontal black lines are the
current limit (solid) and the expected future (dot-dashed) 90\% CL sensitivity for $p \to K^+ \overline \nu$ from DUNE~\cite{DUNECDR}. The dot-dashed red lines are contours of $m_h$ evaluated using {\tt FeynHiggs~2.14.1}~\cite{FeynHiggs} (legends on the right axes), and the horizontal shaded band shows where the $m_h$ calculation agrees with the experimental measurement within the estimated uncertainty of $\pm 3$~GeV. 
}}
\label{fig:strips}
\end{figure}

In Fig.~\ref{fig:strips}, we show the 
proton lifetime calculated for 
$c = 0$ (black) and $c \ne 0$ (green) as
functions of $m_{1/2}$ along selected stop coannihilation strips in the CMSSM and super-GUT models, with the corresponding  scales on the left axes. 
The shaded bands surrounding the curves show the $ \pm 1\sigma_{\rm tot}$
uncertainties in our calculations. The red dot-dashed curves show the Higgs mass calculated with {\tt FeynHiggs~2.14.1}~\cite{FeynHiggs}, with the corresponding  scales on the right axes. The horizontal shaded region shows the estimated $\pm 3$~GeV theoretical uncertainty in the calculated Higgs mass. The two horizontal lines
show the current limit on the proton lifetime (solid) and expected 20-yr DUNE~\cite{DUNECDR} sensitivity limit (dot-dashed). 

The upper left panel of Fig.~\ref{fig:strips} is computed in the CMSSM using $\tan \beta = 5$, $A_0/m_0 = 3$, $M_{\rm in} = M_{\rm GUT}$, and $\mu > 0$ as in the upper left panel of Fig.~\ref{fig:CMSSM}.
In this case, we see that the current proton lifetime limit already excludes $m_{1/2} \lesssim 8$ TeV when $c = 0$,
but there is a portion of the parameter space extending to $m_{1/2} \gtrsim 15$~TeV that is also consistent with the experimental value of the Higgs mass. 
When $c \ne 0$, the current proton lifetime limit allows the range of $m_{1/2} \gtrsim 5$~TeV where $m_h > 122$ GeV remains allowed. 
We also see that DUNE~\cite{DUNECDR} should be able to explore the entire $m_{1/2}$ range shown.
The upper right panel assumes the same CMSSM input parameters, but with $\mu < 0$ as in the lower left panel of Fig.~\ref{fig:CMSSM}. In this case, when $c = 0$ the calculated Higgs mass exceeds 131 GeV when the proton lifetime is sufficiently long. In contrast, with $c \ne 0$, as in the case of $\mu > 0$, a range of $m_{1/2} \gtrsim 3$~TeV and $\lesssim 7$~TeV is compatible with the measurement of $m_h$ as well as the current proton lifetime limit.  Here too, DUNE~\cite{DUNECDR} should be able to explore the entire range of $m_{1/2}$ allowed by $m_h$.

\begin{figure}[htb!]
\begin{minipage}{8in}
\includegraphics[height=3.in]{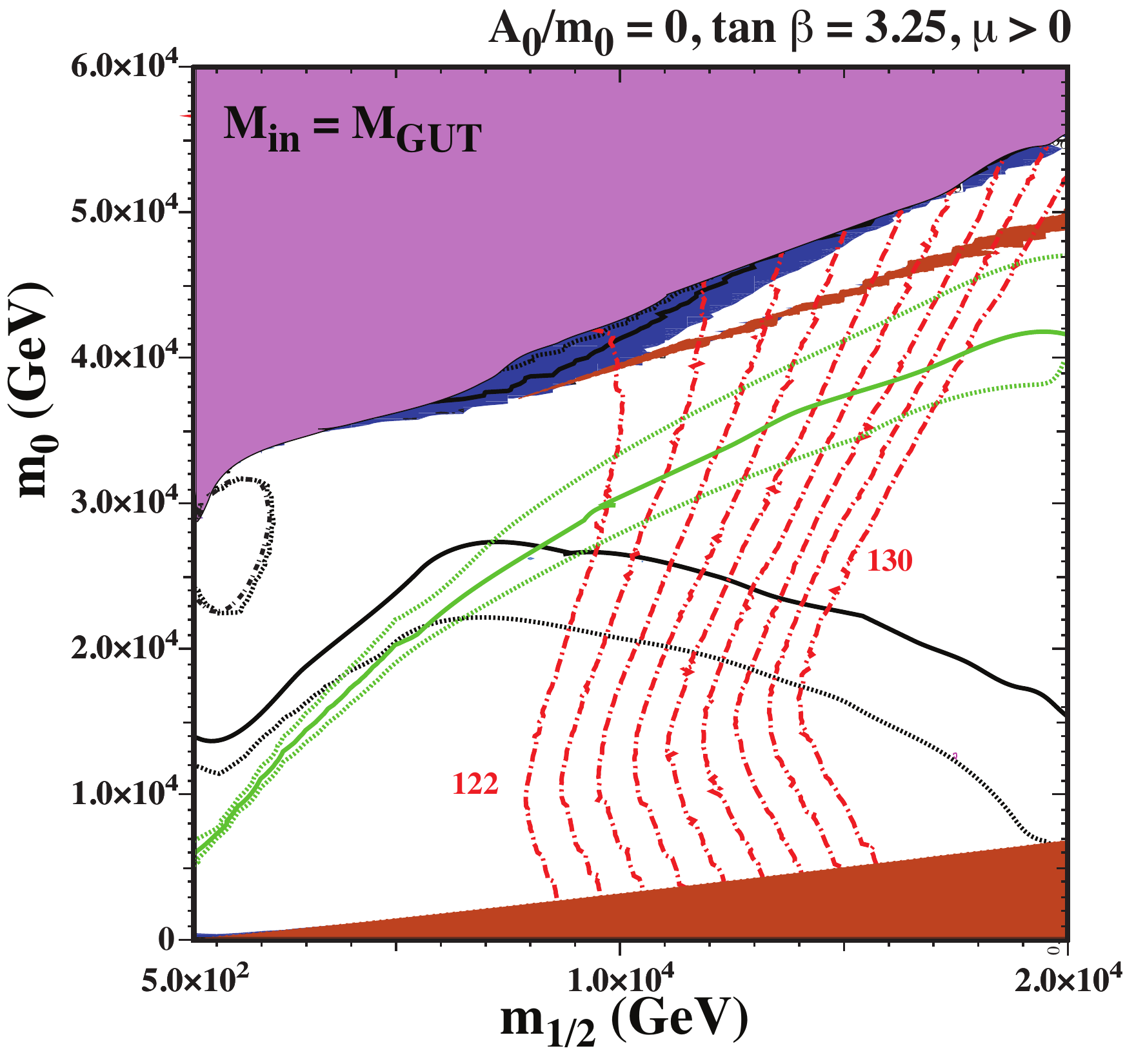}
\includegraphics[height=3.in]{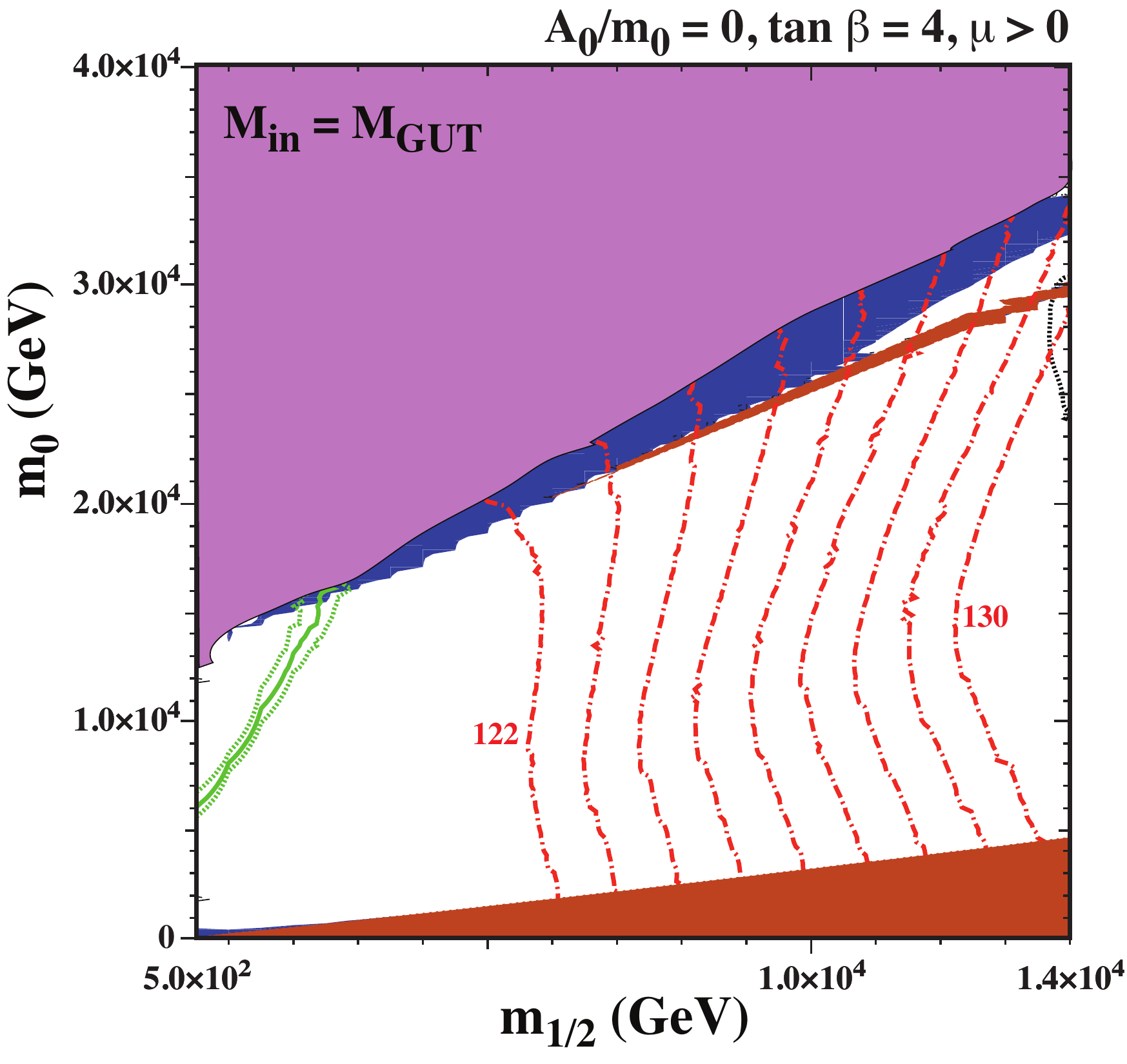}
\hfill
\end{minipage}
\caption{
{\it
Two $(m_{1/2}, m_0)$ planes in the focus-point region of the CMSSM for $\tan \beta = 3.25$ (left panel), $\tan \beta = 4$ (right panel), $\mu > 0$ and $A_0/m_0 = 0$. The black lines are contours of the $p \to K^+ \bar{\nu}$ lifetime, as calculated varying the GUT phases to minimize this decay rate, using the central parameter values and their combined 1-$\sigma$ variations.
The green lines are corresponding results including the dimension-5 contribution discussed in the text. The red dot dashed lines are the indicated contours of $m_h$.
}}
\label{fig:CMSSMFP}
\end{figure}

A different region of the CMSSM parameter space where the relic density is acceptable is found when $A_0/m_0 = 0$ and $m_0$ is large, namely the focus-point region where $\mu \to 0$~\cite{fp}. 
The value of $m_0$ at a focus point is sensitive to $\lambda$. For $\lambda = 0.6$, the value of $m_0$ needed to drive $\mu$ close to zero is so large that the RGE running becomes unstable. Therefore, in our analysis of the focus-point region we take $\lambda=0.1$ and $\lambda^\prime=0.0001$. Two examples of ($m_{1/2}, m_0$) planes exhibiting the focus-point region are shown in Fig. \ref{fig:CMSSMFP}, where we have chosen $\tan \beta = 3.25$ (left panel) and $\tan \beta = 4.0$ (right panel). 
In both of these examples, the LSP is Higgsino-like along the focus-point strip, and its mass is therefore
close to 1.1 TeV everywhere along the strip \cite{OSi}. The strip in these panels appears relatively thick because we have (as in previous plots) shaded the region where $0.01 < \Omega_\chi h^2 < 2$. We note also the appearance of a red shaded strip below the focus point, where the chargino is the LSP.

In contrast to the previous examples, the proton lifetime constraints are not monotonic. Consider for example, the left panel of Fig. \ref{fig:CMSSMFP} with $c=0$. We see two solid black contours corresponding to the proton lifetime limit of $0.066 \times 10^{35}$ years. One of the two spans the figure at $m_0$ between 10 and 20 TeV. The 2nd contour is found inside the blue shaded region, just below the
boundary where there is no radiative EWSB. Between the two, the proton lifetime is found to be greater than the limit. Also within the blue shaded region, there is the 1-$\sigma$ limit on the lifetime contour (dotted black). Above this dotted line, the proton lifetime is too short. The unusual suppression of the proton lifetime near the focus point is due to the reduction in $M_{H_C}$ as $\mu$ decreases~\cite{evno}, which enhances the decay rate along the focus-point strip. In addition, the cancellation between the Higgsino- and Wino-mediated pieces (important at values of $m_0$ between the solid black lines) disappears due to the suppression of the Higgsino contribution, leading to a shorter proton lifetime. 
These two effects combined give a proton lifetime that is too short along the focus-point strip.
(See, for example, the entry in Table~\ref{tab:re} for this Figure.) 

Another interesting feature in this Figure is the fact that the $c\ne0$ proton lifetime constraint is sometimes stronger than that for $c=0$. This is because we have fixed the values of $\lambda$ and $\lambda'$, which effectively fix the coloured Higgs mass. If this value is smaller than the value determined by Eq.~\eqref{mhc} for $c=0$, then $\epsilon$ in Eq.~\eqref{defepsilon} suppresses the proton lifetime. In addition, for the regions with smaller $m_{1/2}$ there is an enhancement in $M_{H_C}$ for $c\ne0$ relative to the case with $c=0$, so the corresponding constraint (shown in green) is weaker.  Additionally, in the upper left corner of Fig.~\ref{fig:CMSSMFP}, both the Wino and Higgsino mass go to zero, further suppressing the proton decay width, whereas the opposite is the case for larger $m_{1/2}$. Finally, we see that for $c=0$, the
contour of $\tau_p - \sigma_{\rm tot}$ is seen at low $m_{1/2}$ and $m_0 = 22 - 32$ TeV. Very near that, we see the DUNE contour (dot-dashed black loop). Outside the loop the lifetime is smaller than the expected DUNE reach. 

In the right panel of Fig.~\ref{fig:CMSSMFP}, which has larger $\tan\beta$, nearly the entire region displayed is excluded by the proton lifetime constraint. The weaker constraint in this case is for $c=0$, for which some region of parameter space is allowed if we consider a $1-\sigma_{\rm tot}$ variation in the lifetime.  Table~\ref{tab:re} gives details for a point corresponding to each panel where $m_h = 125$ GeV and $\Omega_\chi h^2 = 0.12$.

\subsection{Super-GUT Models}
\label{sec:super-GUT}

We next consider super-GUT models
in which $M_{\rm in} > M_{\rm GUT}$, using for illustration $M_{\rm in} = 10^{17}$~GeV. Since the RGEs must now be run above the GUT scale, we must apply all of the boundary conditions discussed in Section~\ref{sec:modelbasics}. In particular, since we fix $\lambda = 0.6$ and $\lambda^\prime = 0.0001$ as in the previous Section, we must take $c \ne 0$ in order to satisfy simultaneously Eqs.~(\ref{eq:matchmhc}--\ref{eq:matchg5}) and Eq.~\eqref{mhc}. The gaugino mass, $M_5$, scalar masses, and $A$-terms are fixed by Eq.~\eqref{eq:inputcond}, which are matched to MSSM parameters at the GUT scale using Eqs.~(\ref{eq:m1match}--\ref{eq:m3match}) and Eq.~\eqref{gutmatch}. As in the CMSSM, we do not specify either of the GUT $B$-terms at $M_{\rm in}$. Instead, again as in the CMSSM, $B$ (and $\mu$) are determined by the minimization of the Higgs potential at the weak scale. These are run up to the GUT scale and 
$B_H, B_\Sigma$ (and $\mu_H$) are given by Eqs.~\eqref{eq:matchingmu} and \eqref{eq:wyukawa} with $\Delta = 0$. We recall that $\mu_\Sigma = \lambda^\prime V/4$, and note that $B_H$ and $B_\Sigma$
are needed in the gaugino matching 
conditions at $M_{\rm GUT}$.

\begin{figure}[!htb]
\begin{minipage}{8in}
\includegraphics[height=3.in]{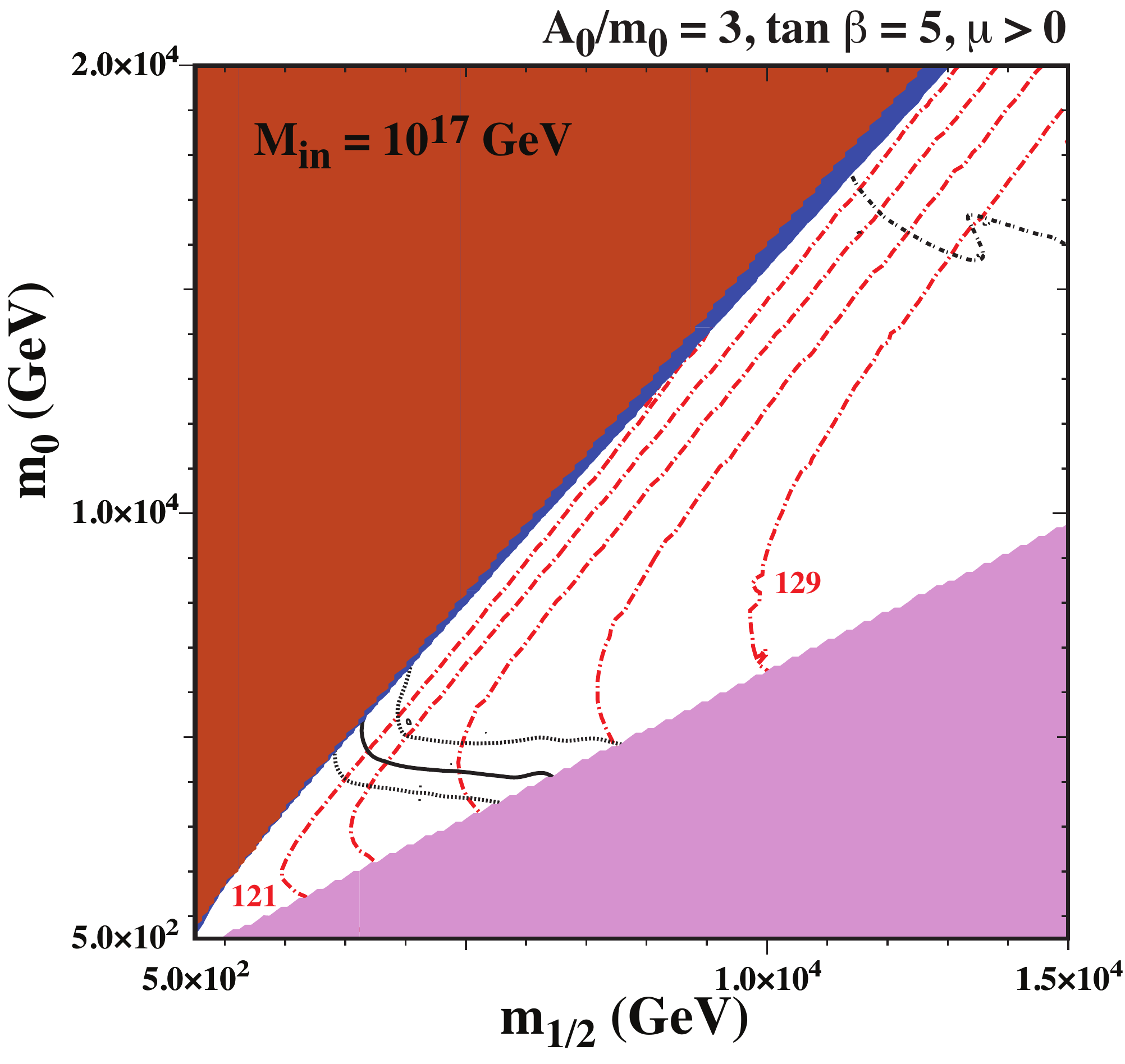}
\includegraphics[height=3.in]{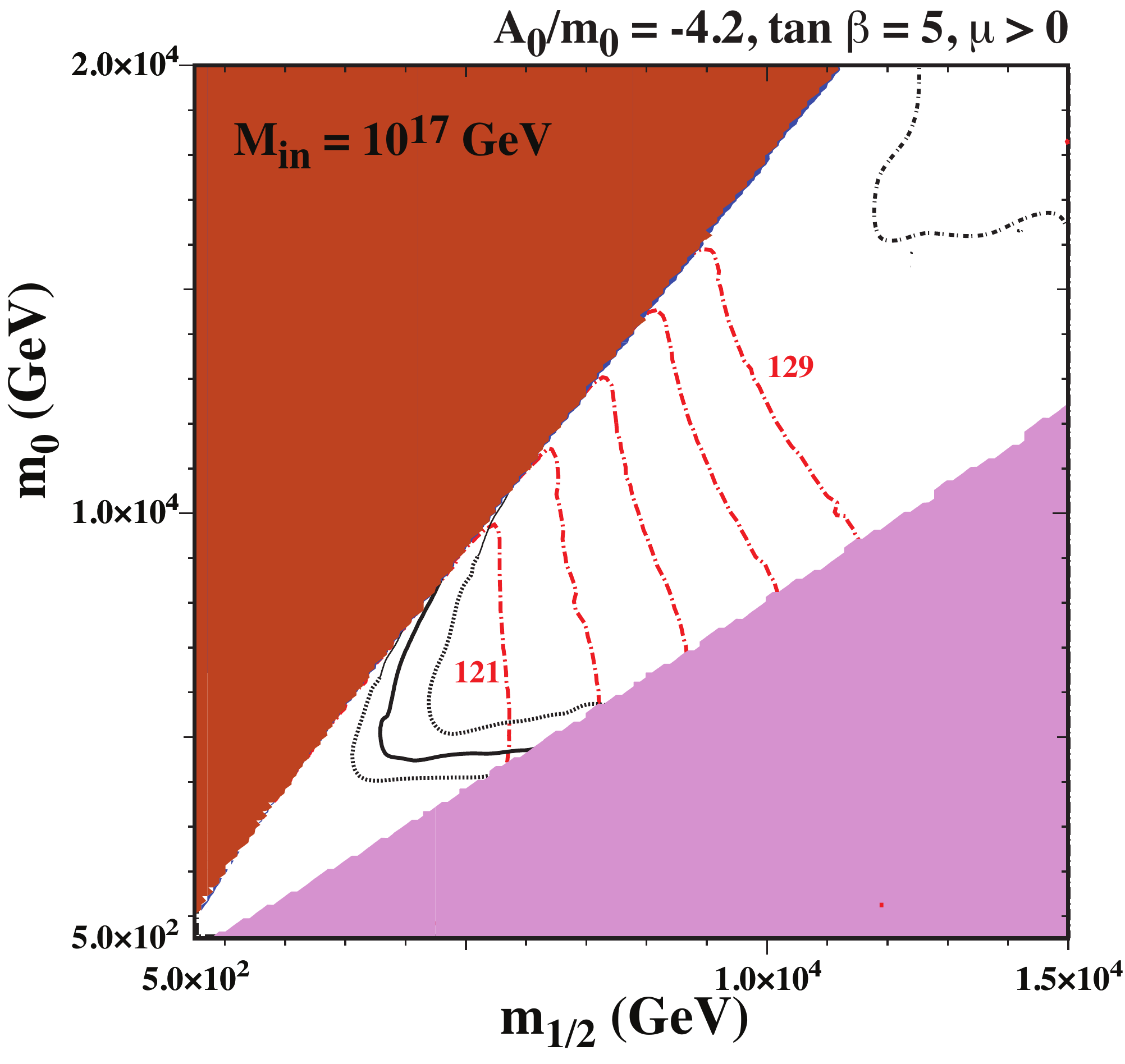}
\end{minipage}
\begin{minipage}{8in}
\includegraphics[height=3.in]{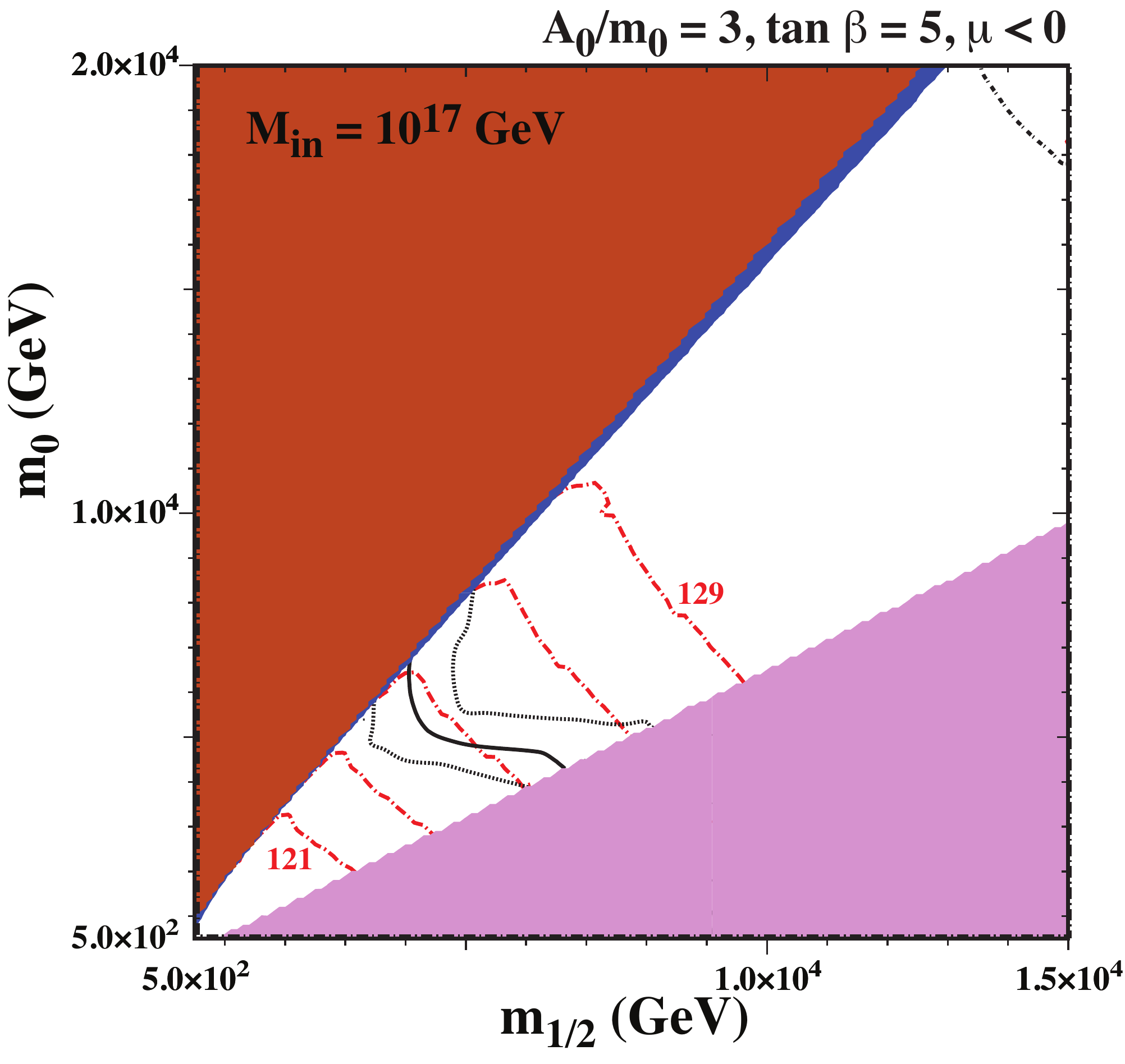}
\includegraphics[height=3.in]{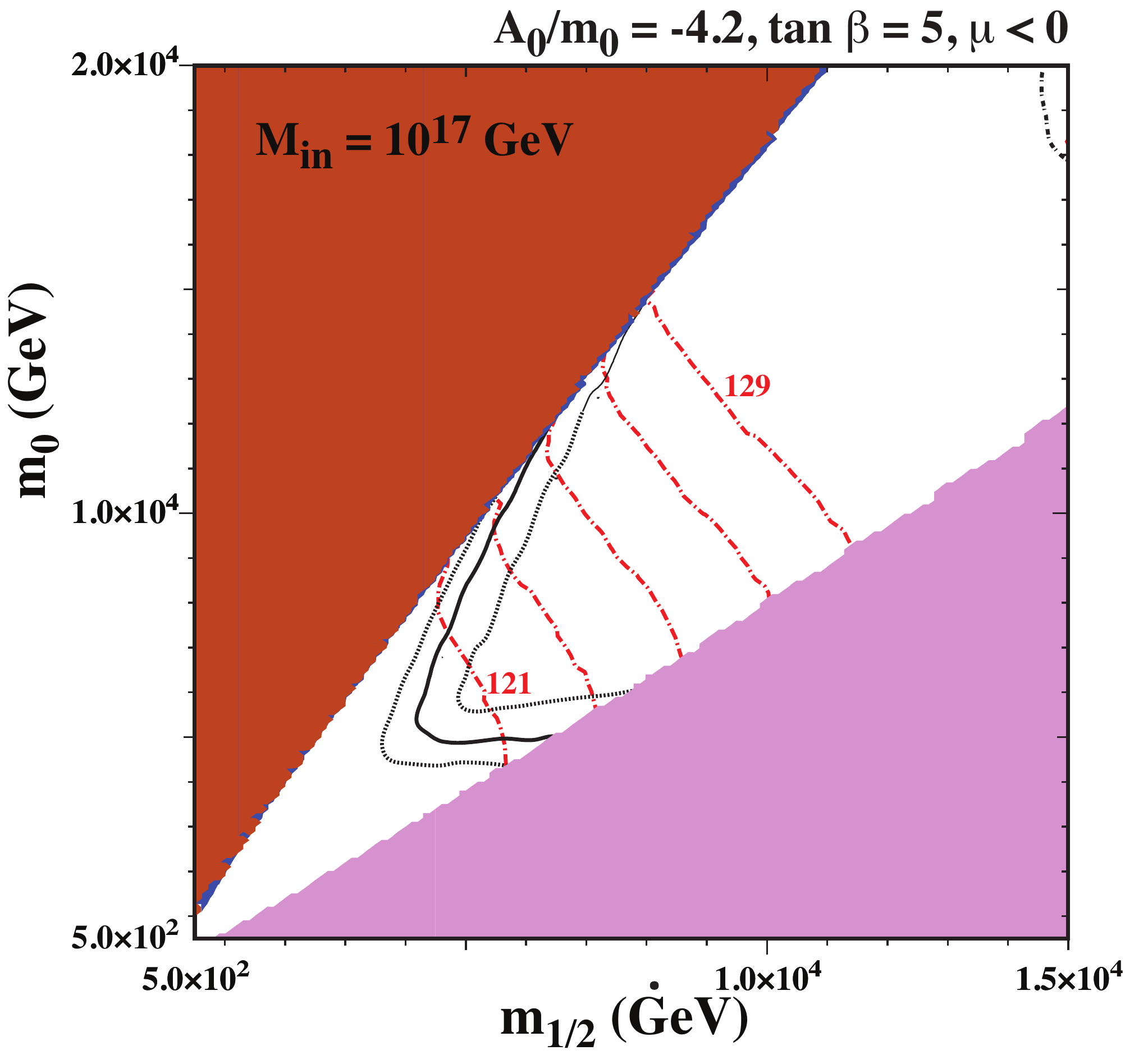}
\end{minipage}
\caption{
{\it
Some $(m_{1/2}, m_0)$ planes in the super-GUT model with $M_{\rm in} = 10^{17}$~GeV for $\tan \beta = 5$, $\mu > 0$ (upper panels), $\mu < 0$ (lower panels), $A_0/m_0 = 3$ (left panels), $A_0/m_0 = - 4.2$ (right panels). The black lines are contours of the $p \to K^+ \bar{\nu}$ lifetime, as calculated varying the GUT phases to minimize this decay rate, using the central parameter values (solid) and their combined 1-$\sigma$ variations (dotted). The DUNE discovery sensitivity is shown by the black dot-dashed curves. 
 The red dot dashed lines are the indicated contours of $m_h$.
}}
\label{fig:super-GUT}
\end{figure}

For better comparison with the CMSSM results in the previous Section,
we choose the same input values of $\tan \beta$ and $A_0$, and our illustrative super-GUT $(m_{1/2}, m_0)$ planes are shown in Fig.~\ref{fig:super-GUT}. In each of the planes, we see pink shaded regions where the minimization of the Higgs potential fails to provide a solution for $\mu$. In the upper left panel, where $\tan \beta = 5$, $A_0/m_0 = 3$ and $\mu > 0$, we see that the proton lifetime sensitivity is relatively weak and in most of the stop
coannihilation strip (shaded blue) the proton lifetime is sufficiently long. 
DUNE will be able to explore much of this strip, extending to $m_0 \sim 17$ TeV if proton decay is not seen. However, for this choice of parameters, the 125 GeV Higgs mass contour does not intersect the stop coannihilation region, which extends beyond the range shown. For this reason, no entry is given in Table \ref{tab:re} for this case.  However, we note that the 123 GeV Higgs mass contour, which is acceptable given the uncertainties in the calculation of the Higgs mass, does intersect the stop coannihilation strip.

In contrast, the 125 GeV Higgs contours do intersect 
the relic density strips in all the three other panels of Fig.~\ref{fig:super-GUT}, and the locations of these intersections are summarized in Table \ref{tab:re}. In the upper right panel, with $A_0/m_0 = -4.2$ and $\mu > 0$, the lifetime is well beyond the current limit, but well within the reach of DUNE. In the lower two panels, with $\mu < 0$,
the current limit on the proton
lifetime intersects the stop coannihilation strip for values
of the Higgs mass that are consistent with experiment, with the calculational uncertainties. 

The proton lifetime profiles
as a function of $m_{1/2}$ along the stop coannihilation strip for the two
super-GUT models with $\tan \beta = 5$ and $A_0/m_0 = 3$ are shown in the two lower panels of  Fig.~\ref{fig:strips}. As $c \ne 0$ is necessary to satisfy the boundary conditions, only one lifetime profile
is shown in each panel. When $\mu > 0$, as noted earlier, the Higgs mass is low for the range of $m_{1/2}$ shown, whereas $m_h$ is consistent with experiment for a range of $m_{1/2} \gtrsim 2$~TeV and $\lesssim 7$~TeV for $\mu < 0$. Much of this range is compatible with the current proton lifetime limit. For both signs of $\mu$, DUNE~\cite{DUNECDR} should be able to explore most of the range of $m_{1/2} \lesssim 15$~TeV. 

\begin{figure}[htb!]
\centering
\begin{minipage}{4.5in}
\hspace*{0.5in}
\includegraphics[height=3in]{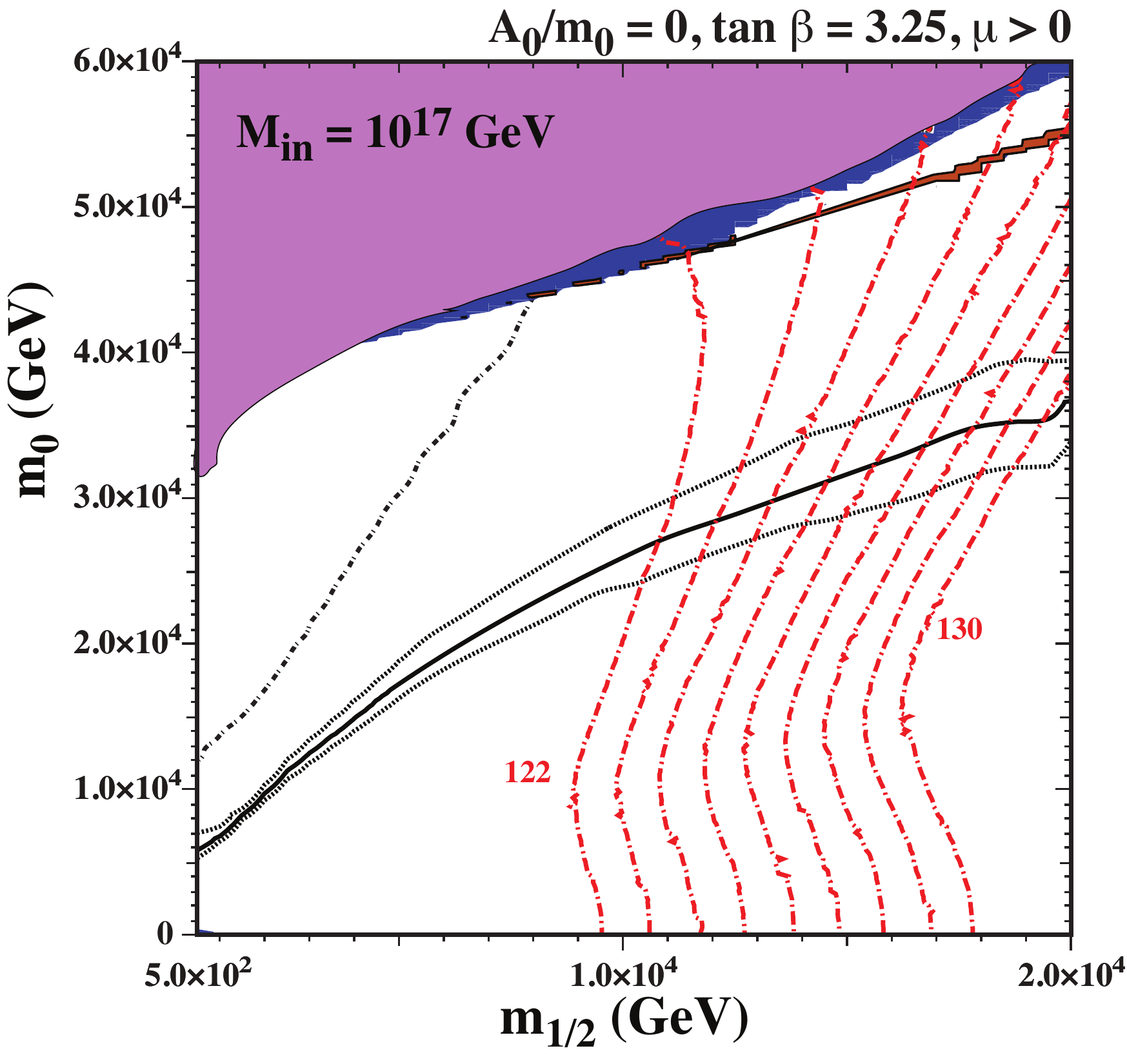}
\hfill
\end{minipage}
\caption{
{\it The $(m_{1/2}, m_0)$ plane in the super-GUT model with $M_{\rm in} = 10^{17}$~GeV for $\tan \beta = 3.25$, $A_0/m_0 = 0$ and $\mu > 0$, which exhibits a focus-point strip at large $m_0$.}}
\label{fig:super-GUTFP}
\end{figure}

We also show in Fig.~\ref{fig:super-GUTFP} one example of a super-GUT model with $\tan \beta = 3.25, A_0 = 0, M_{\rm in} = 10^{17}$~GeV and $\mu > 0$, which exhibits a focus-point strip at very high $m_0$. For this Figure, we again take $\lambda=0.1$ and $\lambda^\prime=0.0001$. The dot-dashed line corresponds to the DUNE discovery sensitivity. The region that cannot be probed by the DUNE experiment is for large sfermion masses and small Higgsino and Wino masses.  Since the proton lifetime scales as either the Higgsino mass or Wino mass squared, divided by the sfermion masses to the fourth power, the lifetime is very long in regions where the Higgsino and Wino masses are  small and the sfermion masses are large. However, the Higgs mass tends to be too light in these regions, even when considering the calculational uncertainties on the Higgs mass. Thus, DUNE should be able to probe all the viable parameter space in this illustrative example. 

\subsection{A Sub-GUT Model}
\label{sec:sub-GUT}

Our final example is a subGUT model with $M_{\rm in} = 10^{11}$ GeV. As in the CMSSM with $M_{\rm in} = M_{\rm GUT}$, in this case it is possible to set the dimension-five coupling $c = 0$, since the universality scale is below the GUT scale. The proton lifetime limit and its uncertainty for $c=0$ are shown by the solid black contours, and the $1-\sigma_{\rm tot}$ line is dotted. For $c \ne 0$, the limit and the $1\sigma_{\rm tot}$ uncertainties are in green and the proton decay constraint is significantly weaker. 

\begin{figure}[!htb]
\centering
\begin{minipage}{4.5in}
\hspace*{0.5in}
\includegraphics[height=3in]{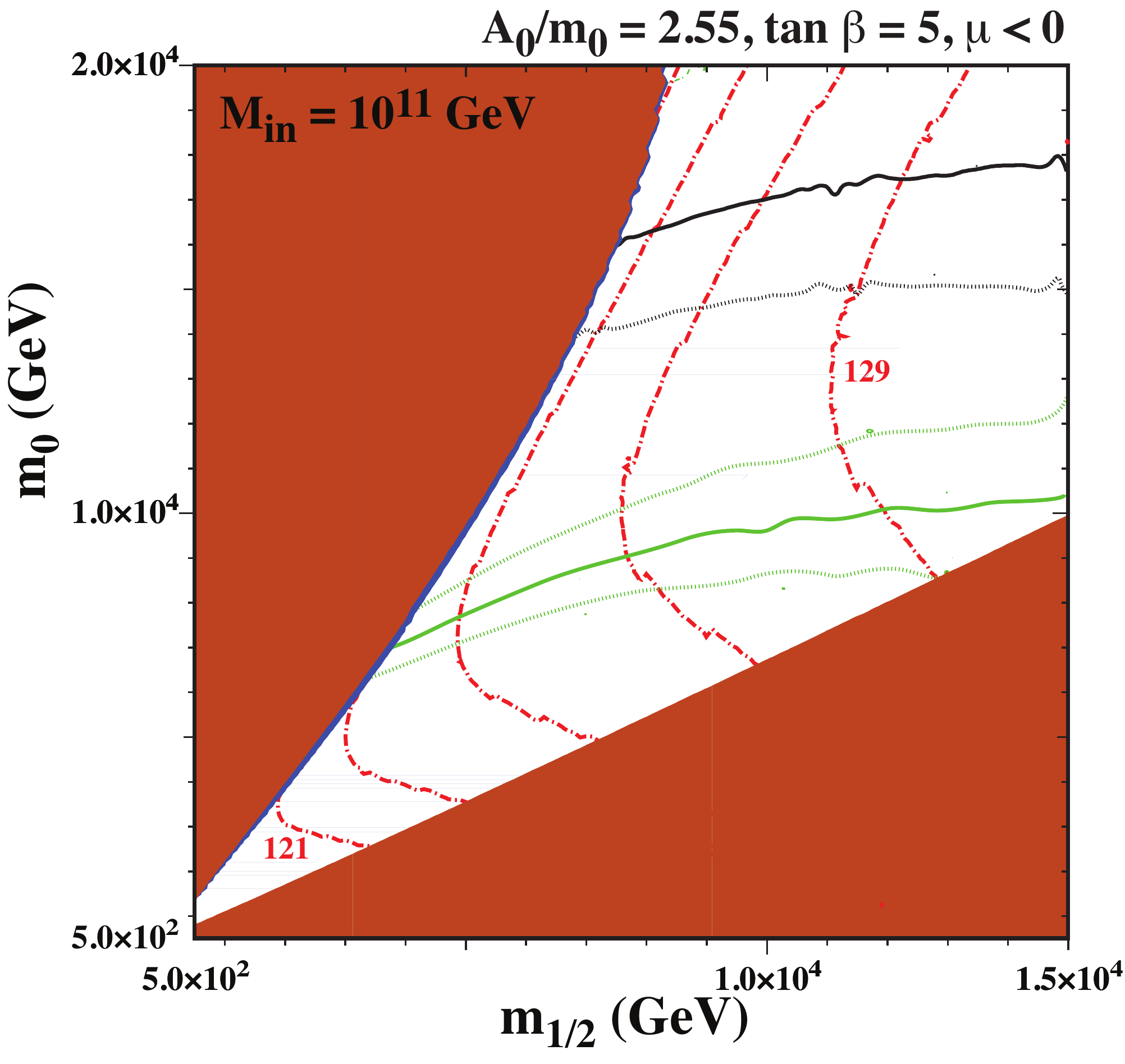}
\hfill
\end{minipage}
\caption{
{\it The $(m_{1/2}, m_0)$ plane in the sub-GUT model with $M_{\rm in} = 10^{11}$~GeV for $\tan \beta = 5$, $A_0/m_0 = 2.55$ and $\mu < 0$.}}
\label{fig:sub-GUT}
\end{figure}

\section{Summary and Discussion}
\label{sec:sowhat}

We have analyzed in this paper the uncertainties associated with various phenomenological inputs in the calculation of the nucleon lifetime. We have used the minimal SU(5) GUT for this analysis, motivated by its relative simplicity, but in full knowledge of its shortcomings and the existence of more attractive alternatives. The considerations we have developed here could also be applied to any other specific GUT model, e.g., flipped SU(5)~\cite{EGNNO5}.

We have found that the largest uncertainties are those associated with lattice calculations of hadronic matrix elements, which have recently found significant changes in central values and reduced errors, and the strong coupling $\alpha_s$. We have also stressed the importance of using the appropriate value of $\sin^2 \theta_W$ in GUT calculations, while noting that its present uncertainty is of lesser importance. The most important quark mass uncertainty is that associated with $m_s$, followed by $m_c$, and we stress the importance of including one-loop mass renormalization effects. The most important CKM mixing uncertainty is that associated with the Wolfenstein parameter $A$. However, much larger uncertainties are associated with the GUT phases that are not observable in electroweak interactions, which can modify not only the dominant $p \to K^+ \overline \nu$ decay rates, but also modify significantly the branching ratios for other decay modes such as $p \to \pi^+ \overline \nu$. However, our overall conclusion is that $p \to K^+ \overline \nu$ is the most promising decay mode for the next generation of underground detectors, particularly DUNE. We have also commented on the ambiguities in the proton decay predictions associated with the discrepancies between the masses of the charged leptons and charged-1/3 quarks, which warrant detailed study in specific models.

\begin{figure}
\centering
\includegraphics[height=6.cm]{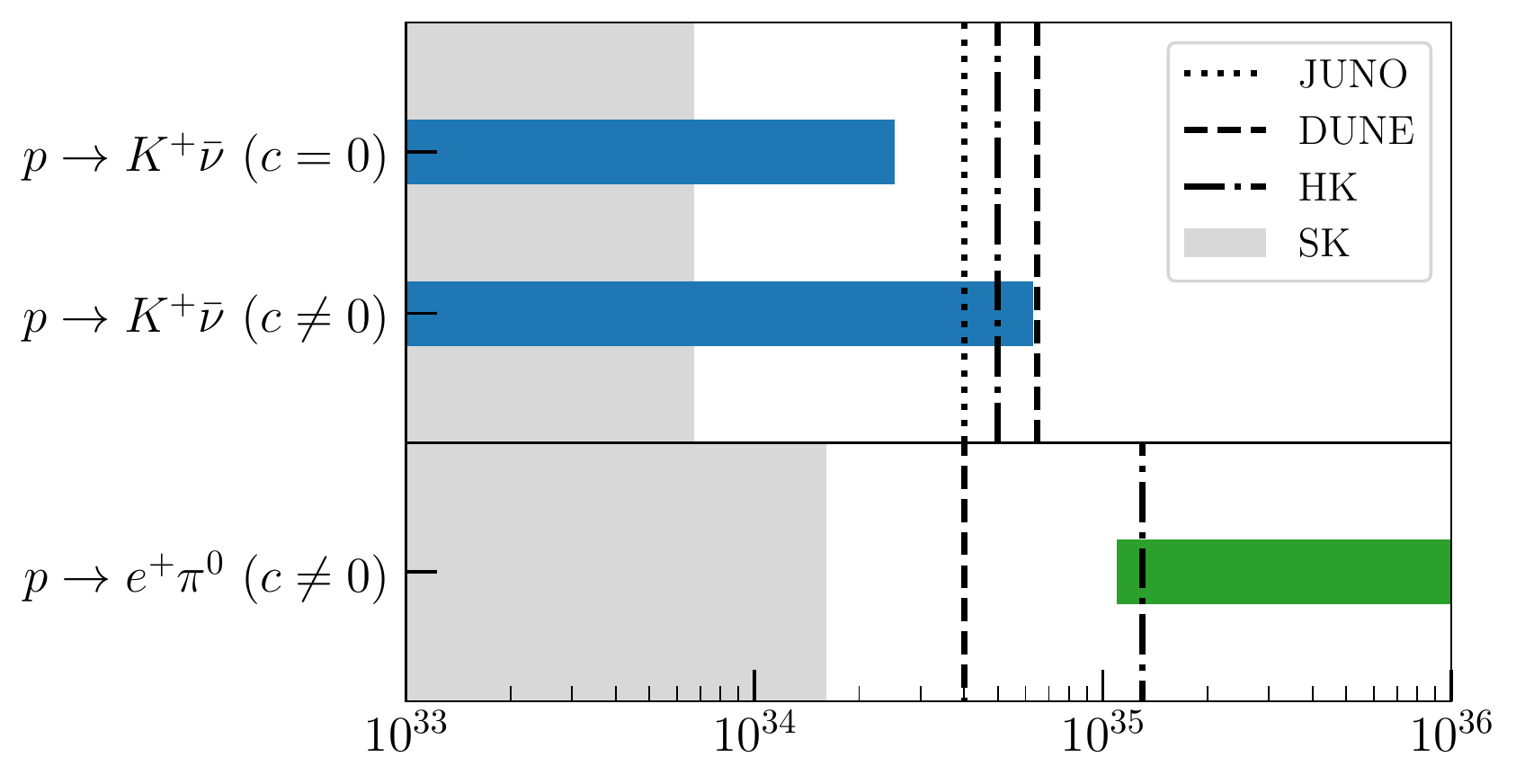}
\caption{
{\it 
The range of $p \to K^+ \overline \nu$ lifetimes found in the CMSSM (see Fig.~\ref{fig:strips}) for the cases $c = 0$ and $c \neq 0$ (blue bands) compared with the sensitivities of the JUNO, Hyper-K and DUNE experiments. We also show results for $p \to \pi^0 e^+$ in the CMSSM (see Fig.~\ref{fig:dim6}) with $c \neq 0$ (green band). The gray shaded areas are excluded by the Super-Kamiokande experiment~\cite{Takhistov:2016eqm,Miura:2016krn}. 
}}
\label{fig:Summary}
\end{figure}

In this paper we have applied our analysis to variants of the minimal supersymmetric GUT with universality of the soft supersymmetry-breaking parameters imposed at the GUT scale (the CMSSM), above it (super-GUTs) and below it (sub-GUTs). The uncertainties reviewed in the previous paragraph, combined with our lack of knowledge of the possible masses of supersymmetric particles, make it impossible to be specific about the nucleon lifetime, even in such well-defined models as those as we have studied. However, our analysis shows that in all these models large regions of model parameter space with sparticle masses $\lesssim {\cal O}(10)$~TeV can be explored with the upcoming generation of underground detectors. This is illustrated in Fig.~\ref{fig:Summary}, where we display the ranges of $p \to K^+ \overline \nu$ lifetimes found in the CMSSM (see Fig.~\ref{fig:strips}) for the cases $c = 0$ and $c \neq 0$ (blue bands) compared with the sensitivities of the JUNO, Hyper-K and DUNE experiments. The gray shaded area is excluded by the Super-Kamiokande experiment. We also show results for $p \to \pi^0 e^+$ in the CMSSM with $c \neq 0$ (green band), where we see that Hyper-K has some sensitivity, as discussed in the Appendix (see Fig.~\ref{fig:dim6}).

It will be interesting to apply the considerations presented here to other GUT models such as flipped SU(5), in which dimension-five baryon decay operators are absent, and the leading baryon-number-violating operators have dimension 6. Some remarks about dimension-6 proton decay are presented in the Appendix. We have also highlighted the ambiguities associated with models of the first- and second-generation quark and lepton masses. Interest in these and other issues in baryon decay will surely increase in the coming years as the start-up dates of JUNO, DUNE and Hyper-Kamiokande get closer. We trust that this paper will serve as a useful contribution to this coming trend.

\section*{Acknowledgements}

J.E. thanks Teppei Katori and Francesca Di Lodovico for useful discussions. The work of J.E. was supported in part by STFC (UK) via research grant ST/L000258/1,
and in part by the Estonian Research Council via a Mobilitas Pluss grant.
The work of N.N. was supported by the Grant-in-Aid for 
Young Scientists B (No.17K14270) and Innovative Areas (No.18H05542).
The work of K.A.O. was supported in part by DOE grant DE-SC0011842 
at the University of Minnesota. K.A.O. also
acknowledges support by the Director, Office of Science, 
Office of High Energy Physics of the U.S. Department of Energy under 
Contract No. DE-AC02-05CH11231, and would like to thank the 
Department of Physics and the high energy theory group
at the University of California, Berkeley as well as the theory group at LBNL
for their hospitality and financial support while finishing this work. 
L. V.  acknowledges the hospitality and the financial support from the 
Fine Theoretical Physics Institute at the University of Minnesota and 
from the Abdus Salam International Centre for Theoretical Physics,  
Italy, during various stages of this project. 

\section*{Appendix: Dimension-Six Proton Decay}
\label{sec:dim6protondecay}

We describe in this appendix the calculation of the rate of proton decay induced by the exchange of the SU(5) gauge bosons, just for completeness. In this case, the relevant effective interactions are expressed by the following dimension-six effective operators:
\begin{equation}
 {\cal L}_{6}^{\rm eff}
=C_{6(1)}^{ijkl}{\cal O}^{6(1)}_{ijkl}
+C_{6(2)}^{ijkl}{\cal O}^{6(2)}_{ijkl}
+{\rm h.c.}
~,
\end{equation}
where
\begin{align}
 {\cal O}^{6(1)}_{ijkl}&\equiv \int d^2\theta d^2\bar{\theta}~
\epsilon_{abc}\epsilon_{\alpha\beta}
\bigl(\overline{U}^\dagger_i\bigr)^a
\bigl(\overline{D}^\dagger _j\bigr)^b
e^{-\frac{2}{3}g^\prime B}
\bigl(e^{2g_3G}Q_k^\alpha\bigr)^cL^\beta_l~,
 \\
{\cal O}^{6(2)}_{ijkl}&\equiv \int d^2\theta d^2\bar{\theta}
\epsilon_{abc}\epsilon_{\alpha\beta}~
Q^{a\alpha}_iQ^{b\beta}_j
e^{\frac{2}{3}g^\prime B}
\bigl(e^{-2g_3G}\overline{U}^\dagger _k\bigr)^c
\overline{E}^\dagger _l~,
\end{align}
and their Wilson coefficients are 
\begin{align}
 C^{ijkl}_{6(1)}&=-\frac{g_5^2}{M_X^2}e^{i\varphi_i}\delta^{ik}
\delta^{jl}~,
\nonumber \\
 C^{ijkl}_{6(2)}&=-\frac{g_5^2}{M_X^2}e^{i\varphi_i}
\delta^{ik}(V^*)^{jl}~.
\label{eq:dim6gutmatch}
\end{align}
We note that these coefficients have the identical phase factor, $e^{i \varphi_i}$. As a result, this phase factor affects only the overall phase of the decay amplitude, and thus the decay rate is independent of this phase. 

At the one-loop level,~\footnote{The two-loop RGEs are given in Ref.~\cite{Hisano:2013ege}.} the RGEs of these coefficients can easily be solved \cite{Munoz:1986kq, Abbott:1980zj}. The coefficients are then matched at the electroweak scale onto the effective operators 
\begin{align}
 {\cal L}(p\to \pi^0 l^+_i)
&=C_{RL}(udul_i)\bigl[\epsilon_{abc}(u_R^ad_R^b)(u_L^cl_{Li}^{})\bigr]
+C_{LR}(udul_i)\bigl[\epsilon_{abc}(u_L^ad_L^b)(u_R^cl_{Ri}^{})\bigr]
~,
\end{align}
where 
\begin{align}
 C_{RL}(udul_i)&=C^{111i}_{6(1)}(M_Z)~, \nonumber \\
 C_{LR}(udul_i)&=V_{j1}\bigl[
C^{1j1i}_{6(2)}(M_Z)+C^{j11i}_{6(2)}(M_Z)
\bigr]~.
\end{align}
We again use the two-loop results given in Ref.~\cite{Nihei:1994tx} for the QCD RGEs. The partial decay width for $p \to e^+ \pi^0$ is then given by
\begin{equation}
 \Gamma (p\to  \pi^0 e^+)=
\frac{m_p}{32\pi}\biggl(1-\frac{m_\pi^2}{m_p^2}\biggr)^2
\bigl[
\vert {\cal A}_L(p\to \pi^0 e^+) \vert^2+
\vert {\cal A}_R(p\to \pi^0 e^+) \vert^2
\bigr]~,
\end{equation}
with 
\begin{align}
 {\cal A}_L(p\to \pi^0 e^+)&=
- \frac{g_5^2}{M_X^2} \cdot
A_1 \cdot \langle \pi^0\vert (ud)_Ru_L\vert p\rangle
~,\nonumber \\
 {\cal A}_R(p\to \pi^0 e^+)&=
- \frac{g_5^2}{M_X^2} (1+|V_{ud}|^2) \cdot
A_2 \cdot \langle \pi^0\vert (ud)_Ru_L\vert p\rangle
~,
\end{align}
where $A_1$ and $A_2$ are the renormalization factors: 
\begin{align}
 A_1 &=
A_L \cdot \biggl[
\frac{\alpha_3(M_{\text{SUSY}})}{\alpha_3(M_{\rm GUT})}
\biggr]^{\frac{4}{9}}
\biggl[
\frac{\alpha_2(M_{\text{SUSY}})}{\alpha_2(M_{\rm GUT})}
\biggr]^{-\frac{3}{2}}
\biggl[
\frac{\alpha_1(M_{\text{SUSY}})}{\alpha_1(M_{\rm GUT})}
\biggr]^{-\frac{1}{18}}
\nonumber \\
&\times
\biggl[
\frac{\alpha_3(M_Z)}{\alpha_3(M_{\rm SUSY})}
\biggr]^{\frac{2}{7}}
\biggl[
\frac{\alpha_2(M_Z)}{\alpha_2(M_{\rm SUSY})}
\biggr]^{\frac{27}{38}}
\biggl[
\frac{\alpha_1(M_Z)}{\alpha_1(M_{\rm SUSY})}
\biggr]^{-\frac{11}{82}} ~, \nonumber \\
 A_2 &=
A_L \cdot \biggl[
\frac{\alpha_3(M_{\text{SUSY}})}{\alpha_3(M_{\rm GUT})}
\biggr]^{\frac{4}{9}}
\biggl[
\frac{\alpha_2(M_{\text{SUSY}})}{\alpha_2(M_{\rm GUT})}
\biggr]^{-\frac{3}{2}}
\biggl[
\frac{\alpha_1(M_{\text{SUSY}})}{\alpha_1(M_{\rm GUT})}
\biggr]^{-\frac{23}{198}}
\nonumber \\
&\times
\biggl[
\frac{\alpha_3(M_Z)}{\alpha_3(M_{\rm SUSY})}
\biggr]^{\frac{2}{7}}
\biggl[
\frac{\alpha_2(M_Z)}{\alpha_2(M_{\rm SUSY})}
\biggr]^{\frac{27}{38}}
\biggl[
\frac{\alpha_1(M_Z)}{\alpha_1(M_{\rm SUSY})}
\biggr]^{-\frac{23}{82}} ~,
\end{align}
with $A_L = 1.25$ the long-distance QCD renormalization factor \cite{Nihei:1994tx}.

Dimension-six proton decay suffers from less uncertainty than the dimension-five proton decay discussed in the main text. First, as mentioned above, the decay rate of this process does not depend on the GUT phases $\varphi_i$. Secondly, the Wilson coefficients at low energies do not depend explicitly on the masses of supersymmetric particles, in contrast to those of the dimension-five proton decay operators. Finally, the mass of the SU(5) gauge boson $M_X$, on which the Wilson coefficients depend directly, can also be determined through the GUT threshold corrections in Eqs.~(\ref{eq:matchmhc}--\ref{eq:matchg5}), since it can be expressed as
\begin{equation}
    M_X = \biggl(\frac{2 g_5}{\lambda^\prime}\biggr)^{\frac{1}{3}} 
    \left(M_X^2 M_\Sigma\right)^{\frac{1}{3}}  ~,
\end{equation}
and the factor $\left(M_X^2 M_\Sigma\right)^{\frac{1}{3}} $ is obtained from Eq.~\eqref{eq:matchmgut}. The error in the hadronic matrix element $\langle \pi^0\vert (ud)_Ru_L\vert p\rangle$ gives approximately 20\% uncertainty in the decay rate, as can be seen from Table~\ref{tab:w0_error}. On the other hand, the uncertainty due to the error in the strong gauge coupling constant can be estimated in the same manner as done in Eq.~\eqref{withe}: 
\begin{align}
    \sigma_{\tau (p \to e^+\pi^0)}
    & \simeq  \tau (p \to e^+\pi^0) \biggl(\frac{4\pi}{9}\biggr) \biggl(\frac{\Delta_{\alpha_s}}{\alpha_s(M_Z)^2}\biggr)
    \nonumber \\
    &= 0.11 \left(\frac{\Delta_{\alpha_s}}{0.0011}\right)\left(\frac{0.1181}{\alpha_s(M_Z)} \right)^2 \tau (p \to e^+\pi^0) ~.
\end{align}

\begin{figure}
\centering
\includegraphics[height=8.cm]{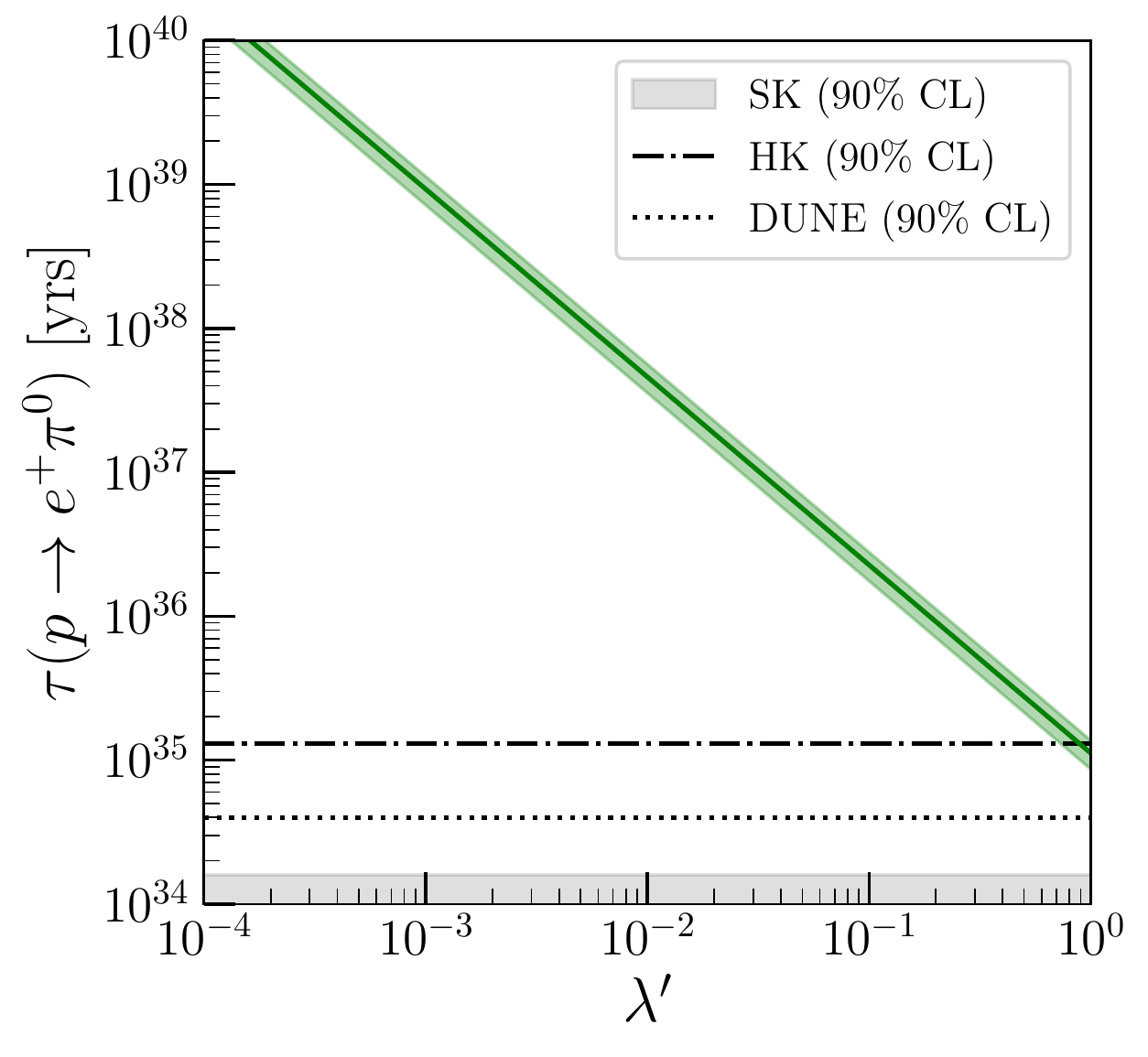}
\caption{
{\it 
Dimension-six proton decay lifetime $\tau (p \to e^+ \pi^0)$ as a function of $\lambda^\prime$ for $m_{1/2} = 9.8$~TeV, $m_0 = 14.1$~TeV, $\tan \beta = 5$, $A_0/m_0 = 3$, $\mu > 0$, with the green band showing the uncertainty. The horizontal dash-dotted and dotted lines show the 20-year 90~\% CL sensitivity of Hyper-K and DUNE, respectively, while the gray shaded area is excluded by Super-Kamiokande.
}}
\label{fig:dim6}
\end{figure}

We show in Fig.~\ref{fig:dim6} the dimension-six proton decay lifetime $\tau (p \to e^+ \pi^0)$ in the CMSSM as a function of $\lambda^\prime$, where we set $m_{1/2} = 9.8$~TeV, $m_0 = 14.1$~TeV, $\tan \beta = 5$, $A_0/m_0 = 3$, $\mu > 0$. The horizontal dash-dotted and dotted lines show the 20-year 90~\% CL sensitivity of Hyper-K and DUNE, respectively, while the gray shaded area is excluded by Super-Kamiokande. We see that the predicted lifetime is well above the current limit imposed by the Super-Kamiokande experiment. Nevertheless, it may be within the reach of the Hyper-Kamiokande experiment if $\lambda^\prime = {\cal O} (1)$. In general, we find that $\tau (p \to e^+ \pi^0)$ is well approximated by
\begin{equation}
    \tau (p \to e^+ \pi^0) \simeq 1.8 \times 10^{35}
    \times \biggl(\frac{M_X}{10^{16}~{\rm GeV}}\biggr)^4 ~.
\end{equation}
This expression shows that $p \to e^+ \pi^0$ can be probed at Hyper-Kamiokande if $M_X \lesssim 10^{16}$~GeV.


\end{document}